\documentclass{LMCS}
\usepackage{latexsym, amsfonts, amstext, amssymb, graphicx, subfigure, amsmath}

\usepackage[mathcal]{euscript}

\usepackage[matrix,arrow,frame,dvips,curve]{xy}
\usepackage{enumerate,hyperref}
\newcommand{\degree}[1]{\ensuremath{deg}({#1})}

\newcommand{\inp}[1]{\ensuremath{\mathsf{in}({#1})}}
\newcommand{\out}[1]{\ensuremath{\mathsf{out}({#1})}}
\newcommand{\init}[1]{\hat{#1}}

\newcommand{\mjoin}{\ensuremath{\uplus}}

\newcommand{\mon}[1]{\ensuremath{{#1}^\oplus}}
\newcommand{\monSub}[2]{\ensuremath{{#1}_{#2}^\oplus}}

\newcommand{\res}[2]{\ensuremath{({#1}\!\downarrow\!{#2})}}

\newcommand{\sres}[2]{\ensuremath{({#1}\!\Downarrow\!{#2})}}

\newcommand{\comp}[1]{\ensuremath{+}_{#1}}

\newcommand{\src}{\ensuremath{\sigma}}
\newcommand{\trg}{\ensuremath{\tau}}

\newcommand{\pre}[1][(\cdot)]{\ensuremath{\!~^\bullet{#1}}}

\newcommand{\post}[1][(\cdot)]{\ensuremath{{#1} {^\bullet}}}

\newcommand{\trans}[1]{\ensuremath{\ [\/{#1}\/\rangle}\ }

\newcommand{\onet}{\ensuremath{\mathbf{ONet}}}

\newcommand{\uonet}{\ensuremath{\mathbf{ONet}^u}}

\newcommand{\net}{\ensuremath{\mathbf{Net}}}

\newcommand{\fo}{\ensuremath{\mathcal{F}}}

\newcommand{\ltr}[3][x]{\ensuremath{\stackrel{{#3}}{\longrightarrow}_{\mathsf{#1},#2}}}

\newcommand{\wltr}[3][x]{\ensuremath{\stackrel{{#3}}{\leadsto}_{\mathsf{#1},#2}}}

\newcommand{\wltrStar}[3][x]{\ensuremath{\stackrel{{#3}}{\leadsto}^*_{\mathsf{#1},#2}}}

\newcommand{\Ltr}[3][x]{\ensuremath{\stackrel{{#3}}{\Longrightarrow}_{\mathsf{#1},#2}}}

\newcommand{\rew}[1]{\Rightarrow^{#1}}

\newcommand{\urlink}[1]{\url{#1}} %{\htmladdnormallink{\url{#1}}{#1}}
\newcommand{\localfile}[1]
{\url{http://www.uni-paderborn.de/cs/ag-engels/Papers/#1}}

\newcommand{\tfspages}[1]{\url{http://tfs.cs.tu-berlin.de}}

\def\doi{4 (4:3) 2008}
\lmcsheading%
{\doi}
{1--41}
{}
{}
{Dec.~23, 2007}
{Oct.~11, 2008}
{}   

\begin{document}

\title[Bisimilarity and Behaviour-Preserving Reconfigurations of
  Petri Nets]{Bisimilarity and Behaviour-Preserving Reconfigurations of Open
  Petri Nets}

\author[P. Baldan]{Paolo Baldan\rsuper a} 
\address{{\lsuper a}Dipartimento di Matematica Pura e Applicata, 
  Universit\`{a} di Padova, Italy}
\email{baldan@math.unipd.it}

\author[A. Corradini]{Andrea Corradini\rsuper b}
\address{{\lsuper b}Dipartimento di Informatica, Universit\`{a} di Pisa, Italy}
\email{andrea@di.unipi.it}

\author[H. Ehrig]{Hartmut Ehrig\rsuper c}
\address{{\lsuper c}Institut f\"ur Softwaretechnik und Theoretische Informatik,
  Technische~Universit\"at Berlin, Germany}
\email{ehrig@cs.tu-berlin.de}

\author[R. Heckel]{Reiko Heckel\rsuper d}
\address{{\lsuper d}Department of Computer Science, University of Leicester, UK}
\email{reiko@mcs.le.ac.uk}

\author[B. K\"onig]{Barbara K\"onig\rsuper e}
\address{{\lsuper e}Abteilung f\"ur Informatik und Angewandte 
  Kognitionswissenschaft,
  Universit{\"{a}t} Duisburg-Essen, Germany}
\email{barbara\_koenig@uni-due.de}

\thanks{Research partially supported by the EU
    IST-2004-16004 {\sc SEnSOria}, the MIUR Project ART, the DFG
    project SANDS, the DFG project Behaviour-GT and CRUI/DAAD
    \textsc{Vigoni} ``Models based on Graph Transformation Systems:
    Analysis and Verification''.}

% \subjclass{MANDATORY list of acm classifications}
% \titlecomment{OPTIONAL comment concerning the title, \eg, if a variant
% or an extended abstract of the paper has appeared elsewehere}

\begin{abstract}
  We propose a framework for the specification of behaviour-preserving
  reconfigurations of systems modelled as Petri nets. The framework is
  based on open nets, a mild generalisation of ordinary
  Place/Transition nets suited to model open systems which might
  interact with the surrounding environment and endowed with a
  colimit-based composition operation. We show that natural notions of
  bisimilarity over open nets are congruences with respect to the
  composition operation.  
  The considered behavioural equivalences differ for the choice
  of the observations, which can be single firings or parallel
  steps. Additionally, we consider weak forms of such
  equivalences, arising in the presence of unobservable actions.
  We also provide an up-to technique for facilitating bisimilarity
  proofs. The theory is used to identify suitable classes of
  reconfiguration rules (in the double-pushout approach to rewriting)
  whose application preserves the observational semantics of the net.
\end{abstract}

\maketitle

\section*{Introduction}

Petri nets are a well-known model of concurrent and distributed
systems, widely used both in theoretical and applicative areas.
In classical approaches, such as~\cite{Rei:PNI}, nets are
intended to represent closed, completely specified systems evolving
autonomously through the firing of transitions.
In order to represent \emph{open} systems, namely systems which can
interact with the surrounding environment or, from a different
perspective, systems which are only partially specified, several
extensions of the basic model of Petri nets have been considered in
the literature. Conceptually, this effort dates back to the early
works on net composition and refinement and to the studies concerning
the development of compositional semantics for Petri nets (a
discussion of the related literature can be found in the concluding
section).

%% ordinary Petri nets do not support directly certain
%% features that are needed to model \emph{open}
%% systems,  namely systems which can interact with the surrounding environment
%% or, in a different view, systems which are only partially specified.

Generally speaking, important issues that must be faced when modelling
open systems can be summarised as follows.
Firstly, a large (possibly still open) system is typically built out
of smaller open components.  Syntactically, an open system is equipped
with suitable interfaces, over which the interaction with the external
environment can take place.  Semantically, openness can be represented
by defining the behaviour of a component as if it were embedded in
general environments, determining any possible interaction over the
interfaces.

Secondly, often the building components of an open system are not
statically determined, but they can change during the evolution of the
system, according to predefined reconfiguration rules triggered by
internal or external solicitations.

% When two or more open components are combined, for each of them it is
% like if the surrounding unknown environment were getting more
% specified (as a kind of refinement).
%
% Then, the composition of two systems would be performed along the
% interfaces, thus restricts their possible internal behaviours and that
% the behaviour of the resulting composed system can be obtained
% compositionally by combining the single behaviours.

The work in this paper outlines a framework where open systems can be
modelled as Petri nets, capturing both the requirements mentioned
above. Observational semantics based on (weak) bisimulation are shown
to be congruences with respect to the composition operation defined
over Petri nets. Building on this, suitable reconfigurations of such
systems can be specified as net rewritings, which preserve the
behaviour of the system.
The relation with other approaches in the literature addressing similar
issues will be discusses in Section~\ref{se:conclusion}.

The framework presented here is based on so-called \emph{open nets}, a
mild generalisation of ordinary Petri nets introduced
in~\cite{BCEH:CMRS,BCEH:CSOP} to answer the first of the requirements
above, i.e., the possibility of interacting with the environment and
of composing a larger net out of smaller open components. An open net
is an ordinary net with a distinguished set of places, designated as
open, through which the net can interact with the surrounding
environment. As a consequence of such interaction, tokens can be
freely generated and removed in open places.
In the mentioned papers open nets are endowed with a composition
operation, characterised as a pushout in the corresponding category,
suitable to model both interaction through open places and
synchronisation of transitions.  
%
% A deterministic process semantics \emph{\`a la} Goltz-Reisig is shown to
% be compositional with respect to such composition operation.

In the first part of the paper, after having extended the existing
theory for open nets to deal with \emph{marked} nets,
% , i.e., open nets with
% a distinguished initial marking. The existing theory of open nets,
% including the pushout-based composition operation, is extended to the
% marked case and a compositionality result for step sequences is proved
% in this context.
% Next
we introduce bisimulation-based observational equivalences for open 
nets.
Following a common intuition about reactive systems (see,
e.g.,~\cite{v:modular-petri,NPS:CBCP} or the recent~\cite{LM:DBCRS})
such equivalences are based on the observation of the interactions
between the given net and the surrounding environment.
The framework treats uniformly \emph{strong
  bisimilarity}, where every transition firing is observed, and
\emph{weak bisimilarity}, where a subset of unobservable transition
labels is fixed (corresponding to $\tau$-actions in process calculi)
and the firings of transitions carrying such labels are considered
invisible.
We also consider \emph{step bisimilarity} (see,
e.g.,~\cite{Vog:BAR,NT:DNDC}), obtained by taking as
observations possibly parallel steps rather than single firings of
transitions, thus capturing, to some extent, the concurrency
properties of the system.

The considered notions of bisimilarity are shown to be congruences
with respect to the composition operation over open
nets. Interestingly enough, this holds also when the set of
non-observable labels is not empty, i.e., for weak bisimilarities: some
natural questions regarding the relation with weak bisimilarity in CCS
are addressed. In addition, we propose an up-to technique for
facilitating bisimilarity proofs.

Exploiting the results in the first part of the paper we next introduce a
framework for open net reconfigurations.
The fact that open net components are combined by means of pushouts
naturally suggests a setting for specifying net
reconfigurations, based on double-pushout (DPO)
rewriting~\cite{Ehr:TIAA}.
Using the congruence result for bisimilarity we identify classes of
transformation rules which ensure that reconfigurations of the system
do not affect its observational behaviour.

In order to understand this paper some basic knowledge of category theory
(see for instance~\cite{p:basic-category-theory}) is required.

\section{Marked Open Nets}
\label{se:open-nets}

An \emph{open net}, as introduced in~\cite{BCEH:CMRS,BCEH:CSOP}, is an
ordinary P/T Petri net with a distinguished set of 
\emph{open places}, which represent the interface through which the
environment can interact with the net. 
% of the net towards the environment, which
% can interact with the net by adding or removing some tokens in the open
% places.  Concretely, a
An open place can be an \emph{input place}, meaning that the
environment can put tokens into it, or an \emph{output place}, 
from which the environment can remove tokens, or both. 
In this section we introduce the basic notions for open nets as
presented in~\cite{BCEH:CSOP}, generalising them to nets with initial
marking:  this will be needed in the treatment of bisimilarity in
Section~\ref{se:bisim}.  

Given a set $X$ we write $\mathbf{2}^{X}$ for the powerset of $X$ and
$\mon{X}$ for the free commutative monoid over $X$, with monoid
operation $\oplus$, whose elements will be referred as
\emph{multisets} over $X$.  Moreover, given a function $h : X \to Y$
we denote by the same symbol $h : \mathbf{2}^{X} \to \mathbf{2}^{Y}$
its extension to sets, and by $\mon{h} : \mon{X} \to \mon{Y}$ its
monoidal extension.
Given a multiset $u \in \mon{X}$, with $u = \bigoplus_{x \in X} u_x
\cdot x$, for $x \in X$ we will write $u(x)$ to denote the coefficient
$u_x$. With little abuse of notation, we will write $x \in u$
iff $u(x) \geq 1$. Given $u, v \in \mon{X}$ we write $u \leq v$ when $u(x) \leq
v(x)$ for any $x \in X$. In this case the \emph{multiset difference}
$v \ominus u$ is the multiset $w$ such that $u \oplus w= v$. 
The symbol $0$ denotes the empty multiset.

\begin{defi}[multiset projection]
  \label{de:multiset-proj}
  Given a function $f : X \to Y$ and a multiset $u \in \mon{Y}$ we
  denote by $\res{u}{f} \in \mon{X}$ the \emph{projection of $u$ along
    $f$}, which is the multiset over $X$ defined as $\res{u}{f} =
  \bigoplus_{x \in X} u_{f(x)} \cdot x$.
\end{defi}
In other words, $\res{\_}{f} : \mon{Y} \to \mon{X}$ is the monoidal
extension of the function $\res{\_}{f} : Y \to \mon{X}$ defined by
$\res{y}{f} = x_1 \oplus \ldots \oplus x_n$ when $f^{-1}(y) = \{ x_1,
\ldots, x_n \}$.
For instance, given $f : \{ s_0, s_1, s_2 \} \to \{ s_1', s_2', s_3'\}$
such that $f(s_0)= f(s_1) = s_1'$ and $f(s_2) = s_2'$, we have $\res{2 s_1'
  \oplus s_2' \oplus s_3'}{f} = 2 s_0 \oplus 2 s_1 \oplus s_2$.
In the following we
will mainly work with injective
functions, for which the projection operation satisfies some expected
properties, such as $\mon{f}(\res{u}{f}) \le u$ and
$\res{\mon{f}(\res{u}{f})}{f} = \res{u}{f}$.

We consider nets where transitions are labelled over a fixed set of
labels $\Lambda$.

\begin{defi}[P/T Petri net]
  A \emph{P/T Petri net} is a tuple $N = (S, T, \src, \trg, \lambda)$
  where $S$ is the set of places, $T$ is the set of transitions (with
  $S \cap T = \emptyset$), $\src, \trg : T \to \mon{S}$ are functions
  mapping each transition to its pre- and post-set and $\lambda : T
  \to \Lambda$ is a labelling function for transitions.
\end{defi}
In the sequel we will denote by $\pre$ and $\post$ the monoidal
extensions of the functions $\src$ and $\trg$ to functions from
$\mon{T}$ to $\mon{S}$.
Moreover, given $s \in S$, the pre- and post-set of $s$ are
defined by $\pre[s] = \{ t \in T : s \in \post[t] \}$ and $\post[s]
= \{ t \in T : s \in \pre[t] \}$.

\begin{defi}[Petri net category]
  Let $N_0$ and $N_1$ be Petri nets. A \emph{Petri net morphism} $f :
  N_0 \to N_1$ is a pair of total functions $f = \langle f_T, f_S
  \rangle$ with $f_T : T_0 \to T_1$ and $f_S : S_0 \to S_1$, such that
  for all $t_0 \in T_0$, $\pre[f_T(t_0)] = \monSub{f}{S}(\pre[t_0])$,
  $\post[f_T(t_0)] = \monSub{f}{S}(\post[t_0])$ and $\lambda_1(f_T(t_0)) =
  \lambda_0(t_0)$.
  The category of P/T Petri nets and Petri net morphisms is denoted by
  $\net$.
\end{defi}
It is worth recalling that category $\net$ is a subcategory of the category
$\mathbf{Petri}$ of~\cite{MM:PNM}, which has the same objects, but
more general morphisms which can map a place to a multiset of
places.

We next introduce the notion of open net. As anticipated above,
differently from~\cite{BCEH:CMRS,BCEH:CSOP}, we work here with marked nets.
% This will be used in the treatment of bisimilarity.

\begin{figure}[t]
  \begin{center}
    \scalebox{.30}{\includegraphics{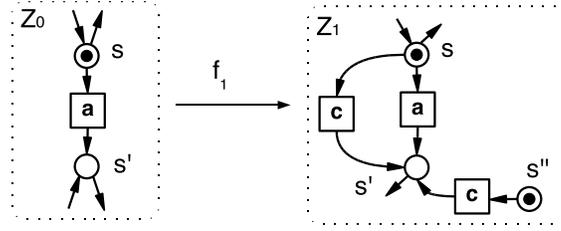}}
  \end{center}

  \caption{Two open nets and an open net morphism.}
  \label{fi:sample}
\end{figure}

\begin{defi}[open net]
  \label{de:open-net}
  An \emph{open net} is a pair $Z = (N_Z, O_Z)$, consisting of a P/T Petri net 
  %
  %\begin{itemize}
  %
  %\item 
  $N_Z = (S_Z, T_Z, \src_Z, \trg_Z, \lambda_Z)$ and a pair
  %
  %\item 
  $O_Z = (O_Z^+, O_Z^-) \in \mathbf{2}^{S_Z} \times \mathbf{2}^{S_Z}$,
  the sets of \emph{input open}, respectively, \emph{output open places}
  of the net.
  % 
  %\end{itemize}
  %
  A \emph{marked} open net is a pair $(Z, \init{u})$ where $Z$ is an
  open net and $\init{u} \in \monSub{S}{Z}$ is the initial marking.
\end{defi}
Hereafter, unless stated otherwise, all open nets will be assumed
implicitly to be marked.  An open net will be denoted simply by $Z$ and
the corresponding initial marking by $\init{u}$. Subscripts carry over to the
net components.
The graphical representation for open nets is similar to that for
standard nets. In addition, the fact that a place is input or
output open is represented by an ingoing or outgoing dangling arc, 
respectively. 
For instance, in net $Z_1$ of Fig.~\ref{fi:sample}, place
$s$ is both input and output open, while $s'$ is only output open.

%% Observe that the sets $O_Z^+$ and $O_Z^-$ are not necessarily
%% disjoint, hence a place can be both an input and an output open place
%% at the same time.

The notion of enabledness for transitions is the usual one, but
besides the changes produced by the firing of the transitions of the
net, we consider also the interaction with the environment which is
modelled by events, denoted by $+_s$ or $-_s$, which produce or
consume a token in an open place $s$. 
Such events corresponds to the pseudo-transitions
of~\cite{v:modular-petri} and to the transition in the universal
context of~\cite{NPS:CBCP}.

\begin{defi}[set of extended events]
  Let $Z$ be an open net.
  The \emph{set of extended events} of $Z$, denoted by $\bar{T}_Z$ and
  ranged over by $\epsilon$ is defined as
  \begin{center}
    $\bar{T}_Z = T_Z \cup \{ +_s : s \in O_Z^+ \} \cup \{-_s : s \in O_Z^- \}$.
  \end{center}
  Defining $\pre[+_s] = 0$ and $\post[+_s] = s$, and symmetrically,
  $\pre[-_s] = s$ and $\post[-_s] = 0$, the notion of pre- and
  post-set extends 
  to multisets of extended events.
\end{defi}

 Given a
marking $u \in \mon{O_Z^+}$, we  denote by $+_u$ the multiset
$\bigoplus_{s \in O_Z^+} u(s) \cdot +_s$.
%(provided that $u$ contains only input open places).  
Similarly, $-_u = \bigoplus_{s \in O_Z^-} u(s) \cdot -_s$ for $u \in
\mon{O_Z^-}$.   

\begin{defi}[firings and steps]
  Let $Z$ be an open net.  A \emph{step} in $Z$ consists of the execution of a
  multiset of (extended) events $A \in \monSub{\bar{T}}{Z}$, i.e.,
  \begin{center}
    $u \oplus \pre[A] \trans{A} u \oplus \post[A]$.
  \end{center}
  A step is called a \emph{firing} when $A$ consists of a single event, i.e.,
  $A = \epsilon \in \bar{T}_Z$.
\end{defi}
A firing can be (i) the execution of a
transition $u \oplus \pre[t] \trans{t} u \oplus \post[t]$, with $u \in
\monSub{S}{Z}$, $t \in T_Z$;
(ii) the creation of a token by the environment $u \trans{+_s} u \oplus s$,
with $u \in \monSub{S}{Z}$, $s \in O_Z^+$;
(iii) the deletion of a token by the environment $u \oplus s \trans{-_s} u$,
with $u \in \monSub{S}{Z}$, $s \in O_Z^-$.
A step is the execution of a multiset of transitions and
interactions with the environment, of the kind $A \oplus
-_{w} \oplus +_{v}$ for  $A \in \monSub{T}{Z}, w \in \mon{O^-_Z}$
and $v \in \mon{O^+_Z}$.

%% Firing sequences and step sequences starting from the initial marking are
%% defined in the obvious way.

% \begin{remark}
%   Observe that in an open net computation, differently from what
%   happens for grammars, the deletion of an item is never necessary for
%   the firing of a transition. Hence we can imagine that all the tokens
%   produced by the environment appear at the beginning of the
%   computation and that all the tokens consumed by the environment are
%   deleted at the end of the considered computation.
% \end{remark}

We now introduce suitable morphisms relating open nets, which are
morphisms between the underlying P/T nets, satisfying certain
conditions on the open places and on the initial marking. In
particular, given an injective morphism $f : Z_1 \to Z_2$, we can
think of $N_{Z_1}$ as a subnet of $N_{Z_2}$.  In this case, we require
that a place of $Z_1$ is input/output open in $Z_2$ only if 
it is so in $Z_1$, and that a transition in $T_{Z_2} - T_{Z_1}$ can
put/remove a token on/from a place of $Z_1$ only if that place is
input/output open in $Z_1$. Furthermore, any place of $Z_1$ must have
the same number of tokens of its image in $Z_2$. This is formalized
by the following definition, which introduces general morphisms,
possibly non-injective.

\begin{defi}[open net category]
  \label{de:open-net-morphism}
  An \emph{open net morphism} $f : Z_1 \to Z_2$ is a Petri net
  morphism $f : N_{Z_1} \to N_{Z_2}$ such that, if we define
  $\inp{f} = \{ s \in S_{Z_1} : \pre[f_S(s)] - f_T(\pre[s]) \neq
  \emptyset \}$
  and
  $\out{f} =  \{ s \in S_{Z_1} : \post[f_S(s)] - f_T(\post[s]) \neq
  \emptyset \}$,
  then

  \begin{enumerate}[(1)]
  \item
    (i) $f_S^{-1}(O_{Z_2}^+) \cup \inp{f} \subseteq O_{Z_1}^+$  and
    (ii)
    $f_S^{-1}(O_{Z_2}^-) \cup \out{f} \subseteq O_{Z_1}^-$.

  \item $\init{u}_1 = \res{\init{u}_2}{f_S}$ (reflection of initial marking).

  \end{enumerate}

  \noindent
  The morphism $f$ is called an \emph{open net embedding} if both
  $f_T$ and $f_S$ are injective.
  We will denote by $\onet$ the category of open nets and
  open net morphisms.
\end{defi}
%======================

%% Given an embedding $f : Z_1 \to Z_2$, we can think of $N_{Z_1}$ as a
%% subnet of $N_{Z_2}$.  In this case, condition 1 requires that a place
%% of $Z_1$ is input/output open in $Z_2$ only if it is so in $Z_1$, and
%% that a transition in $T_{Z_2} - T_{Z_1}$ can put/remove a token
%% on/from a place of $Z_1$ only if that place is input/output open in
%% $Z_1$.
%

Conceptually, condition 1 formalizes the
intuition that each open net can interact with the environment only
through open places.  In fact, given an embedding $f : Z_1 \to Z_2$,
if $s$ is a place of $Z_1$ which is
open in $Z_2$, then an interaction of the environment with $Z_2$
through $s$ would also affect $Z_1$: therefore $s$ must be open in
$Z_1$ as well. That is, input/output open places must be reflected by
the embedding, as stated by the first part of conditions 1.(i) and
1.(ii).  Furthermore, if a transition in $T_{Z_2} - T_{Z_1}$ can put a
token in a place $s$ of $Z_1$, this is seen from $Z_1$ as an
interaction with the environment, and therefore $s$ must be (input)
open in $Z_1$: this is formalized by the second part of conditions
1.(i) and 1.(ii).  
Finally, condition~2 requires the marking of $Z_1$ to be the
projection of the marking of $Z_2$: any place $s_1 \in S_{Z_1}$ must
carry the same number of tokens as its image $f(s_1) \in S_{Z_2}$,
i.e., $\init{u}_1(s_1) = \init{u}_2(f(s_1))$ for any $s_1 \in
S_{Z_1}$.

%% Intuitively, an embedding $f : Z_1 \to Z_2$ ``inserts'' net $Z_1$ into a larger
%% net $Z_2$, which might constrain the behaviour of $Z_1$.
%% %
%% Conditions~1.(i) and~1.(ii) first require that open places are reflected
%% and hence that places which are ``internal'' to $Z_1$ cannot be
%% promoted to open places in $Z_2$.
%% %
%% Furthermore, they ensure that the context in which $Z_1$ is inserted
%% can interact with $Z_1$ only through the open places. Note that
%% intuitively $\inp{f}$ is the ``input boundary'' of the morphism $f$ within
%% the net $Z_2$, i.e., it contains all the places which are images of a
%% place in $Z_1$ and are furthermore in the post-set of a transition of
%% $Z_2$ which is not in the image of $Z_1$. From the perspective of
%% $Z_1$ the environment can generate tokens in $s$ and thus
%% Condition~1.(i) requires $s$ to be an input place.  Condition~1.(ii)
%% is analogous for output places.  Finally, Condition~2 requires that
%% the marking of $Z_1$ is the projection of the marking of $Z_2$: any
%% place $s_1 \in S_{Z_1}$ must carry the same number of tokens as its
%% image $f(s_1) \in S_{Z_2}$, i.e., $\init{u}_1(s_1) =
%% \init{u}_2(f(s_1))$ for any $s_1 \in S_{Z_1}$.

% All morphisms $f_1$, $f_2$, $\alpha_1$ and $\alpha_2$ in Fig.~\ref{fi:sample}
% are examples of open net embeddings (the mappings on places and transitions
% are those suggested by the shape and labelling of the nets). 
%
Consider, for instance, morphism $f_1 : Z_0 \to Z_1$ in
Fig.~\ref{fi:sample}: the mapping of places and transitions
is suggested by the shape and labelling of the nets.
Note that in
$Z_1$ a ``new'' $c$-labelled transition is attached to the places $s$
and $s'$. This is legal since the corresponding places in $Z_0$ are
output open and input open, respectively. Note also that the
number of tokens in places in $Z_0$ and in their image through $f_1$
is the same.  Instead, the number of tokens in the place $s''$ in
$Z_1$ is not constrained since it is not in the image of $f_1$: the
place is marked, but $f_1$ would have been a legal morphism also if
$s''$ were not marked.

It is worth observing that most of the constructions in the paper will
be defined for open net embeddings, hence readers can limit their
attention to embeddings if this helps the intuition. Still, on the
formal side, working in a larger host category with more general
morphisms is essential to obtain a characterisation of the composition
operation in terms of pushouts. Specifically, non-injective open net
morphisms are needed as mediating morphisms (recall, for example, that
the category of sets with injective functions does not have all
pushouts).

%% On the intuitive level, non-injective morphisms ``merge'' resources,
%% i.e., two different types of resources, represented by two places, are
%% subsumed into one type of resource. In such a case the same number of
%% tokens has to be present in two such places in the source net in
%% order to properly reproduce this notion of subsumption on the markings
%% (compare with the notion of multiset projection in
%% Definition~\ref{de:multiset-proj} and with Fig.~\ref{fi:non-inj}).

Observe that the constraints characterising open nets morphisms have
an intuitive graphical interpretation:

\begin{enumerate}[$\bullet$]
\item
  The connections of transitions to their pre-set and post-set have
  to be preserved. New connections cannot be added.

\item
  In the larger net, a new arc may be attached to a place only if
  the corresponding place of the subnet has a dangling arc in the
  same direction. Dangling arcs may be removed, but cannot be added
  in the larger net.

\item 
  The number of tokens in each place in the source net must be preserved in
  the target. Instead, there are no restrictions on the marking of places of
  the target net which are not in the image of the source net.
\end{enumerate}

In the sequel, given an open net morphism $f = \langle f_S, f_T
\rangle : Z_1 \to Z_2$, to lighten the notation we will omit the
subscripts ``$S$'' and ``$T$'' in its place and transition components,
writing $f(s)$ for $f_S(s)$ and $f(t)$ for $f_T(t)$.
Moreover we will write $\mon{f} : \monSub{\bar{T}}{{Z_1}} \to
\monSub{\bar{T}}{{Z_2}}$ to denote the monoidal function defined on the
generators by $\mon{f}(t) = f(t)$ for $t \in T_{Z_1}$ 
and, for $x \in \{ +, -\}$,
$\mon{f}(x_{s}) = x_{f(s)}$, if $f(s) \in O_{Z_2}^x$ and
$\mon{f}(x_{s})$ undefined, otherwise. Note that  
$\mon{f}$ can be partial since open places can be mapped to closed
places.

The next proposition explicitly shows that category $\onet$, as
introduced in Definition~\ref{de:open-net-morphism}, is well
defined. To prove this fact we will use the well-definedness of the
category of unmarked open nets, introduced in~\cite{BCEH:CSOP}.
This category, denoted here by $\uonet$, has (unmarked) open nets as
objects and mappings satisfying only condition~1 in
Definition~\ref{de:open-net-morphism} as morphisms.  These will be
referred to as \emph{unmarked open net morphisms}.

\begin{prop}
  Open net morphisms are closed under composition.
\end{prop}

\proof
  Let $f_1 : Z_1 \to Z_2$ and $f_2 : Z_2 \to Z_3$ be open net morphisms.  Then
  $f_1$ and $f_2$ are unmarked open net morphisms and thus, since $\uonet$ is
  a well-defined category, also $f_2 \circ f_1$ is an unmarked open net
  morphism. In order to prove that $f_2 \circ f_1$ is a well defined open net
  morphism it remains to show that it satisfies also condition~2 in
  Definition~\ref{de:open-net-morphism}, i.e., that it reflects the initial
  marking.
  But this fact follows easily from the definition. In fact, for any $s_1 \in
  S_{Z_1}$,
  \begin{quote}
    \begin{tabular}{ll}
      $\init{u}_3(f_2(f_1(s_1))) =$  &\\
      \ \quad $= \init{u}_2(f_1(s_1))$ &  [since $f_2$ is an open net morphism]\\
      \ \quad $= \init{u}_1(s_1)$ & [since $f_1$ is an open net
  morphism]\hbox to137 pt{\hfill\qEd}
    \end{tabular}
  \end{quote}\medskip

\begin{figure}[t]
  \subfigure[]{
    \scalebox{.28}{\includegraphics{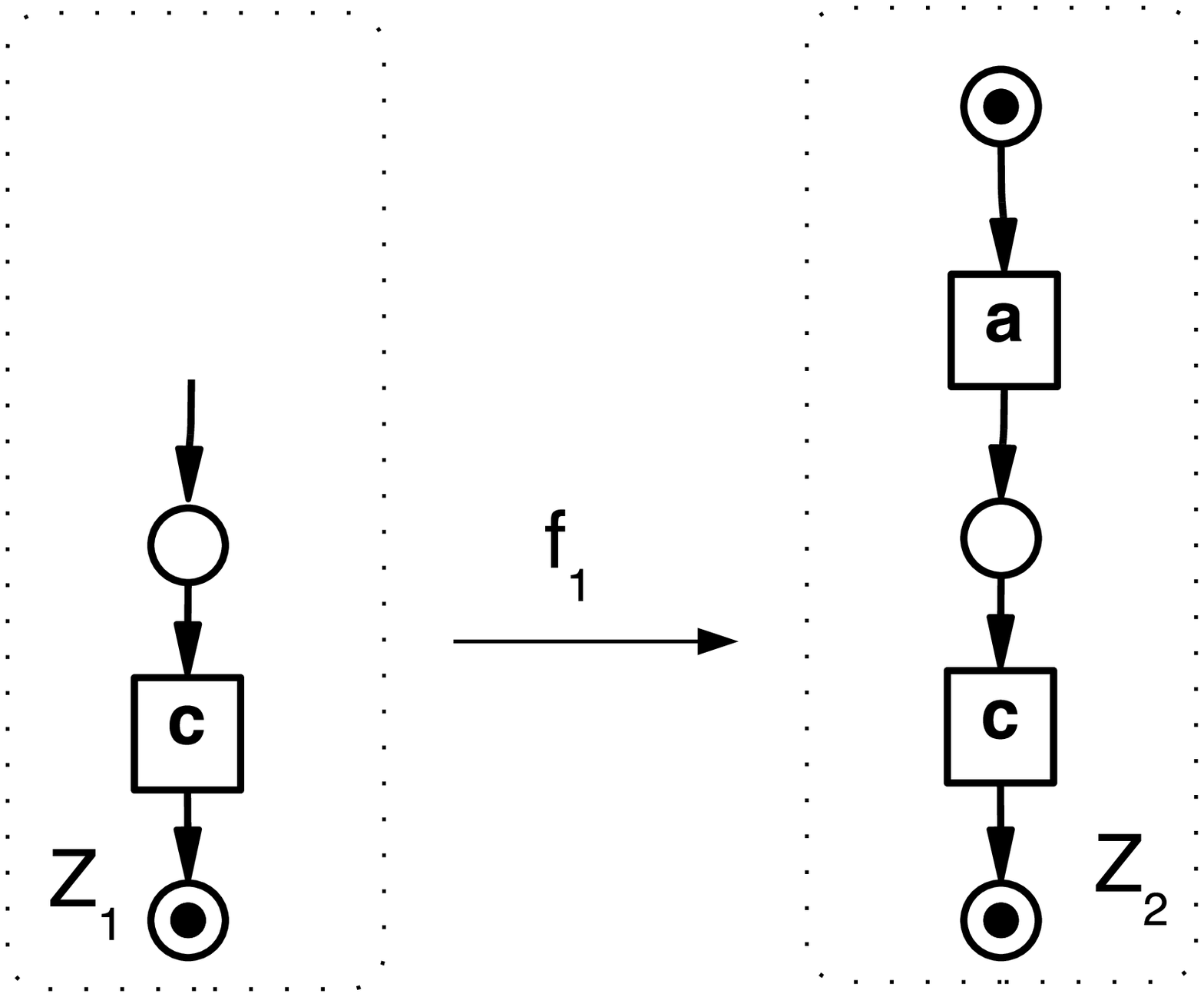}}
    \label{fi:non-sim}
  }
  \subfigure[]{
    \scalebox{.28}{\includegraphics{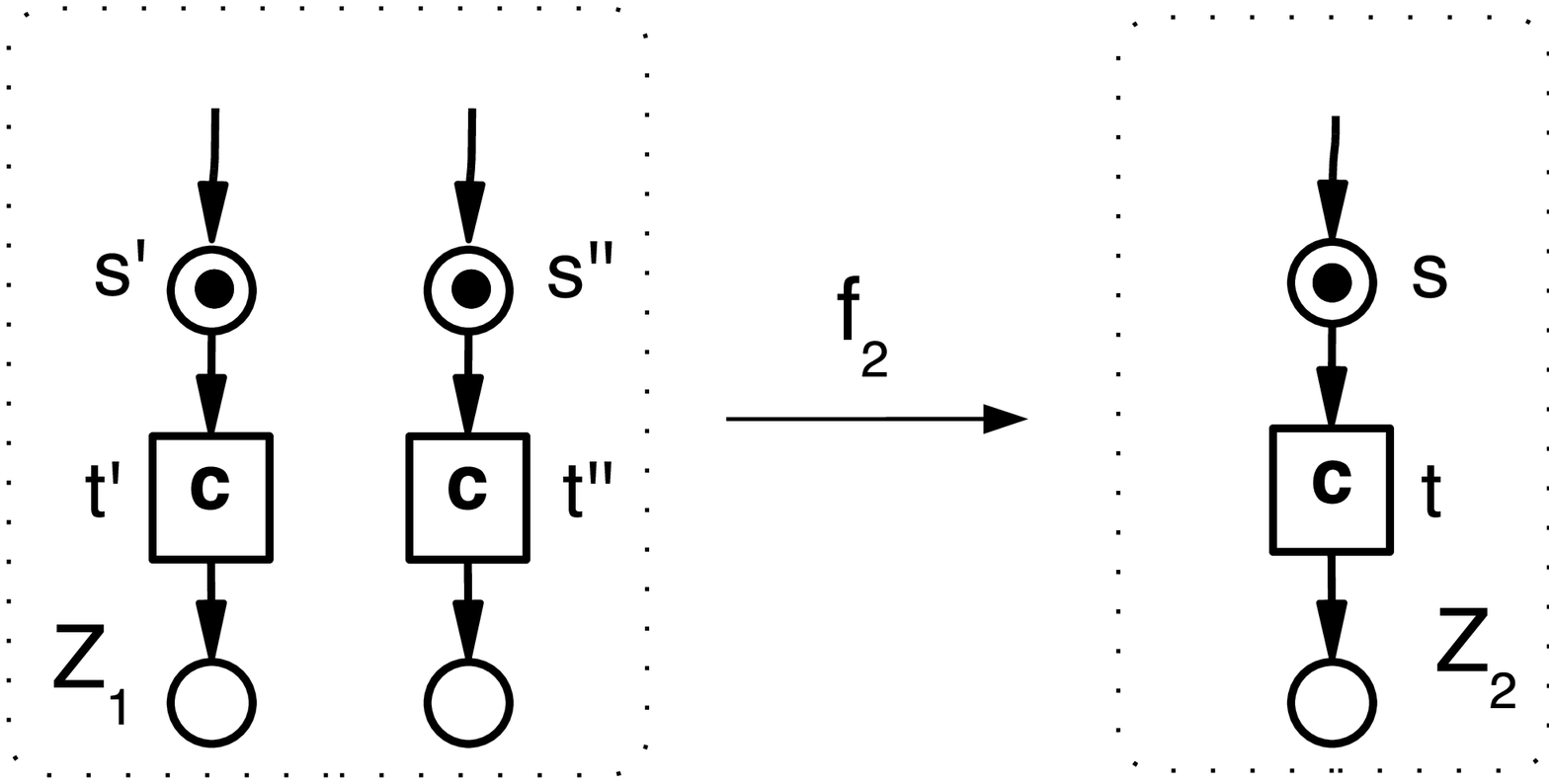}}
    \label{fi:non-inj}
  }
  \caption{(a) Open net morphisms are not simulations and (b) an example of non-injective open net morphism.}
\end{figure}

Unlike most of the morphisms considered over Petri
nets in the literature, open net morphisms are \emph{not} simulations.
As an example, consider the open net embedding in
Fig.~\ref{fi:non-sim}. While the transition labelled $c$ in the net
$Z_1$ can fire infinitely many times, its image in the second net
$Z_2$ can fire only once.

Instead, since open net embeddings are designed to capture the idea of
inserting a net into a larger one, they are expected to reflect the
behaviour, in the sense that given an embedding $f : Z_0 \to Z_1$, the
behaviour of $Z_1$ can be projected along $f$ to the
behaviour of $Z_0$.
%
% This is consistent with the fact that we perform system composition by
% colimits: any component will have a morphism into the full system and
% we can not expect that the larger system is able to simulate a (less
% specified) component of itself.  
%
The target net of a morphism is in general more ``instantiated'' and
thus more constrained than the source net (e.g., a place which is open
in the source net can be closed in the target).
We will come back to this fact in the conclusions.

Although the paper will mainly use open net embeddings, a remark about
non-injective morphisms is in order. Consider the open net morphism
$f_2$ in Fig.~\ref{fi:non-inj}, where $f_2(t') = f_2(t'') = t$ and
$f_2(s') = f_2(s'') = s$. As, intuitively, the two transitions of
$Z_1$ become the same transition in $Z_2$, in this case by reflection
of behaviour we mean that the firing of $t$ in $Z_2$ must be reflected
to the parallel firing of $t'$ and $t''$ in $Z_1$.  Note that this is
the case, e.g., for the initial markings: $s$ enables $t$ and its
projection $\res{s}{f_2} = s' \oplus s''$ enables $t' \oplus
t''$. 

In the rest of this section we formalize the intuition that an open
net embedding $f: Z \to Z'$ reflects the behaviour by showing that
each step of $Z'$ can be projected along $f$ to a step of $Z$. 
It could be shown that the behaviour of an open net is reflected
along non-injective morphisms as well,  but this would require
some technical complications which we prefer to avoid, as it will not
be used in the rest of the paper.

%% The
%% result could be proved for general morphisms, but this would require
%% some technical complications in the definition of step projection
%% which we prefer to avoid, as it will not be useful in the rest of the
%% paper.
%

We start by defining the projection of multisets of extended events
along open net embeddings.

\begin{defi}[projecting extended events]
  Given an open net embedding $f: Z \to Z'$, the \emph{projection of
    extended events along $f$}, denoted  $\sres{\_}{f} :
    \bar{T}_{Z'} \to \monSub{\bar{T}}{Z}$, is defined as
    follows. For each  $\epsilon' \in \bar{T}_{Z'}$,

  \begin{enumerate}[$\bullet$]
    
  \item if $\epsilon' = t' \in T_{Z'}$ is a transition, then
    \[\sres{t'}{f} = 
    \begin{cases}
          t & \hbox{if $t \in T_Z$ and $f(t) = t'$}\cr
          -_{\res{\pre[t']}{f}} \oplus +_{\res{\post[t']}{f}}
            & \hbox{if $t'\not\in f(T_Z)$}
    \end{cases}\]
    \smallskip
  \item if $\epsilon' = x_{s'}$, with $x \in \{+,-\}$,
    % is an interaction with the environment, 
    then $\sres{x_{s'}}{f} =
    x_{\res{s'}{f}}$.
  \end{enumerate}
  The monoidal extension of $\sres{\_}{f}$ to \emph{multisets}
  of extended events will be denoted by the same symbol 
  $\sres{\_}{f} : \monSub{\bar{T}}{{Z'}} \to \monSub{\bar{T}}{Z}$.
\end{defi}  

In words, if we think of the embedding $f : Z \to Z'$ as an inclusion,
then given a transition $t'$, the projection $\sres{t'}{f}$ is the
transition itself if $t'$ is in $Z$. Otherwise, if $t'$ is not in $Z$
but it consumes or produces tokens in places of $Z$, the projection of
$t'$ contains the corresponding extended events, expressing the
interactions over open places.
Similarly, the projection of an extended event $+_{s'}$ is
the event itself if $s'$ is in $Z$, and it is the empty multiset
otherwise: in fact, in this case $\res{s'}{f} = 0$.

It is easily checked that the projection operation is
well-defined, in the sense that, e.g., if $+_s \in \sres{\epsilon}{f}$
then $s \in O_Z^+$. In fact, if $+_s \in \sres{t'}{f}$ then $s \in
\inp{f}$, while if $+_s \in \sres{+_{s'}}{f}$, then $s' \in O_{Z'}^+$
and $f(s) = s'$. In both cases $s \in O_{Z}^+$ by condition~1.(i) of
Definition~\ref{de:open-net-morphism}.

The projections of multisets of places and extended events enjoy nice
properties which are summarized by the next lemma.

\begin{lem}[properties of projection]
  \label{le:proj-prop}
  Let $f : Z \to Z'$ be an open net embedding. Then
  
  \begin{enumerate}[\em(1)]

  \item for $u_1, u_2 \in \monSub{S}{{Z'}}$ we have
    \[\res{(u_1 \oplus u_2)}{f} = \res{u_1}{f} \oplus  \res{u_2}{f}
      \quad\hbox{and}\quad
      \res{0}{f} = 0
    \]
    and for $u \in \monSub{S}{{Z}}$ 
    \[\res{\mon{f}(u)}{f} = u\]
    
  \item for $x_1, x_2 \in \monSub{\bar{T}}{{Z'}}$ we have
    \[\sres{(x_1 \oplus x_2)}{f} = \sres{x_1}{f} \oplus  \sres{x_2}{f}
      \quad\hbox{and}\quad
      \sres{0}{f} = 0
    \]
    and for $x \in \monSub{\bar{T}}{{Z}}$, 
    if $\mon{f}(x)$ is defined we have
    \[\sres{\mon{f}(x)}{f} = x\] 

  \item given $A' \in \monSub{\bar{T}}{{Z'}}$
    \[\res{\pre[A']}{f} =  \pre[\sres{A'}{f}]
      \quad\hbox{and}\quad
      \res{\post[A']}{f} =  \post[\sres{A'}{f}]
    \]

  \item for $u \in \monSub{S}{{Z'}}$ we have 
    \[\mon{f}(\res{u}{f}) \leq u\]
  \end{enumerate}
\end{lem}

\begin{proof}
  Proofs are routine. We prove explicitly only the third point. Since
  $\pre$ and $\post$ are monoidal functions it is sufficient to prove
  the result only on the generators.  We concentrate on $\pre$, since
  the proof for $\post$ is completely analogous.

  We distinguish various cases:

  \begin{enumerate}[$\bullet$]

  \item $A' = t' \in T_{Z'}$\\
    If there exists $t \in T_Z$ such that $f(t) = t'$, then $\sres{t'}{f} =
    t$. Since $f$ is an open net morphism $\mon{f}(\pre[t]) = \pre[t']$ and
    thus, as desired
    \begin{quote}
      $\pre[\sres{t'}{f}] = \pre[t] = \res{\mon{f}(\pre[t])}{f} =
      \res{\pre[t']}{f}$
    \end{quote}
    where the second equality is justified by point (1).
    
    If, instead, $t' \not\in f(T_{Z})$ we have that 
    $\sres{t'}{f} =  -_{\res{\pre[t']}{f}} \oplus +_{\res{\post[t']}{f}}$.
    Hence, in this case the result is obvious since
    \begin{quote}
      $\pre[\sres{t'}{f}] = \pre[(-_{\res{\pre[t']}{f}} \oplus
      +_{\res{\post[t']}{f}})] = \res{\pre[t']}{f}$
    \end{quote}

  \item $A' = +_{s'}$ or $A' = -_{s'}$\\
    Suppose, e.g., that $A' = -_{s'}$. In this case $\sres{A'}{f} =
    -_{\res{s'}{f}}$ and the result trivially holds.
  \end{enumerate}
\end{proof}

We are now ready to present the main result of this section.

\begin{lem}[reflection of behaviour]
  \label{le:reflect}
  Let $f: Z \to Z'$ be an open net embedding.  For every step $u' \trans{A'}
  v'$ in $Z'$ there is a step $\res{u'}{f} \trans{\sres{A'}{f}}
  \res{v'}{f}$ in $Z$, called the \emph{projection of $u'
  \trans{A'} v'$ along $f$}.
\end{lem}

\begin{proof}
  Let $f: Z \to Z'$ be an open net embedding and assume that $u' \trans{A'}
  v'$ is a step in $Z'$. Therefore
  \begin{quote}
    $u' = u'' \oplus \pre[A']$ \quad and \quad $v' = u'' \oplus \post[A]$
  \end{quote}
  Now, we have
  \begin{quote}
    \begin{tabular}{lll}
      $\res{u'}{f} =$ \\
      \ \quad $= \res{u''}{f} \oplus \res{\pre[A']}{f}$ & \ \quad &  [by Lemma~\ref{le:proj-prop}.(1)]\\
      \ \quad $=  \res{u''}{f} \oplus \pre[\sres{A'}{f}]$ & &
      [by Lemma~\ref{le:proj-prop}.(3)]
    \end{tabular}
  \end{quote}
  and similarly
    \begin{quote}
      $\res{v'}{f} = \res{u''}{f} \oplus \post[\sres{A'}{f}]$
    \end{quote}
    Therefore, as desired, there is the step
    \begin{quote}
      $\res{u'}{f} = \res{u''}{f} \oplus \pre[\sres{A'}{f}]
      \trans{\sres{A'}{f}} 
      \res{u''}{f} \oplus \post[\sres{A'}{f}] = \res{v'}{f}$.
    \end{quote}
\end{proof}

Observe that there is an obvious forgetful functor $\fo : \onet \to
\net$, defined by $\fo(Z) = N_Z$ and $\fo(f : Z_0 \to Z_1) = f :
N_{Z_0} \to N_{Z_1}$. Since functor $\fo$ acts on arrows as the identity,
with abuse of notation, given an open net morphism $f : Z_0 \to Z_1$
we will often write $f : \fo(Z_1) \to \fo(Z_2)$ instead of $\fo(f) :
\fo(Z_1) \to \fo(Z_2)$.

\section{Composing Open Nets}
\label{se:composing-nets}

We introduce next a basic mechanism for composing open nets which is
characterised as a pushout construction in category $\onet$. A pushout
is a canonical way of describing a gluing construction.
The case of unmarked nets was already discussed in~\cite{BCEH:CSOP}. Here we
extend the theory to deal with marked open nets. 
This will allow later to define reconfigurations of open nets, where
the applicability of a reconfiguration rule can depend on the marking.
Intuitively, two open nets $Z_1$ and $Z_2$ are composed by specifying a
common subnet $Z_0$, and then by joining the two nets along $Z_0$.
%
% The categorical characterisation will be useful for proving the
% results about behaviour-preserving transformations in
% Section~\ref{se:bisim}, but the concrete characterisation in terms of a
% join along a common subnet can help the intuition.

Let us start with a technical definition which will be useful below. 

\begin{prop}[composition of multisets]
  \label{pr:sum-mark}
  Consider a pushout diagram in the category of sets as below, where all
  morphisms are injective. 

  \noindent
  \begin{minipage}[c]{0.63\linewidth}
    Given $u_1 \in \monSub{S}{1}$ and $u_2 \in \monSub{S}{2}$ such
    that $\res{u_1}{f_1} = \res{u_2}{f_2} = u_0$, there is a (unique)
    multiset $u_3 \in \monSub{S}{3}$ such that $\res{u_3}{\alpha_i} =
    u_i$, for $i \in \{ 1, 2\}$.  Such a multiset $u_3$ will be
    denoted by $u_3 = u_1 \mjoin_{u_0} u_2$ or simply by $u_1 \mjoin
    u_2$ when making $u_0$ explicit is not needed.
  \end{minipage}
  \begin{minipage}[c]{0.36\linewidth}
    \vspace{-5mm}
    \[
    \xymatrix@C=12mm@R=0mm{
      & S_0 \ar[dl]_{f_1} \ar[dr]^{f_2} &\\
      S_1 \ar[dr]_{\alpha_1} & & S_2 \ar[dl]^{\alpha_2} \\
      & S_3 & }
    \]
  \end{minipage}
  
  \smallskip

  Additionally, if $u_3= u_1 \mjoin_{u_0} u_2$ and $u_3' = u_1'
  \mjoin_{u_0'} u_2'$, then $u_3 \oplus u_3' = (u_1 \oplus u_1')
  \mjoin_{(u_0 \oplus u_0')} (u_2 \oplus u_2')$.
\end{prop}

\begin{proof} 
Define $u_3 \in \monSub{S}{3}$ as follows: for
  each $s \in S_3$, 
\begin{quote}
$u_3(s) = \left\{
\begin{array}{lll}
u_1(s_1) & \quad & \text{if $\exists s_1 \in S_1$ such that
    $\alpha_1(s_1) = s$}\\
u_2(s_2) & \quad & \text{if $\exists s_2 \in S_2$ such that
    $\alpha_2(s_2) = s$}
\end{array}
\right.
$
\end{quote}
\smallskip
Let us start checking that $u_3$ is well-defined. In fact, firstly, the
definition assigns a coefficient to every $s \in S_3$
because $\alpha_1$ and $\alpha_2$ are jointly surjective. Secondly, if
there are $s_1 \in S_1$ and $s_2 \in S_2$ such
that $\alpha_1(s_1)= \alpha_2(s_2)$, since the square is a pushout and
all functions are injective we have ${f_1}^{-1}(s_1) = \{s_0\}$ and
${f_2}^{-1}(s_2) = \{s_0\}$ for some $s_0 \in S_0$: thus, 
since $\res{u_1}{f_1} = \res{u_2}{f_2}$ by hypothesis, we obtain $u_1(s_1) =
u_1(f_1(s_0)) = \res{u_1}{f_1}(s_0) = \res{u_2}{f_2}(s_0) =
u_2(f_2(s_0)) = u_2(s_2)$.

Now, in order to prove (for $i \in \{1, 2\}$) that
$\res{u_3}{\alpha_i} = u_i$, notice that, since $\alpha_i$ is injective,
this amounts to show that for any $s \in S_i$ we have $u_i(s) =
u_3(\alpha_i(s))$, which is immediate by the definition of $u_3$.

    Concerning the second part of the statement, let $u_3= u_1 \mjoin_{u_0} u_2$ and
  $u_3' = u_1' \mjoin_{u_0'} u_2'$. Then just observe that by
  Lemma~\ref{le:proj-prop}.(1), we have for $i \in \{1,2\}$
  \begin{quote}
    $\res{(u_3 \oplus u_3')}{\alpha_i} =  
    \res{u_3}{\alpha_i} \oplus \res{u_3'}{\alpha_i} =
    u_i \oplus u'_i$
  \end{quote}
  hence the result $u_3 \oplus u_3' = (u_1 \oplus u_1') \mjoin_{(u_0 \oplus u_0')}
  (u_2 \oplus u_2')$ follows by the defining property of the
  composition of markings. 
\end{proof}

Intuitively, the multiset $u_1 \mjoin_{u_0} u_2$ can be seen as the
``least upper bound'' of the images of the two multisets in
$\monSub{S}{3}$.

As in~\cite{BCEH:CMRS,BCEH:CSOP}, two embeddings $f_1
: Z_0 \to Z_1$ and $f_2 : Z_0 \to Z_2$ are called \emph{composable} if the places
which are used as interface by $f_1$, i.e., the places $\inp{f_1}$ and
$\out{f_1}$, are mapped by $f_2$ to input and output open places of $Z_2$,
respectively, and also the symmetric condition holds.

%% The next definition introduces the condition, informally explained
%% above, which ensures the composability of a span of embeddings in $\onet$.

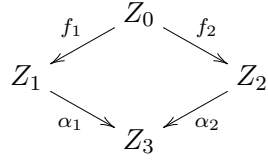
\begin{figure}[t]
  
  \[
  \xymatrix@R=3mm{
    {} & {Z_0} \ar[dr]^{f_2} \ar[dl]_{f_1} & {}\\
    {Z_1} \ar[dr]_{\alpha_1} & {} & {Z_2} \ar[dl]^{\alpha_2} \\
    {} & {Z_3} & {}
  }
  \]
  
  \caption{Pushout in $\onet$.}
  \label{fi:pushout-onet}
\end{figure}

\begin{defi}[composability of embeddings]
  \label{de:composable}
  Let $f_1 : Z_0 \to Z_1$, $f_2 : Z_0 \to Z_2$ be embeddings in
  $\onet$ (see Fig.~\ref{fi:pushout-onet}).%
  We say that $f_1$ and $f_2$ are \emph{composable} if
  \begin{enumerate}[(1)]
  \item
    %
    % 1.
    $f_2(\inp{f_1}) \subseteq O_{Z_2}^+$  and
    $f_2(\out{f_1}) \subseteq O_{Z_2}^-$;

  \item 
    %and
    %2. 
    $f_1(\inp{f_2}) \subseteq O_{Z_1}^+$ and
    $f_1(\out{f_2}) \subseteq O_{Z_1}^-$.
  \end{enumerate}
\end{defi}

Composability is necessary and sufficient to ensure that the pushout
of $f_1$ and $f_2$ can be computed in $\net$ and then lifted to
$\onet$.

\begin{prop}[pushouts in $\onet$]
  \label{pr:push-onet}
  Let $f_1 : Z_0 \to Z_1$, $f_2 : Z_0 \to Z_2$ be
  embeddings in $\onet$ 
  (see Fig.~\ref{fi:pushout-onet}).
  Compute the pushout of the corresponding diagram in category $\net$
  obtaining net $N_{Z_3}$ and morphisms $\alpha_1$ and
  $\alpha_2$,\footnote{The pushout in $\net$ is computed componentwise
    on places and transitions, by defining the pre- and post-set
    functions, for any $t_i \in T_{Z_i}$, $i\in\{1,2\}$, as
    $\sigma_{Z_3}(\alpha_i(t_i)) = \mon{\alpha_i}(\sigma_{Z_i}(t_i))$
    and $\tau_{Z_3}(\alpha_i(t_i)) =
    \mon{\alpha_i}(\tau_{Z_i}(t_i))$. It is routine to show that this definition is well given.}
  and then take as open places, for
  $x \in \{ +, -\}$,
  \begin{center}
    $O_{Z_3}^x = \{ s_3 \in S_{Z_3} : \alpha_1^{-1}(s_3) \subseteq
    O_{Z_1}^x\ \wedge\ \alpha_2^{-1}(s_3) \subseteq O_{Z_2}^x \}$
  \end{center}
  and as initial marking  $\init{u}_3 = \init{u}_1 \mjoin_{\init{u}_0}
  \init{u}_2$,  defined according to
  Proposition~\ref{pr:sum-mark}.
  Then $(\alpha_1, Z_3, \alpha_2)$ is the pushout in $\onet$ of $f_1$
  and $f_2$ if and only if $f_1$ and $f_2$ are composable.
  In this case we write $Z_3 = Z_1 \comp{f_1,f_2} Z_2$.
\end{prop}

\begin{proof}
  We know by~\cite{BCEH:CSOP} (Proposition~6) that the above result
  holds for unmarked nets, i.e., in the category $\uonet$.
  Here we must additionally show that (i) the $\alpha_i$ are
  marked morphisms and that (ii) if we take any other net $Z_3'$, with
  $\alpha_i' : Z_i \to Z_3'$ making the diagram commute, then the
  mediating morphism $\gamma : Z_3 \to Z_3'$ (which exists uniquely as
  an unmarked net morphism by the result in~\cite{BCEH:CSOP}) respects the
  condition on the marking.
  
  Now, (i) is immediate since Proposition~\ref{pr:sum-mark} tells us that
  $\res{\init{u}_3}{\alpha_i} = \init{u}_i$ for $i \in \{1,2\}$.
  Property (ii) can be proved along the same lines.
\end{proof}

\begin{figure}[t]
  \begin{center}
    \scalebox{.28}{\includegraphics{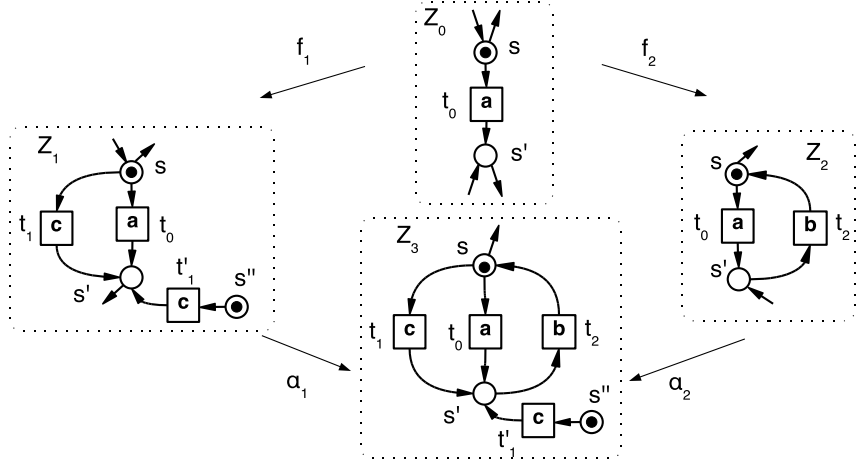}}
  \end{center}

  \caption{An example of a pushout in $\onet$.}
  \label{fi:pushout}
\end{figure}

As an example, the open net embeddings $f_1$ and $f_2$ in
Fig.~\ref{fi:pushout} are composable. In fact, $\inp{f_1} = \{ s'\}$,
$\out{f_1} = \{ s \}$ and $\inp{f_2} = \{ s \}$, $\out{f_2}= \{ s'
\}$, and thus it is easy to see that the conditions of
Definition~\ref{de:composable} are satisfied. The net $Z_3$ is the
resulting pushout object.

\section{Composing Steps}

In this section we analyse the behaviour of an open net $Z_3$ arising
as the composition of two nets $Z_1$ and $Z_2$ along an interface
$Z_0$.  More specifically,
%  concentrating on steps
we show that steps of the component
nets $Z_1$ and $Z_2$ can be ``composed'' to give a step of $Z_3$ when
they agree on the interface and satisfy suitable compatibility
conditions.

For instance, concerning the example pushout in Fig.~\ref{fi:pushout},
note that net $Z_1$ can fire the transition labelled $a$ and the lower
transition labelled $c$. If this is ``mimicked'' in $Z_2$ by firing
$a$ and putting a token into the lower place $s'$ (via an interaction
$+_{s'}$ with the environment), then such steps are compatible in a
sense made precise below and can be combined into a step of the
composed net $Z_3$.

We start with a technical lemma which will be pivotal in the paper.
Assume that the first component makes a step and the second component
imitates this step, acting only on the places of the common interface,
without firing any internal transition. Then the two local steps can
be combined to a step of the composed net.

\begin{lem}
  \label{le:downup}
  Let $Z_3 = Z_1 \comp{f_1,f_2} Z_2$ be the pushout of two composable
  embeddings $f_1 : Z_0 \to Z_1$ and $f_2 : Z_0 \to Z_2$ in $\onet$ 
  (see Fig.~\ref{fi:pushout-onet}).
  %
  % Let $u_1 \trans{A_1} v_1$ be a step in $Z_1$ such that 
  % $\monSub{f}{2}(\sres{A_1}{f_1}) \stackrel{\mbox{\tiny def}}{=} A_2$
  % is defined, and let $u_2 \trans{A_2} v_2$ be a step in $Z_2$
  % such that $\res{u_1}{f_1} = \res{u_2}{f_2} \stackrel{\mbox{\tiny
  %     def}}{=} u_0$. 
  %
  % Then, $\res{v_1}{f_1} = \res{v_2}{f_2} \stackrel{\mbox{\tiny
  %     def}}{=} v_0$,
  % $\monSub{\alpha}{1}(A_1) \stackrel{\mbox{\tiny
  %     def}}{=} A_3$ is defined, and
  %
  Let $u_1 \trans{A_1} v_1$ and $u_2 \trans{A_2} v_2$ be steps in
  $Z_1$ and $Z_2$, respectively, such that $\res{u_1}{f_1} =
  \res{u_2}{f_2}$ and $A_2 = \monSub{f}{2}(\sres{A_1}{f_1})$.

  Then $\res{v_1}{f_1} = \res{v_2}{f_2}$ and, if we define $A_3 =
  \monSub{\alpha}{1}(A_1)$,

  \begin{center}
    $u_1 \mjoin u_2 \trans{A_3} v_1 \mjoin v_2$.
  \end{center}
\end{lem}

\begin{proof}
  Let us start showing that $A_3 = \monSub{\alpha}{1}(A_1)$ is
  defined, i.e., that for $x \in \{ +, -\}$ if $x_s \in A_1$ then
  $\alpha_1(s) \in O_{Z_3}^{x}$. In fact $x_s \in A_1$ implies that $s
  \in O_{Z_1}^{x}$; now either $s \not \in f_1(S_{Z_0})$ and then
  $\alpha_1(s) \in O_{Z_3}^{x}$ by
  Proposition~\ref{pr:push-onet}. Otherwise, since $f_1$ is an
  embedding, there is exactly one place in $S_{Z_0}$ which is mapped
  to $s$. With a little abuse of notation let such place be denoted
  $f_1^{-1}(s)$. Then clearly $f^{-1}_1(s) \in O_{Z_0}^{x}$ because
  $f_1$ is a morphism, and $f_2(f_1^{-1}(s)) \in O_{Z_2}^{x}$ because
  $\monSub{f}{2}(\sres{A_1}{f_1}) = A_2$ is defined by hypothesis;
  thus again $\alpha_1(s) \in O_{Z_3}^{x}$ by
  Proposition~\ref{pr:push-onet}.

  Next observe that, since $A_2 = \monSub{f}{2}(\sres{A_1}{f_1})$ is
  defined, by Lemma~\ref{le:proj-prop}.(2),
  \begin{center}
    $\sres{A_2}{f_2} = \sres{A_1}{f_1}$.
  \end{center}
  Let $A_0 =  \sres{A_i}{f_i}$, for $i \in \{ 1,2\}$, be the
  common projection. As a consequence, we have $\pre[\sres{A_2}{f_2}]=
  \pre[\sres{A_1}{f_1}]$ and thus, by Lemma~\ref{le:proj-prop}.(3)
  \begin{center}
    \res{\pre[A_1]}{f_1} = \res{\pre[A_2]}{f_2}
  \end{center}
  so that we can consider the composition of markings $\pre[A_1] \mjoin_{\pre[A_0]}
  \pre[A_2]$. We claim that
  \begin{equation}
    \label{eq:steps}
    \pre[A_3] = \pre[A_1] \mjoin_{\pre[A_0]} \pre[A_2]
  \end{equation}
  and symmetrically, since $\res{\post[A_1]}{f_1} =
  \res{\post[A_2]}{f_2}$, that 
  \begin{center}
    $\post[A_3] = \post[A_1] \mjoin_{\post[A_0]} \post[A_2]$
  \end{center}

  Let us concentrate on $\pre$, as the other case is analogous.
  To prove~(\ref{eq:steps}), by Proposition~\ref{pr:sum-mark} we can
  show that $\res{\pre[A_3]}{\alpha_1} = \pre[A_1]$ and
  $\res{\pre[A_3]}{\alpha_2} = \pre[A_2]$. In fact we have 
  \begin{quote}
    \begin{tabular}{lll}
      $\res{\pre[A_3]}{\alpha_1} =$ & \quad & 
      \\
      \ \quad $= \pre[\sres{A_3}{\alpha_1}]$ & \quad & 
      [by Lemma~\ref{le:proj-prop}.(3)]\\ 
      \ \quad $= \pre[\sres{\monSub{\alpha}{1}(A_1)}{\alpha_1}]$ & \quad & 
      [by definition of $A_3$]\\ 
      \ \quad $= \pre[A_1]$ & & [by Lemma~\ref{le:proj-prop}.(2)]
    \end{tabular}
  \end{quote}
  and
  \begin{quote}
    \begin{tabular}{lll}
      $\res{\pre[A_3]}{\alpha_2} =$ & \quad & 
      \\
      \ \quad $= \pre[\sres{A_3}{\alpha_2}]$ & \quad & 
      [by Lemma~\ref{le:proj-prop}.(3)]\\ 
      \ \quad $= \pre[\sres{\monSub{\alpha}{1}(A_1)}{\alpha_2}]$ & &
      [by definition of $A_3$]
    \end{tabular}
  \end{quote}
  Thus to conclude we must show that $\pre[\sres{\monSub{\alpha}{1}(A_1)}{\alpha_2}] =
  \pre[A_2]$, and this is proved by showing
  \begin{equation}
    \label{eq:steps1}
    \sres{\monSub{\alpha}{1}(A_1)}{\alpha_2} = 
    \monSub{f}{2}(\sres{A_1}{f_1}) [ = A_2]
  \end{equation}
  Since $\sres{\_}{\_}$ is monoidal in the first argument by
  Lemma~\ref{le:proj-prop}.(1), it is sufficient to
  show~(\ref{eq:steps1}) on generators: 

%  To this aim, we proceed by induction on the size\footnote{Maybe
%    better ``cardinality'' or ``structure''?} of $A_1$:

  \begin{enumerate}[$\bullet$]

%    \medskip

%%   \item $A_1 = 0$\\
%%     In this case $A_2 = \monSub{f}{2}(\sres{A_1}{f_1}) = 0 =
%%     \sres{\monSub{\alpha}{1}(A_1)}{\alpha_2}$, as desired.

    \medskip

  \item $A_1 = t_1$\\
    We distinguish two subcases. If $\sres{t_1}{f_1} = t_0 \in T_{Z_0}$ then
    $A_2 = f_2(t_0) = \sres{\alpha_1(t_1)}{\alpha_2}$, as desired, by
    construction of the pushout.

    If, instead, $\sres{t_1}{f_1} = -_{\res{\pre[t_1]}{f_1}} \oplus
    +_{\res{\post[t_1]}{f_1}}$, then 
%%    by hypothesis, $A_2 =
%%    \monSub{f}{2}(\sres{A_1}{f_1})$ and thus
    \begin{quote}
      $A_2 = -_{\monSub{f}{2}(\res{\pre[t_1]}{f_1})} \oplus
      +_{\monSub{f}{2}(\res{\post[t_1]}{f_1})}$
    \end{quote}
    On the other hand, we have
    \begin{quote}
      $\sres{\monSub{\alpha}{1}(A_1)}{\alpha_2} =
      \sres{\alpha_1(t_1)}{\alpha_2} =
      -_{\res{\pre[\alpha_1(t_1)]}{\alpha_2}} \oplus
      +_{\res{\post[\alpha_1(t_1)]}{\alpha_2}}
      $
    \end{quote}
      
    Now, by exploiting the fact that $Z_3$ is a pushout, it is easy to see that 
    $\monSub{f}{2}(\res{\pre[t_1]}{f_1}) = \res{\pre[\alpha_1(t_1)]}{\alpha_2}$
    and similarly $\monSub{f}{2}(\res{\post[t_1]}{f_1}) =
    \res{\post[\alpha_1(t_1)]}{\alpha_2}$. Hence we conclude that $A_2 =
    \sres{\monSub{\alpha}{1}(A_1)}{\alpha_2}$, as desired.

    \medskip

  \item $A_1 = +_{s_1}$ or $A_1 = -_{s_1}$\\
    Assume, for instance, that $A_1 = +_{s_1}$ (the other case is completely
    analogous). 
    Therefore
    \begin{quote}
      $A_2 = \monSub{f}{2}(\sres{A_1}{f_1}) = +_{\monSub{f}{2}(\res{s_1}{f_1})}$
    \end{quote}
    On the other hand
    \begin{quote}
      $\sres{\monSub{\alpha}{1}(A_1)}{\alpha_2} = \sres{+_{\alpha_1(s_1)}}{\alpha_2}
      = +_{\res{\alpha_1(s_1)}{\alpha_2}}$      
    \end{quote}
    and, again, by the fact that $Z_3$ is a pushout, we deduce easily
    that $\monSub{f}{2}(\res{s_1}{f_1}) = \res{\alpha_1(s_1)}{\alpha_2}$, hence
    the desired equality.

%%   \item 
%%     $A_1 = A_1' \oplus A_1''$, with $A_1', A_1'' \neq 0$.\\
%%     In this case we have
%%     \begin{quote}
%%       \begin{tabular}{lll}
%%         $\res{\monSub{\alpha}{1}(A_1)}{\alpha_2} =$ \\
%%         %
%%         \ \quad $= \sres{(\monSub{\alpha}{1}(A_1') \oplus
%%           \monSub{\alpha}{1}(A_1''))}{\alpha_2} =$ & \quad &
%%         [by Lemma~\ref{le:proj-prop}.(1)]\\
%%         %
%%         \ \quad $= \sres{\monSub{\alpha}{1}(A_1')}{\alpha_2} 
%%         \oplus   \sres{\monSub{\alpha}{1}(A_1'')}{\alpha_2}$
%%       \end{tabular}
%%     \end{quote}
%%     On the other hand
%%     \begin{quote}
%%       \begin{tabular}{lll}
%%         $A_2 = \monSub{f}{2}(\sres{A_1}{f_1}) =$
%%         & \quad & [by Lemma~\ref{le:proj-prop}.(1)]\\
%%         %
%%         \ \quad  $= \monSub{f}{2}(\sres{A_1'}{f_1} \oplus \sres{A_1''}{f_1}) =$\\
%%         %
%%         \ \quad  $= \monSub{f}{2}(\sres{A_1'}{f_1}) \oplus
%%         \monSub{f}{2}(\sres{A_1''}{f_1}) =$ & \quad & [by inductive hypothesis]\\
%%         %
%%         \ \quad  $= \sres{\monSub{\alpha}{1}(A_1')}{\alpha_2} 
%%         \oplus   \sres{\monSub{\alpha}{1}(A_1'')}{\alpha_2}$
%%       \end{tabular}
%%     \end{quote}
%%     Hence, the desired equality follows.
   \end{enumerate}
  This concludes the proof of (\ref{eq:steps1}), from which (\ref{eq:steps})
  follows.
  
  \bigskip
  
  Now, by exploiting (\ref{eq:steps}) we can easily conclude. In fact, the
  steps in $Z_1$ and $Z_2$ are of the kind
  \begin{quote}
    $u_i = u_i' \oplus \pre[A_i] \trans{A_i} u_i' \oplus \post[A_i] = v_i$
  \end{quote}
  for $i \in \{1,2\}$. First observe that, since $\res{u_1}{f_1} =
  \res{u_2}{f_2}$ and $\res{\pre[A_1]}{f_1} = \res{\pre[A_2]}{f_2}$, we
  immediately get: 
  \begin{quote}
    $\res{u_1'}{f_1} = \res{u_2'}{f_2}$
  \end{quote}
  Let $u_0' = \res{u_i'}{f_i}$, for $i \in  \{ 1,2\}$, be the common projection.
  Since $v_i = u_i' \oplus
  \post[A_i]$, for $i \in \{1,2\}$, by the fact that $\res{\post[A_1]}{f_1} =
  \res{\post[A_2]}{f_2}$, we deduce that, as desired
  \begin{quote}
    $\res{v_1}{f_1} = \res{v_2}{f_2}$
  \end{quote}
  Hence, if $v_0 = \res{v_i}{f_i}$ is the common projection, we can define
  $v_3 = v_1 \mjoin_{v_0} v_2$.

  Now, if we set $u_3' = u_1' \mjoin_{u_0'} u_2'$ we have
    \begin{quote}
      \begin{tabular}{lll}
        $u_3 = u_1 \mjoin_{u_0} u_2 =$\\
        \ \quad $= (u_1' \oplus \pre[A_1])  \mjoin_{u_0' \oplus \pre[A_0]} (u_2'
        \oplus \pre[A_2]) =$
        & \quad & \\
        \ \quad $= (u_1'  \mjoin_{u_0'} u_2')  \oplus (\pre[A_1]  \mjoin_{\pre[A_0]}
        \pre[A_2]) =$ 
        & \quad & [by Proposition~\ref{pr:sum-mark}]\\
        \ \quad  $= u_3' \oplus \pre[A_3]$ & &
        [by (\ref{eq:steps})]
      \end{tabular}
    \end{quote}  
    Therefore we have the step
    \begin{quote}
      $u_3 \trans{A_3} u_3' \oplus \post[A_3]$.
    \end{quote}
    By a sequence of passages analogous to those used above, we can show that
    $u_3' \oplus \post[A_3] = v_3$ and thus, as desired,
    $u_3 \trans{A_3} v_3$.

    The fact that such step projects to $u_i \trans{A_i} v_i$ for $i \in \{
    1,2\}$ immediately follows by construction.
\end{proof}

We are now able to show how steps of the component nets can be ``joined''
to a step of their composition, provided that the steps satisfy a
suitable compatibility condition, that we are going to introduce.
Roughly, we must be able to split each of the two steps $A_1,A_2$ into
an internal part $A_i^I$ and an external part $A_i^E$, with the
intuition that the external part can include only firings of
transitions in the interface and interactions with the environment
induced by the internal part of the other step.

Put more precisely, from the point of view of $Z_1$ the events can be
of four different kinds: (1)~transitions that are local to $Z_1$
(2)~transitions that occur also in $Z_0$ (3)~interactions with $Z_2$
(of the form $+_s,-_s$) (4)~interactions with the environment of both
nets (also of the form $+_s,-_s$). Now if one splits the set $A_1$
into $A_1^I$ and $A_1^E$, it is necessary to put all events of
type~(1) into $A_1^I$ and all events of type~(3) into $A_1^E$. For the
remaining two types we have a choice, but whenever we put an event of
$Z_1$ into $A_1^E$, we have to put the corresponding event in $Z_2$
into $A_2^I$ (and vice versa).

For reasons of simplicity we have chosen to work with a split into
only two sets instead of four, even if this split is non-unique.

\begin{defi}[compatible steps]
  \label{de:compatible-steps}
  Let $Z_3 = Z_1 +_{f_1,f_2} Z_2$ be a pushout in $\onet$.
  %% (see Fig.~\ref{fi:pushout-onet}).
  %
  We say that two steps $u_i \trans{A_i} v_i$ ($i \in \{1,2\}$) are
  \emph{compatible} if $\res{u_1}{f_1} = \res{u_2}{f_2}$ and we can
  decompose the steps as $A_i = A_i^I \oplus A_i^E$ ($i \in \{1,2\}$)
  such that
  \begin{center}
    $A_2^E = \monSub{f}{2}(\sres{A_1^I}{f_1})$ \ \ and \ \
    $A_1^E = \monSub{f}{1}(\sres{A_2^I}{f_2})$
  \end{center}
\end{defi}

It is immediate to see that if $A_1$ and $A_2$ are compatible, then
$\sres{A_1}{f_1} = \sres{A_2}{f_2}$.

For instance, let us consider again the pushout in
Fig.~\ref{fi:pushout}. Two compatible steps can be $A_1 = t_0 \oplus
t_1'$ and $A_2 = t_0 \oplus +_{s'}$. The compatibility is
witnessed by the decomposition $A_1^I = A_1$, $A_1^E=0$ and $A_2^I =
0$, $A_2^E = A_2$. As mentioned above such decompositions are not uniquely determined: 
alternative ones are given by $A_1^I = t_1'$, $A_1^E=t_0$ and $A_2^I =
t_0$, $A_2^E = +_{s'}$.
Note that since transition $t_0$ also belongs to the interface,
it can be considered either internal to $Z_1$ or internal to $Z_2$,
while $t_1'$ has to be considered internal to $Z_1$, and the interaction
$+_{s'}$ on the open place $s'$ has to be considered external to $Z_2$.

Another simple example of compatible steps is given by $A_1 = -_s$
and $A_2 = -_s$. In this case, we have the choice to consider the only
event $-_s$ internal to $Z_1$ and external to $Z_2$ or vice versa.

\begin{lem}[composing steps]
  \label{le:compose-step}
  Let $f_1 : Z_0 \to Z_1$ and $f_2 : Z_0 \to Z_2$ be composable
  embeddings in $\onet$ and let $Z_3 = Z_1 \comp{f_1,f_2} Z_2$.  Let
  $u_1 \trans{A_1} v_1$ and $u_2 \trans{A_2} v_2$ be compatible steps
  and let $A_i = A_i^I \oplus A_i^E$, for $i \in \{1,2\}$, be a
  corresponding decomposition (see
  Definition~\ref{de:compatible-steps}).
  Then there exists a unique step $u_3 \trans{A_3} v_3$, with $A_3 =
  \monSub{\alpha}{1}(A_1^I) \oplus \monSub{\alpha}{2}(A_2^I)$, which is
  projected to $u_i \trans{A_i} v_i$ along $\alpha_i$ for $i \in \{
  1,2\}$.

  Vice versa, any step $u_3 \trans{A_3} v_3$ projects over two
  compatible steps $u_1 \trans{\sres{A_3}{\alpha_1}} v_1$ of $Z_1$ and
  $u_2 \trans{\sres{A_3}{\alpha_2}} v_2$ of $Z_2$, whose composition
  gives back the original step.
\end{lem}
% OLD VERSION
% \begin{lem}[composing steps]
%   \label{le:compose-step}
%   Let $f_1 : Z_0 \to Z_1$ and $f_2 : Z_0 \to Z_2$ be composable
%   embeddings in $\onet$ and let $Z_3 = Z_1 \comp{f_1,f_2} Z_2$.  Let $u_1
%   \trans{A_1} v_1$ and $u_2 \trans{A_2} v_2$ be compatible steps
  
%   Then there exists a unique step $u_3 \trans{A_3} v_3$ which is projected to
%   $u_i \trans{A_i} v_i$ along $\alpha_i$ for $i \in \{ 1,2\}$.
% \end{lem}

\begin{proof}
  Concerning the first part, by definition of compatibility, we know
  that $A_1$ and $A_2$ can be decomposed as $A_i = A_i^I \oplus A_i^E$
  ($i \in \{1,2\}$) such that
  \begin{center}
    $A_2^E = \monSub{f}{2}(\sres{A_1^I}{f_1})$ \quad and \quad
    $A_1^E = \monSub{f}{1}(\sres{A_2^I}{f_2})$.
  \end{center}
  Moreover, $\res{u_1}{f_1} = \res{u_2}{f_2}$.

  Now, since $u_i \trans{A_i^I \oplus A_i^E} v_i$, we can find markings 
  $u_i^I$, $u_i^E$, $v_i^I$, $v_i^E$ such that
  \begin{quote}
    \begin{tabular}{lll}
      $u_1^I \trans{A_1^I} v_1^I$ \ \ \ \ &  $u_2^E \trans{A_2^E} v_2^E$, \ \ & $\res{u_1^I}{f_1} = \res{u_2^E}{f_2}$\\
    $u_1^E \trans{A_1^E} v_1^E$, & $u_2^I \trans{A_2^I} v_2^I$, & $\res{u_1^E}{f_1} = \res{u_2^I}{f_2}$
  \end{tabular}
\end{quote}
  In fact, just observe that, since $u_i \trans{A_i} v_i$, the marking $u_i$
  must be of the kind $w_i \oplus \pre[A_i^I] \oplus \pre[A_i^E]$ and 
  similarly $v_i = w_i \oplus \post[A_i^I] \oplus \post[A_i^E]$.
  Thus
  we could choose 
  \begin{quote}
    $u_1^I = \pre[A_1^I]$, \quad $v_1^I = \post[A_1^I]$,  \quad
    $u_2^E = \pre[A_2^E]$, \quad $v_2^E = \post[A_2^E]$,
  \end{quote}
  and dually
  \begin{quote}
    $u_1^E= \pre[A_1^E] \oplus w_1$, \quad $v_1^E = \post[A_1^E] \oplus w_1$ \quad
    $u_2^I = \pre[A_2^I] \oplus w_2$, \quad $v_2^I = \post[A_2^I] \oplus w_2$
  \end{quote}
  Therefore, we can use Lemma~\ref{le:downup} and, defining
  $u_3' = u_1^I \mjoin u_2^E$, $u_3'' = u_1^E \mjoin u_2^I$,
  $v_3' = v_1^I \mjoin v_2^I$, $v_3'' = v_1^E \mjoin v_2^E$,
  we conclude
  \begin{quote}
    $u_3' \trans{\monSub{\alpha}{1}(A_1^I)} v_3'$ and 
    $u_3'' \trans{\monSub{\alpha}{2}(A_2^I)} v_3''$ 
  \end{quote}
  Therefore
  \begin{quote}
    $u_3' \oplus u_3'' \trans{\monSub{\alpha}{1}(A_1^I) \oplus
    \monSub{\alpha}{2}(A_2^I)} v_3' \oplus v_3''$
  \end{quote}

  By exploiting Proposition~\ref{pr:sum-mark}, we easily see that
  $u_3' \oplus u_3'' = (u_1^I \oplus u_1^E) \mjoin (u_2^E \oplus u_2^I) = u_1 \mjoin u_2$, where $u_0$ denotes
  the common projection of $u_1$ and $u_2$ over $Z_0$. Similarly,
  $v_3' \oplus v_3'' = v_1 \mjoin v_2$ and thus
  \begin{quote}
    $u_1 \mjoin u_2
    \trans{\monSub{\alpha}{1}(A_1^I) \oplus \monSub{\alpha}{2}(A_1^E)}
    v_1 \mjoin v_2$
  \end{quote}
  is the desired step. The fact that it projects over the steps we started
  from in $Z_1$ and $Z_2$ follows by construction.

  \bigskip 

  For the second part, consider any step $u_3 \trans{A_3} v_3$. Let
  $A_1 = \sres{A_3}{\alpha_1}$ and $A_2 = \sres{A_3}{\alpha_2}$.
  Decompose $A_3$ as
  \begin{quote}
    $A_3 = A_3^1 \oplus A_3^2 \oplus A_3^0 \oplus A_3^{open}$
  \end{quote}
  where $A_3^j$, for $j \in \{ 1, 2\}$ includes only transitions in
  $\alpha_i(T_{Z_i} - f_i(T_{Z_0}))$, $A^0_3$ includes only transitions in
  $\alpha_i(f_i(T_{Z_0}))$ and finally $A_3^{open}$ includes only
  interactions with the environment.

  Then, if we define
  \begin{quote}
    \begin{tabular}{lll}
      $A_1^I = \sres{A_3^1}{\alpha_1}$ & \hspace{10mm} & $A_1^E = A_1 \ominus A_1^I$\\
      $A_2^I = \sres{(A_3^2 \oplus A_3^0 \oplus A_3^{open})}{\alpha_2}$ & \quad & $A_2^E = A_2 \ominus A_2^I$
    \end{tabular}
  \end{quote}
  it is easy to show that the decomposition satisfies the requirements
  in Definition~\ref{de:compatible-steps}, hence the two steps are
  compatible, and their composition is immediately seen to give back
  the original step.
\end{proof}
Note that, in the decomposition of steps $A_1$ and $A_2$ considered in
the proof above, all firings of transitions in the interface $Z_0$ are
included in the internal part of $A_2$, i.e., no such transition is
included in $A_1^I$.  The possibility of having a decomposition with
these properties will be useful later, in the proof of the congruence
results.

\section{Bisimilarity of Open Nets}
\label{se:bisim}

In this section we study various notions of bisimilarity for open
nets, proving that they are congruences with respect to the
colimit-based composition operation. The considered behavioural
equivalences will differ for the choice of the observations, which can
be single firings or parallel steps. Additionally, we will consider
weak forms of such equivalences, arising in the presence of
unobservable actions.

\subsection{A High Level View on the Congruence Results}
\ \\

\noindent
A first step consists of defining suitable \emph{labelled transition
  systems} (\textsc{lts}s) associated with an open net.  Generally
speaking, net transitions carry a label which is observed when they
fire. Additionally, in the labelled transition systems we
also observe what happens at the open places. This corresponds to
observing the potential interactions with the surrounding environment,
as open places act as gluing points in the composition operation, and
it is pivotal for the mentioned congruence results.

Given an open net $Z$, the labeled transition systems we
shall consider will have all markings of the net, $\monSub{S}{Z}$, as
states, but they will differ concerning the transitions and their
labels. For example, in the \emph{firing} \textsc{lts} the transitions
are generated by the firings of $Z$, and correspondingly they are
labelled over the set
\begin{center}
  $\Lambda_Z = \Lambda \cup \{ +_s : s \in O_Z^+ \} \cup \{ -_s
  : s \in O_Z^- \}$.
\end{center}
%
%% More precisely, given an open net $Z$, first possible labelled
%% transition system, called \emph{firing} \textsc{lts}, has the markings
%% of the net as states.  Transitions are generated by firings of
%% $Z$ and correspondingly labelled over the set
%
As discussed in the conclusions, the firing \textsc{lts} resembles the
labelled transition system arising from the view of Petri nets as
reactive systems in~\cite{Mil:BRS,SS:CPN}.
Analogous \textsc{lts}s are also obtained in~\cite{v:modular-petri}
with the use of pseudo-transitions and in~\cite{NPS:CBCP} by
inserting a net in a universal context.

Instead, in the \emph{step} \textsc{lts} the transitions are generated by
the steps of $Z$, and they are labeled over   $\monSub{\Lambda}{Z}$.
%
%% Alternatively, since we are dealing with a concurrent model, it can be
%% natural to observe the parallel execution of events, i.e., to consider
%% labels corresponding to the executions of steps, rather than to single
%% firings, thus obtaining the so-called \emph{step} \textsc{lts}. 
%
The
corresponding notion of bisimilarity will capture, to some extent, the
concurrency properties of the system (see,
e.g.,~\cite{Vog:BAR,NT:DNDC}).

%% \begin{itemize}

%% \item Label $-_s$: environment action removes a token in an open
%%   output place;

%% \item Label $+_s$: environment action adds a token in an  input open
%%   place;

%% \item Label $a$: firing of a transition $t$ labelled by $a =
%%   \lambda_Z(t)$.

%% \end{itemize}

For notational convenience we extend the labelling function
$\lambda_Z$ to the set of extended events $\bar{T}_Z$, by defining
$\lambda_Z(x) = x$ for $x \in \bar{T}_Z -T_Z$ (i.e., for $x = +_s$ or
$x= -_s$ with $s \in S_Z$).

\begin{defi}[step and firing lts for an open net]
  The \emph{step} \textsc{lts} associated to an open net $Z$ is the
  pair $\langle \monSub{S}{Z}, \to_{\mathsf{S},Z} \rangle$, where states are
  markings $u_Z \in \monSub{S}{Z}$ and the transition relation
  $\to_{\mathsf{S},Z}\, \subseteq \monSub{S}{Z} \times \monSub{\Lambda}{Z} \times
  \monSub{S}{Z}$ includes all transitions
  \[    
  u_Z  \ltr[S]{Z}{\monSub{\lambda}{Z}(A)} u_Z'
  \]
  for all markings $u_Z, u_Z' \in \monSub{S}{Z}$ and $A \in
  \mon{\bar{T}_Z}$ such that there is a step $u_Z \trans{A} u_Z'$ in
  $Z$.
  The \emph{firing} \textsc{lts} $\langle \monSub{S}{Z},
  \to_{\mathsf{F},Z} \rangle$ is defined 
  similarly: the transition  relation
  $\to_{\mathsf{F},Z}\, \subseteq \monSub{S}{Z} \times \Lambda_Z \times
  \monSub{S}{Z}$ includes all transitions
  \[    
  u_Z  \ltr[F]{Z}{\lambda_Z(\epsilon)} u_Z'
  \]
  such that there is a firing $u_Z \trans{\epsilon} u_Z'$ in
  $Z$, with $\epsilon \in \bar{T}_Z$.
\end{defi}

As we have done above for the transition relations, in the sequel the
subscripts ``\textsf{S}'' and ``\textsf{F}'' will be used for distinguishing
notions based on the step and on the firing behaviour,
respectively, of a net.

When observing the behaviour of a system, usually only a subset of
events is considered visible.  Here this is formalised by selecting
a subset of labels representing internal firings, playing a role
similar to $\tau$-actions in process calculi, and then considering a
corresponding notion of weak bisimilarity.
Let $\Lambda_\tau \subseteq \Lambda$ be a subset of
\emph{unobservable} labels, fixed for the rest of the paper.

\begin{defi}[weak transition systems]
  For $\mathsf{x} \in \{ \mathsf{S}, \mathsf{F} \}$ we write $v
  \wltr{Z}{\ell} v'$ if $v, v' \in \monSub{S}{Z}$ are markings such
  that $v \ltr{Z}{\ell'} v'$ with $\ell =
  \res{\ell'}{(\Lambda-\Lambda_\tau)}$.
  Then the \emph{weak} (step or firing) \textsc{lts} is defined by
  letting

  \begin{enumerate}[$\bullet$]
    
  \item $v \Ltr{Z}{0} v'$ whenever $v \wltrStar{Z}{0} v'$.
    
    % \item $v \Ltr{Z}{\ell} v'$ when $v (\wltrtau{Z})^* \wltr{Z}{\ell}
    %   (\wltrtau{Z})^* v'$.
    
  \item $v \Ltr{Z}{\ell} v'$ whenever $v\
    \wltrStar{Z}{0}\ \wltr{Z}{\ell}\
    \wltrStar{Z}{0}\ v'$ \qquad $\ell \neq 0$.
    
  \end{enumerate}
\end{defi}

Transitions labelled with $0$ will be often referred to as
$\tau$-transitions or silent transitions.

Weak step and firing bisimilarity is now defined in a standard way,
but note that when the set of unobservable labels is empty, this
actually  corresponds to strong bisimilarity. 
Only, in order to be able to relate the extended events of the two
nets, we need to specify for each open place of one net which is the
corresponding open place in the other net; therefore bisimulations
between two nets are parametrised by a bijection between their open
places.
Given two open nets $Z_1$ and $Z_2$ a \emph{correspondence}
$\eta = \langle \eta^+, \eta^- \rangle$ between $Z_1$ and $Z_2$ is a
pair of bijections $\eta^+ : O_{Z_1}^+ \to O_{Z_2}^+$ and $\eta^- : O_{Z_1}^- \to
O_{Z_2}^-$. In order to simplify the notation, in the following, given an
open place $s_1 \in O_{Z_1}^+ \cup O_{Z_1}^-$ we will write simply $\eta(s_1)$
to denote its image through the appropriate component of $\eta$, i.e.,
a correspondence $\eta = \langle \eta^+, \eta^- \rangle$ will be
identified with the function $\eta^+ \cup \eta^- : O_{Z_1}^+ \cup O_{Z_1}^-
\to O_{Z_2}^+ \cup O_{Z_2}^-$.

\begin{defi}[(weak) step and firing bisimilarity]
  \label{de:weak-bisim}
  Let $Z_1$, $Z_2$ be open nets and $\eta : O_{Z_1} \leftrightarrow
  O_{Z_2}$ be a correspondence between $Z_1$ and $Z_2$.
  A (weak) $\eta$-$\mathsf{x}$-bisimulation (with $\mathsf{x} \in \{
  \mathsf{S}, \mathsf{F} \}$ - $\mathsf{S}$ for \emph{step} and $\mathsf{F}$ for \emph{firing})
  between $Z_1$ and $Z_2$ is a relation over markings $\mathcal{R}
  \subseteq \monSub{S}{1} \times \monSub{S}{2}$ such that if $(u_1,
  u_2) \in \mathcal{R}$ then

  \begin{enumerate}[$\bullet$]
%   \item if $u_1 \wltrtau{Z_1} u_1'$, then there exists $u_2'$ such
%     that $u_2 Ltr{Z_2}{0} u_2'$ and $(u_1', u_2') \in
%     \mathcal{R}$;

  \item if $u_1 \wltr{Z_1}{\ell} u_1'$ in $Z_1$, then there exists
    $u_2'$ such that $u_2 \Ltr{Z_2}{\mon{\eta}(\ell)} u_2'$ in
    $Z_2$ and $(u_1', u_2') \in \mathcal{R}$;

  \item the symmetric condition holds;

  \end{enumerate}
  where $\eta(+_s) = +_{\eta(s)}$, $\eta(-_s) = -_{\eta(s)}$, and
  $\eta(\ell) = \ell$ for any $\ell \in \Lambda$.

  Two open nets $Z_1$ and $Z_2$ are \emph{(weakly)
    $\eta$-$\mathsf{x}$-bisimilar}, denoted $Z_1 \approx_\eta^\mathsf{x} Z_2$, if
  $\eta : O_{Z_1} \leftrightarrow O_{Z_2}$ is a correspondence and there
  exists a (weak) $\eta$-bisimulation $\mathcal{R}$ over $Z_1$ and
  $Z_2$ such that $(\init{u}_1,\init{u}_2) \in \mathcal{R}$.  We will
  say that $Z_1$ and $Z_2$ are (weakly) $\mathsf{x}$-bisimilar,
  written $Z_1 \approx^\mathsf{x} Z_2$, if $Z_1
  \approx_\eta^\mathsf{x} Z_2$ for some correspondence $\eta$.
\end{defi}
Clearly, step bisimilarity is finer than firing bisimilarity, i.e., if $Z_1 \approx^{\mathsf{S}} Z_2$ then $Z_1 \approx^{\mathsf{F}} Z_2$.

Observe that in the definition of step bisimilarity, whenever $v
\Ltr[S]{Z}{\ell} v'$ and thus $v\ \wltrStar[S]{Z}{0}\ \wltr[S]{Z}{\ell}\
\wltrStar[S]{Z}{0}\ v'$, one can assume that the step inducing
$\wltr[S]{Z}{\ell}$ does not include any $\tau$-transition (since, if
this is not the case, the $\tau$-transitions can be anticipated or
postponed).

\begin{figure}[t]
  \begin{center}
  \subfigure[Travel agency A.]{
    \label{fi:step-example-a}
    \scalebox{.42}{\includegraphics{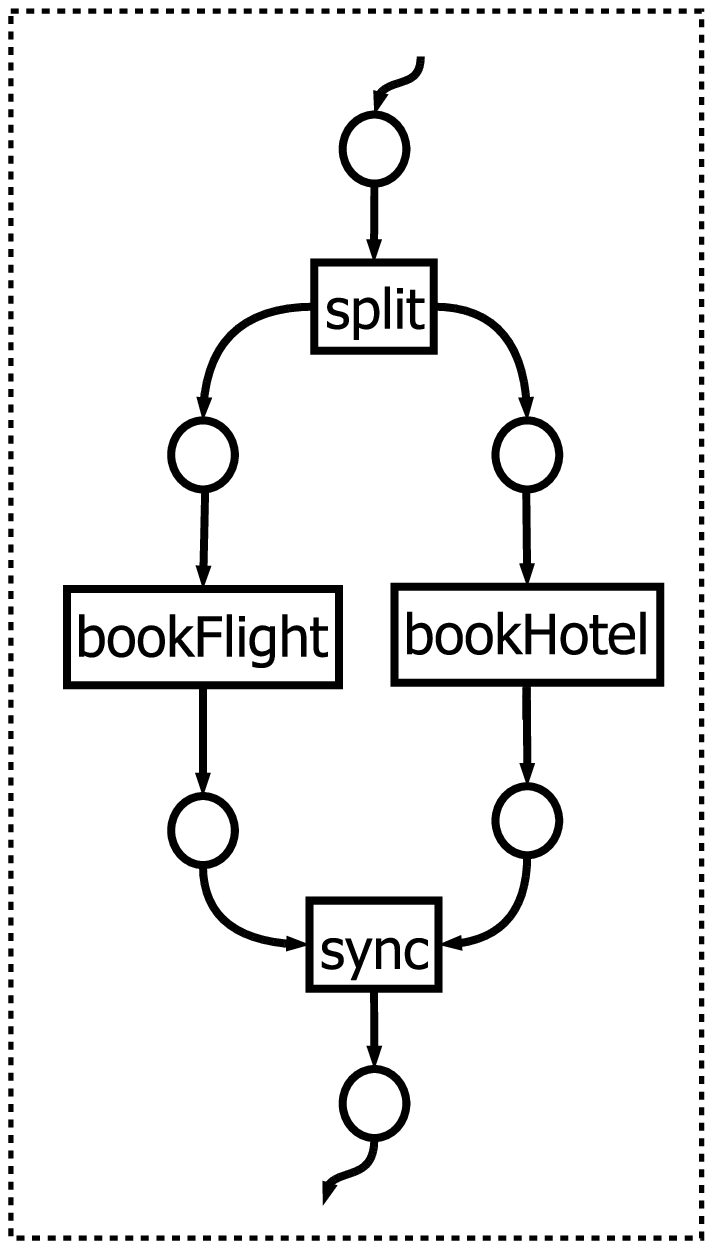}}}
  \hspace{2cm}
  \subfigure[Travel agency B.]{
    \label{fi:step-example-b}
    \scalebox{.42}{\includegraphics{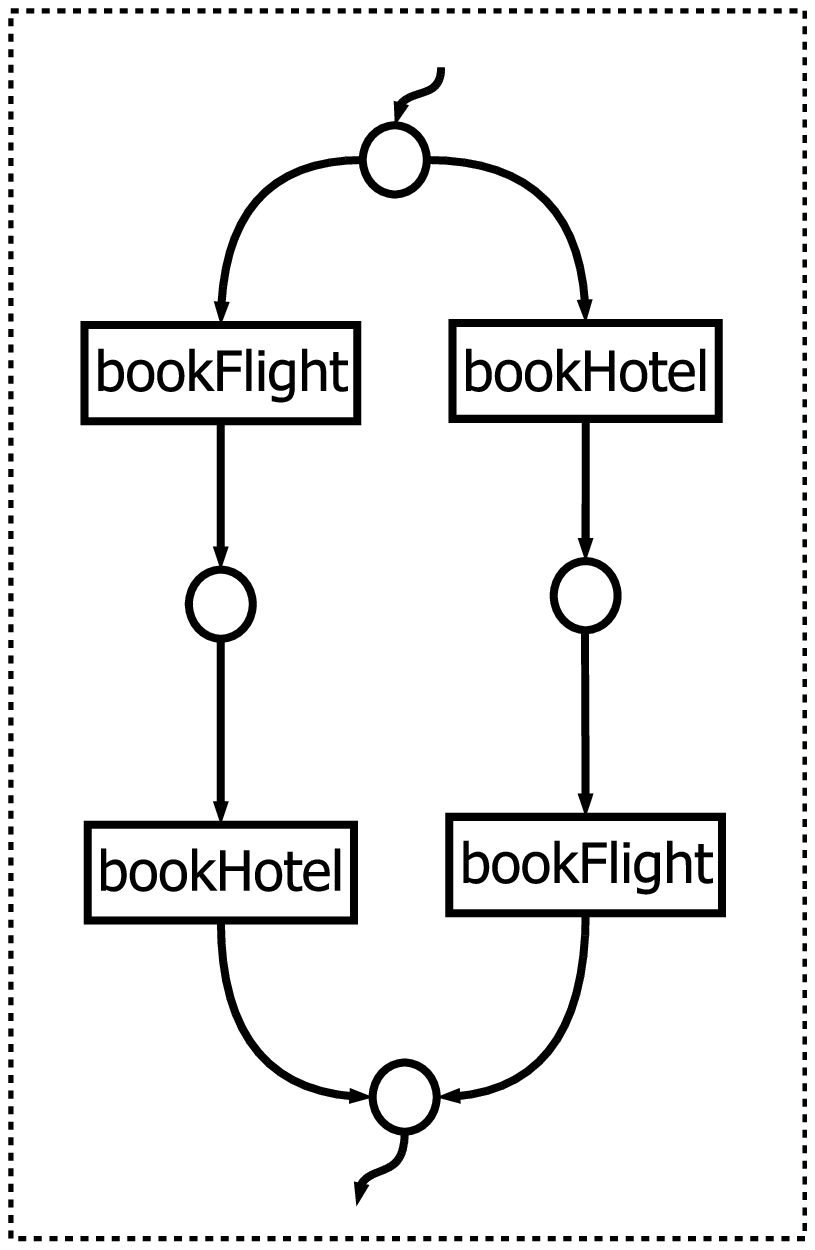}}}
\end{center}
\caption{Two open nets which are firing bisimilar but not step bisimilar.}
  \label{fi:step-example}
\end{figure}

As an example, consider the open nets in Fig.~\ref{fi:step-example},
which can be seen as the representation of (part of) the booking
process in a travel agency. The bookings of the flight
(\textsf{bookFlight}) and of the hotel (\textsf{bookHotel}) are
independent and could be performed in parallel. However, this is
possible only for agency A (Fig.~\ref{fi:step-example-a}), while in
agency B (Fig.~\ref{fi:step-example-b}), where a single person takes care
of all bookings, the two actions will be executed sequentially.
Now, it is easy to check that, assuming that only the actions
\textsf{bookFlight} and \textsf{bookHotel} are visible, the two nets
are firing bisimilar, but they are not step bisimilar.
Hence, as already mentioned, step bisimilarity discriminates
also according to the degree of parallelism that is possible in a
computation.

As already mentioned, weak bisimilarity boils down to the notion of
strong bisimilarity when all labels are observable, i.e., when
$\Lambda_\tau = \emptyset$.
For convenience of the reader we make explicit the notion of
strong bisimilarity.

\begin{defi}[strong bisimilarity]
  \label{de:strong}
  When $Z_1$ and $Z_2$ are weakly $\eta$-$\mathsf{x}$-bisimilar open
  nets, with $\Lambda_\tau = \emptyset$ we say that $Z_1$ and $Z_2$
  are \emph{strongly $\eta$-$\mathsf{x}$-bisimilar} and write $Z_1
  \sim_\eta^\mathsf{x} Z_2$ or simply $Z_1 \sim^{\mathsf{x}}
  Z_2$. Explicitly, a \emph{strong $\eta$-$\mathsf{x}$-bisimulation}
  over $Z_1$ and $Z_2$ is a relation over their markings $\mathcal{R}
  \subseteq \monSub{S}{1} \times \monSub{S}{2}$ such that if $(u_1,
  u_2) \in \mathcal{R}$ then
  
  \begin{enumerate}[$\bullet$]
  \item if $u_1 \ltr{Z_1}{\ell} u_1'$ in $Z_1$, then there
    exists $u'_2$ such that $u_2 \ltr{Z_2}{\eta(\ell)} u_2'$ in
    $Z_2$ and $(u_1', u_2') \in \mathcal{R}$;
      
  \item the symmetric condition holds.
    
  \end{enumerate}
  
\end{defi}

We can finally state the congruence property for the considered
behavioural equivalences with respect to the composition operation on
open nets. The result will be proved separately for the various cases
in the next subsection.

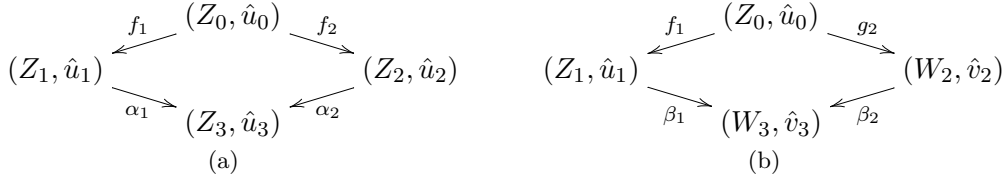
\begin{figure}[t]
  \subfigure[]{
    \xymatrix@R=1mm{
      \label{fi:th-congruence-a}
      {} & {(Z_0,\init{u}_0)} \ar[dr]^{f_2} \ar[dl]_{f_1} & {}\\
      {(Z_1,\init{u}_1)} \ar[dr]_{\alpha_1} & {} & {(Z_2,\init{u}_2)} \ar[dl]^{\alpha_2} \\
      {} & {(Z_3,\init{u}_3)} & {}
    }
  }
  \hspace{5mm}
  \subfigure[]{
    \label{fi:th-congruence-b}
    \xymatrix@R=1mm{
      {} & {(Z_0,\init{u}_0)} \ar[dr]^{g_2} \ar[dl]_{f_1} & {}\\
      {(Z_1,\init{u}_1)} \ar[dr]_{\beta_1} & {} & {(W_2,\init{v}_2)} \ar[dl]^{\beta_2} \\
      {} & {(W_3,\init{v}_3)} & {} } }
  \caption{Pushouts in $\onet$.}
  \label{fi:th-congruence}
\end{figure}

\begin{thm}[bisimilarity is a congruence]
  \label{th:congruence}
  Let $Z_0$, $Z_1$, $Z_2$, $W_2$ be open nets.
  Let $Z_2 \approx_\eta^{\mathsf{x}} W_2$, for some correspondence $\eta$ and
  ${\mathsf{x}} \in \{ \mathsf{S}, \mathsf{F} \}$. Consider the nets $Z_3 = Z_1
  \comp{f_1,f_2} Z_2$ and $W_3 = Z_1 \comp{f_1,g_2} W_2$, as in
  Fig.~\ref{fi:th-congruence} where $f_1$, $f_2$ and $g_2$ are
  embeddings, $f_1$ and $f_2$ are composable, and $f_1$ and $g_2$ are
  composable as well.

  If $g_2|_{O_0} = \eta \circ (f_2|_{O_0})$ (i.e., $f_2$ and $g_2$ are
  consistent with $\eta$ on open places) then $Z_3 \approx_{\eta'}^{\mathsf{x}}
  W_3$, where $\eta': O_{Z_3} \leftrightarrow O_{W_3}$ is the
  correspondence defined as follows:
  for all $s \in O_{Z_3}$, $\eta'(s) = \beta_1(s')$ if $s =
  \alpha_1(s')$, and  $\eta'(s) = \beta_2(\eta(s'))$ if $s =
  \alpha_2(s')$.
\end{thm}

\subsection{Proofs of the Congruence Results}
\ \\

\noindent
In order to prove the congruence results it is convenient to proceed
as follows: we first consider strong step bisimilarity which can be
more easily handled than its weak variant.  Next the proof of the
congruence result for the weak variant can adapted from the strong
case. Finally, as firing bisimulation can (almost) be considered as a
special case of step bisimulation, the proof of the corresponding
congruence result easily follows from that of step bisimilarity. It is
worth stressing that the complexity of the proof is mainly due to the
fact that we consider steps instead of single firings. 

We start with a technical lemma which will play a central role 
later.
It states that for given \emph{composable}
embeddings $f_1 :Z_0 \to Z_1$ and $f_2: Z_0 \to Z_2$, any step in
$Z_2$ where interactions with the environment only occur on places
which are open also in $Z_1 +_{Z_0} Z_2$, can be projected along $f_2$
to $Z_0$ and then simulated in $Z_1$.

\begin{lem}
  \label{le:move-step}
  Let $f_1 : Z_0 \to Z_1$ and $f_2 : Z_0 \to Z_2$ be composable
  embeddings in $\onet$, let $Z_3 = Z_1 \comp{f_1,f_2} Z_2$ and let
  $u_i \in \mon{S_{Z_i}}$ ($i \in \{ 1, 2\}$) be markings such that
  $\res{u_1}{f_1} = \res{u_2}{f_2}$.  Let $u_2 \trans{A_2} v_2$ be a
  step such that if $x_s \in A_2$, for $x \in \{ +, - \}$ then
  $\alpha_2(s) \in O_{Z_3}^x$. Then $u_1
  \trans{\monSub{f}{1}(\sres{A_2}{f_2})} v_1$ and $u_1 \mjoin u_2
  \trans{\monSub{\alpha}{2}(A_2)} v_1 \mjoin v_2$.
\end{lem}

\begin{proof}
  Let $A_1 = \monSub{f}{1}(\sres{A_2}{f_2})$. 
  First note that $A_1$ is well-defined, i.e., $A_1 \in \monSub{\bar{T}}{{Z_1}}$.
  For instance, let us show that if $+_{s_1} \in A_1$ then $s_1$ is
  input open, i.e., $s_1 \in O_{Z_1}^+$.  By definition of $A_1$ we
  deduce that there is $+_{s_0} \in \sres{A_2}{f_2}$ with $f_1(s_0)
  =s_1$. Now, by the assumptions on $A_2$, there are two
  possibilities:
  \begin{enumerate}[$\bullet$]
  
  \item $+_{s_0} \in \sres{t_2}{f_2}$ with $t_2 \in T_{Z_2}$\\
    By the definition of projection for steps, this implies that
    $f_2(s_0) \in \pre[t_2]$, with $t_2 \not\in f_2(T_{Z_0})$ and thus
    $s_0 \in \inp{f_2}$.  Since $f_1$ and $f_2$ are composable, we
    have that $s_1 = f_1(s_0) \in f_1(\inp{f_2}) \subseteq O_{Z_1}^+$,
    as desired.

  \item $+_{s_0} \in \sres{+_{s_2}}{f_2}$ with $\alpha_2(s_2) \in O_{Z_3}^+$\\
    Since the diagram in Fig.~\ref{fi:pushout-onet} commutes, we have
    that $\alpha_1(s_1) = \alpha_2(s_2)$. Since $\alpha_2(s_2) \in
    O_{Z_3}^+$, by condition (1) in the definition of open net
    morphism (Definition~\ref{de:open-net-morphism}), $s_1 \in
    O_{Z_1}^+$, as desired.
  \end{enumerate}
  Now observe that
  \begin{quote}
    \begin{tabular}{lll}
      $\pre[A_1] = \pre[(\monSub{f}{1}(\sres{A_2}{f_2}))]$\\
      \ \quad $= \monSub{f}{1}(\pre[\sres{A_2}{f_2}])$  \quad & [by def. of open
      net morphism]\\
      \ \quad $= \monSub{f}{1}(\res{\pre[A_2]}{f_2})$ & [by Lemma~\ref{le:proj-prop}.(3)]
    \end{tabular}
  \end{quote}

\noindent
Since the step $u_2
  \trans{A_2} v_2$ is enabled, we know that $\pre[A_2] \leq u_2$, and thus
  \begin{quote}
    \begin{tabular}{lll}
      $\pre[A_1] = \monSub{f}{1}(\res{\pre[A_2]}{f_2})$\\
      \ \quad $\leq \monSub{f}{1}(\res{u_2}{f_2})$\\
      \ \quad $=  \monSub{f}{1}(\res{u_1}{f_1})$  
      & \quad & [since $\res{u_2}{f_2} = \res{u_1}{f_1}$]\\
      \ \quad $\leq u_1$& 
      \quad & [by Lemma~\ref{le:proj-prop}.(4)]
    \end{tabular}
  \end{quote}
  Hence, the step $u_1 \trans{A_1} v_1$ can be performed. Clearly, the two
  steps in $Z_1$ and $Z_2$ are compatible, and thus we conclude with
  Lemma~\ref{le:compose-step}.
\end{proof}

\subsubsection{Strong Step Bisimilarity}

\begin{thm}
  \label{th:congruence-strong-step}
  Strong step bisimilarity is a congruence.
\end{thm}

\begin{proof}
  Let $Z_0$, $Z_1$, $Z_2$, $W_2$ be open nets, with $Z_2
  \sim_\eta^{\mathsf{S}} W_2$, for some correspondence $\eta$.
  Let  $Z_3 = Z_1
  \comp{f_1,f_2} Z_2$ and $W_3 = Z_1 \comp{f_1,g_2} W_2$, as in
  Fig.~\ref{fi:th-congruence}, where $f_1$, $f_2$ and $g_2$ are
  embeddings, with $f_1$, $f_2$ and  $f_1$, $g_2$
  composable and $g_2|_{O_{Z_0}} = \eta \circ (f_2|_{O_{Z_0}})$.

  To simplify the notation, assume, without loss of generality, that
  all the morphisms in the diagrams of Fig.~\ref{fi:th-congruence} are
  inclusions and $\eta = id$.  Hence $f_2|_{O_{Z_0}} = g_2|_{O_{Z_0}}$.

  Now let $\mathcal{R}$ be a $\eta$-$\mathsf{S}$-bisimulation over $Z_2$ and
  $W_2$ such that $(\init{u}_2, \init{v}_2) \in \mathcal{R}$, which exists by
  hypothesis.  Consider the relation $\mathcal{R}'$ over $Z_3$ and
  $W_3$ defined as

 \begin{center}
   $\mathcal{R}' = \{ (u_1 \mjoin_{u_0} u_2, v_1 \mjoin_{v_0} v_2) : (u_2, v_2)
   \in \mathcal{R}\ \land\ u_1 \ominus u_0
    = v_1 \ominus v_0 \}$
  \end{center}
  The condition above on $u_1$ and $v_1$ means that the markings can
  differ, but only for the number of tokens in places of the interface
  net $Z_0$ (notice that the marking of $Z_0$ is completely determined
  by the marking of components $Z_2$ and $W_2$).

  We claim that $\mathcal{R}'$ is a $\eta'$-$\mathsf{S}$-bisimulation
  over $Z_3$ and $W_3$, where $\eta'$ is again the identity on open
  places.  Since, by the construction of the pushout, $(\init{u}_3,
  \init{v}_3) = (\init{u}_1 \mjoin_{\init{u}_0} \init{u}_2, \init{u}_1
  \mjoin_{\init{u}_0} \init{v}_2) \in \mathcal{R}'$, this provides the
  desired result.

  \medskip

  In order to prove that $\mathcal{R}'$ is a
  $\eta'$-$\mathsf{S}$-bisimulation, assume that $u_3
  \ltr[S]{Z_3}{\ell} u_3'$.
  Therefore
  \begin{center}
    $u_3 \trans{A_3} u_3'$ \quad with $\ell = \monSub{\lambda}{Z_3}(A_3)$
  \end{center}
  and by Lemma~\ref{le:compose-step} we can project the step $A_3$ over
  the components $Z_1$ and $Z_2$ thus getting for $i \in \{ 1, 2\}$
  the following steps in $Z_i$:
  \begin{equation}
    \label{eq:single-steps}
    u_i \trans{A_i} u_i'   
  \end{equation}
  Since by the same lemma such steps are compatible, according to
  Definition~\ref{de:compatible-steps}, we can find partitions
  \begin{center}
    $A_i = A_i^I \oplus A_i^E$ with  $i \in \{ 1, 2\}$
  \end{center}
  such that
  \begin{equation}
    \label{eq:compat}
    A_1^E = \monSub{f}{1}(\sres{A_2^I}{f_2}) \qquad     
    A_2^E = \monSub{f}{2}(\sres{A_1^I}{f_1})
  \end{equation}
  and 
  \begin{equation}
    \label{eq:A3}
    A_3 = \monSub{\alpha}{1}(A_1^I) \oplus \monSub{\alpha}{2}(A_2^I)
  \end{equation}
  Additionally, as shown in the proof of Lemma~\ref{le:compose-step},
  we can assume, w.l.o.g., that $A_2^E$ consists only of interactions
  with the environment, i.e., $A_2^E \in \mon{\{x_s \mid x \in \{ +,
    -\},\ s \in O_{Z_2}^x\}}$, or, equivalently, that $A_1^I$ does not
  contain firings of transitions of $Z_0$.

  \medskip

  Now, since $(u_2,v_2) \in \mathcal{R}$, the step
  (\ref{eq:single-steps}) of $Z_2$ can be simulated by $W_2$, i.e.,
  there is
  \begin{equation}
    \label{eq:trans2}
    v_2 \trans{B_2} v_2'
  \end{equation}
  with $\mon{\lambda}(B_2) = \mon{\lambda}(A_2)$ and $(u_2', v_2') \in
  \mathcal{R}$.

  \bigskip

  We can now split $B_2$ in an ``internal'' and an ``external'' part,
  according to the splitting of $A_1$, i.e., we define
  \begin{equation}
    \label{eq:B2E}
    B_2^E = A_2^E \qquad \qquad B_2^I = B_2 \ominus B_2^E
  \end{equation}
  Notice that we can legally define $B_2^E = A_2^E$ since $A_2^E$
  consists only of interactions with the environment, which are
  necessarily also in $B_2$ since $\mon{\lambda}(B_2) =
  \mon{\lambda}(A_2)$ (and recall that places in the interface have
  the same name in $Z_2$ and $W_2$).

  Now, define
  \begin{eqnarray}
    \label{eq:B2}
    v_2^I = \pre[B_2^I] & \quad & {v_2^I}' = \post[B_2^I]\\
    \label{eq:v2}
    v_2^E = v_2 \ominus v_2^I &  &  {v_2^E}' = v_2' \ominus {v_2^I}'
  \end{eqnarray}
  and thus we have
  \begin{equation}
    \label{eq:step-in}
    v_2^I \trans{B_2^I} {v_2^I}'
  \end{equation}
  \begin{equation}
    \label{eq:step-ext}
    v_2^E \trans{B_2^E} {v_2^E}'
  \end{equation}

  Now, the idea is to construct a step in $W_3$ by using
  separately the internal part of the step in $W_2$ and the internal
  part of the step in $Z_1$ (which plays the role of a context).

  \medskip

  In order to apply Lemma~\ref{le:move-step} to the step
  in~(\ref{eq:step-in}), we note that if $+_s \in B_2^I$ then $s \in
  O_{W_3}^+$ (and the same holds for $-_s$).
  In fact, if $+_s \in B_2^I$, then by construction of $B_2^I$ and
  since $\mon{\lambda}(A_2) = \mon{\lambda}(B_2)$, we must have $+_s
  \in A_2^I$. Now, if $s \not\in Z_0$ then, given that $s \in
  O_{Z_2}^+$ we have that $s \in O_{Z_3}^+ = O_{W_3}^+$. Otherwise, if
  $s \in Z_0$ then, by~(\ref{eq:compat}), we have that
  $\monSub{f}{1}(\res{+_s}{f_2}) = +_s \in A_1^E$,  thus $s \in O_{Z_1}^+$, and
  hence $s \in O_{Z_3}^+ = O_{W_3}^+$.

  Therefore if we define:
  \begin{equation}
    \label{eq:v1E}
    v_1^E = \monSub{f}{1}(\res{v_2^I}{g_2}) \qquad
    B_1^E =  \monSub{f}{1}(\sres{B_2^I}{g_2})
  \end{equation}
  since clearly $\res{v_1^E}{f_1} = \res{v_2^I}{g_2}$,
  we can apply Lemma~\ref{le:move-step} to deduce that
  \begin{center}
    $v_1^E \trans{B_1^E} {v_1^E}'$
  \end{center}
  and
  \begin{equation}
    \label{eq:int1}
    v_1^E \mjoin v_2^I \trans{\mon{\beta_2}(B_2^I)} {v_1^E}' \mjoin {v_2^I}'
  \end{equation}
  Note that $v_1^E \leq v_1$. In fact $v_2^I \leq v_2$. Therefore 
  $\res{v_2^I}{g_2} \leq \res{v_2}{g_2}$ and thus
  \begin{quote}
    $v_1^E = \monSub{f}{1}(\res{v_2^I}{g_2}) \leq \monSub{f}{1}(\res{v_2}{g_2}) \leq v_1$
  \end{quote}
  
  \bigskip

  Let us now construct the other part of the step in $W_3$,
  arising as the composition of an internal step in $Z_1$ and the
  external part of the step in $W_2$. As mentioned before, since the
  component $Z_1$ plays the role of a context (it is the same in both
  composed nets) we can simply define:
  \begin{equation}
    \label{eq:B1I}
    B_1^I=A_1^I
  \end{equation}
  If we let
  \begin{equation}
    \label{eq:v1I}
    v_1^I = v_1 \ominus v_1^E
  \end{equation}
  then we can see that
  \begin{equation}
    \label{eq:step-BI}
    v_1^I \trans{B_1^I} {v_1^I}'
  \end{equation}
  We can show that indeed $\pre[B_1^I] \leq v_1^I$, with a long, but
  easy calculation.  In fact, since $A_1^I = B_1^I$ by (\ref{eq:B1I})
  \begin{equation}
    \label{eq:B1Ileq}
    \pre[B_1^I] = \pre[A_1^I] = \res{\pre[A_1^I]}{(S_{Z_1} - S_{Z_0})} \oplus \res{\pre[A_1^I]}{S_{Z_0}}
  \end{equation}
  In the last expression, $\res{\pre[A_1^I]}{(S_{Z_1} - S_{Z_0})}$ and $
  \res{\pre[A_1^I]}{S_{Z_0}}$ stands for the projections along the
  inclusions of $S_{Z_1}-S_{Z_0}$ and $S_{Z_0}$, respectively, into $S_{Z_1}$. Now,
  let us consider the two summands separately. Concerning the first
  one:
  \begin{quote}
    \begin{tabular}{ll}
      $\res{\pre[A_1^I]}{(S_{Z_1}-S_{Z_0})}  \leq u_1 \ominus u_0 =$ &
      [since  $A_1^I$ enabled in $u_1$ by~(\ref{eq:single-steps})]\\
      \ \quad $= v_1 \ominus v_0 =$ & 
      [by construction of $\mathcal{R}'$]\\
      \ \quad $= v_1 \ominus \monSub{f}{1}(\res{v_2}{g_2})$
    \end{tabular}
  \end{quote}
  Let us consider the second one:
  \begin{quote}
    \begin{tabular}{ll}
      $\res{\pre[A_1^I]}{S_{Z_0}}  = \monSub{f}{1}(\res{\pre[A_1^I]}{f_1}) =$\\
      \ \quad $= \monSub{f}{1}(\res{\pre[A_2^E]}{f_2}) =$ & 
      [since  $\sres{A_1^I}{f_1} = \sres{A_2^E}{f_2}$ by~(\ref{eq:compat})]\\
      \ \quad $= \monSub{f}{1}(\res{\pre[B_2^E]}{g_2}) \leq$ &
      [by~(\ref{eq:B2E}) and the fact that $g_2$, $f_2$ agree on $O_{Z_0}$]\\
      \ \quad $\leq \monSub{f}{1}(\res{v_2^E}{g_2})$ &
      [since by~(\ref{eq:step-ext})  $\pre[B_2^E\ \leq v_2^E]$]
    \end{tabular}
  \end{quote}
  Putting together the two summands, from~(\ref{eq:B1Ileq}) we have
  \begin{quote}
    \begin{tabular}{ll}
      $\pre[B_1^I]  \leq v_1 \ominus \monSub{f}{1}(\res{v_2}{g_2}) \oplus \monSub{f}{1}(\res{v_2^E}{g_2}) =$\\
      \ \quad $= v_1 \ominus \monSub{f}{1}(\res{v_2 \ominus v_2^E}{g_2}) =$ &
      [since $v_2^E \leq v_2$ and $f$ injective]\\
      \ \quad $= v_1 \ominus \monSub{f}{1}(\res{v_2^I}{g_2}) =$ &
      [since  $v_2^I = v_2 \ominus v_2^E$ by~(\ref{eq:v2})]\\
      \ \quad $= v_1 \ominus v_1^E =$ &
      [by~(\ref{eq:v1E})]\\
      \ \quad $= v_1^I$ &
      [by~(\ref{eq:v1I})]
    \end{tabular}
  \end{quote}

  In order to apply Lemma~\ref{le:move-step} to the
  step~(\ref{eq:step-BI}), we can prove that if $+_s \in B_1^I$ then
  $s \in O_{W_3}^+$ (and the same for $-_s$) as in the previous case.
  Additionally, we have
  \begin{quote}
    \begin{tabular}{ll}
      $\res{v_1^I}{f_1}  = \res{(v_1 \ominus v_1^E)}{f_1} =$ & 
      [by def. of $v_1^I$ in~(\ref{eq:v1I})]\\
      \ \quad $= \res{v_1}{f_1} \ominus \res{v_1^E}{f_1} =$ & \\
      \ \quad $= \res{v_2}{g_2} \ominus \res{v_1^E}{f_1} =$ &
      [since  $\res{v_1}{f_1} = \res{v_2}{g_2}$ by hypothesis]\\
      \ \quad $= \res{v_2}{g_2} \ominus \res{v_2^I}{g_2} =$ &
      [since  $\res{v_1^E}{f_1} = \res{v_2^I}{g_2}$ by~(\ref{eq:v1E})]\\
      \ \quad $= \res{(v_2 \ominus v_2^I)}{g_2} =$ &\\
      \ \quad $= \res{v_2^E}{g_2}$ &
      [by def. of $v_2^E$ in~(\ref{eq:v2})]
    \end{tabular}
  \end{quote}
  and moreover
  \begin{center}
    $B_2^E  = \mon{g_2}(\sres{B_1^I}{f_1})$.
  \end{center}
  In fact
  \begin{quote}
    \begin{tabular}{lll}
      $B_2^E = A_2^E$ & & 
      [by~(\ref{eq:B2E})]\\
      \ \quad $= \monSub{f}{2}(\sres{A_1^I}{f_1}) =$ & &
      [by~(\ref{eq:compat})]\\
      \ \quad $= \monSub{f}{2}(\sres{B_1^I}{f_1}) =$ & &
      [by~(\ref{eq:B1I})]\\
      \ \quad $= \mon{g_2}(\sres{B_1^I}{f_1}) =$ & &
      [since $g_2$ and $f_2$ ``agree'' on $O_{Z_0}$ ]\\
    \end{tabular}
  \end{quote}
  
\noindent
  Therefore, by Lemma~\ref{le:move-step}, we have that 
  \begin{equation}
    v_2^E \trans{B_2^E} {v_2^E}'
  \end{equation}
  and
  \begin{equation}
    \label{eq:int2}
    v_1^I \mjoin v_2^E \trans{\mon{\beta_1}(B_1^I)} {v_1^I}' \mjoin {v_2^E}'
  \end{equation}
 
  \bigskip
\noindent
  Now, by Proposition~\ref{pr:sum-mark}, we can join the
  steps~(\ref{eq:int1}) and~(\ref{eq:int2}) and obtain
  \begin{center}
    $(v_1^E \mjoin v_2^I) \oplus (v_1^I \mjoin v_2^E) 
    \trans{\mon{\beta_1}(B_1^I) \oplus \mon{\beta_2}(B_2^I)} 
    ({v_1^I}' \mjoin {v_2^E}') \oplus ({v_1^E}' \mjoin {v_2^I}')$
  \end{center}
  i.e., the desired step which can be used to simulate $u_3
  \ltr[S]{Z_3}{\ell} u_3'$. In fact the label is
  \begin{quote}
    \begin{tabular}{ll}
      $\monSub{\lambda}{{W_3}}(\mon{\beta_1}(B_1^I) \oplus \mon{\beta_2}(B_2^I)) =$\\
      \ \ \ = $\monSub{\lambda}{{W_3}}(\mon{\beta_1}(B_1^I)) \oplus \monSub{\lambda}{{W_3}}(\mon{\beta_2}(B_2^I))$ & [since the diagram in Fig.~\ref{fi:th-congruence-b} commutes]\\
      \ \ \ = $\monSub{\lambda}{{Z_1}}(B_1^I) \oplus \monSub{\lambda}{{W_2}}(B_2^I)$ &  
      [since $A_1^I = B_1^I$ by~(\ref{eq:B1I})  and\\
      & \phantom{[}$\monSub{\lambda}{{W_2}}(B_2^I) = \monSub{\lambda}{{Z_2}}(A_2^I)$ by construction~(\ref{eq:trans2})]\\
      \ \ \ = $\monSub{\lambda}{{Z_1}}(A_1^I) \oplus \monSub{\lambda}{{Z_2}}(A_2^I)$ &  
       [since the diagram in Fig.~\ref{fi:th-congruence-a} commutes]\\\\
      \ \ \ = $\monSub{\lambda}{{Z_3}}(\monSub{\alpha}{1}(A_1^I)) \oplus \monSub{\lambda}{{Z_3}}(\monSub{\alpha}{2}(A_2^I))$\\
      \ \ \ = $\monSub{\lambda}{{Z_3}}(\monSub{\alpha}{1}(A_1^I) \oplus \monSub{\alpha}{2}(A_2^I))$ &  [by (\ref{eq:A3})]\\
      \ \ \ = $\monSub{\lambda}{{Z_3}}(A_3)$\\
      \ \ \ $= \ell$
    \end{tabular}
  \end{quote}
  
\noindent
  Moreover, using~(\ref{eq:v2}), we have 
  \begin{center}
    $(v_1^E \mjoin v_2^I) \oplus (v_1^I \mjoin v_2^E) = (v_1^I \oplus
    v_1^E) \mjoin (v_2^I \oplus v_2^E) = v_1 \mjoin v_2 =
    v_3$.
  \end{center}
  And, if we define
  \begin{center}
    $v_1' = {v_1^I}' \oplus {v_1^E}'$ 
  \end{center}
  recalling that, by~(\ref{eq:v2}), $v_2' = {v_2^I}' \oplus {v_2^E}'$,
  we have that the target state of the step is
  \begin{center}
    $v_3' = v_1' \mjoin v_2'$
  \end{center}
  
  Now, $(u_2', v_2') \in \mathcal{R}$ by construction. Moreover, the
  fact that $u_1' \ominus u_0' = v_1' \ominus v_0'$ immediately follows from
  the fact that this property holds of the starting markings and we executed
  the same internal step in $Z_1$.

  Hence $(u_3',v_3') \in \mathcal{R}'$ as desired.
\end{proof}

\subsubsection{Weak Step Bisimilarity}

\begin{thm}
  \label{th:congruence-weak-step}
  Weak step bisimilarity is a congruence.
\end{thm}

\begin{proof}
  In order to show the desired result, we build on the proof of the
  strong case (Theorem~\ref{th:congruence-strong-step}).
  Let us use the same notation and define the relation $\mathcal{R}'$ in the
  same way. In order to prove that $\mathcal{R}'$ is an
  $\mathsf{S}$-weak bisimulation we proceed as follows.

  Let $u_3 \wltr[S]{Z_3}{\ell_3} u_3'$ and let us focus on the case
  $\ell_3 \neq 0$ (the case in which $\ell_3 = 0$ is completely
  analogous). This transition is induced by a step $u_3 \trans{A_3}
  u_3'$, which can be projected over $Z_1$ and $Z_2$, thus getting,
  for $i \in \{ 1, 2\}$
  \begin{center}
    $u_i \trans{A_i} u_i'$
  \end{center}
  Now, since $(u_2,v_2) \in \mathcal{R}$, the transition $u_2
  \wltr[S]{Z_2}{\ell_2} u_2'$, induced by $u_2 \trans{A_2} u_2'$ can
  be simulated in $W_2$, by $v \Ltr[S]{W_2}{\ell_2} v'$.  Let the weak
  transition in $W_2$ arise from the sequence of steps
  \begin{center}
    $v_2 = v_2^0 \trans{B_2^1} v_2^1 \ldots v_2^{h} \trans{B_2^h} v_2^{h+1}  \ldots v_2^k \trans{B_2^k} v_2^{k+1} = v_2'$
  \end{center}
  where $\monSub{\lambda}{{W_2}}(B_2^i) = 0$ for $i \neq h$ and
  $\monSub{\lambda}{{W_2}}(B_2^h) = \ell$ (and as remarked after
  Definition~\ref{de:weak-bisim} we can assume that no transition in
  $B_2^h$ has an unobservable label).

  Now, any $\tau$-step $v_2^i \trans{B_2^i} v_2^{i+1}$ ($i < h$)
  consists only of firings of transitions of $W_2$. Hence, as in the
  strong case, by using Lemma~\ref{le:move-step} we can conclude that
  there is a ``corresponding'' step $v_1^i \trans{B_1^{i}} v_1^{i+1}$,
  consisting only of interactions with the environment, and their
  composition is a $\tau$-step in $W_3$ of the kind $v_1^i
  \trans{\monSub{\alpha}{2}(B_2^{i})} v_3^{i+1}$, with
  $\monSub{\lambda}{{Z_1}}(B_1^i) = 0$.

  Note that since $v_1^i \trans{B_1^{i}} v_1^{i+1}$ consists only of
  interactions with the environment, $u_1 \ominus u_0 = v_1^{i+1}
  \ominus v_0^{i+1}$ for $i<h$.

For the ``visible'' step $v_2^h \trans{B_2^h} v_2^{h+1}$, we can apply
the same argument as in the strong case, to get steps $v_1^h
\trans{B_1^{h}} v_1^{h+1}$ and $v_3^h \trans{B_3^{h}} v_3^{h+1}$, with
$\monSub{\lambda}{{W_3}}(B_3^h) = \ell$.  Additionally, $u_1' \ominus u_0' =
v_1^{h+1} \ominus v_0^{h+1}$.

Repeating the same argument for the remaining $\tau$-steps, $v_2^i
\trans{B_2^i} v_2^{i+1}$ ($i > h$), i.e., using again
Lemma~\ref{le:move-step}, we can prove that there are steps $v_1^i
\trans{B_1^{i}} v_1^{i+1}$, consisting only of interactions with the
environment, correspondingly $\tau$-steps in $W_3$ of the kind $v_1^i
\trans{\monSub{\alpha}{2}(B_2^{i})} v_3^{i+1}$, with
$\monSub{\lambda}{{Z_1}}(B_1^i) = 0$, for $i>h$.
Such sequence of further $\tau$-steps in $W_3$ leads to a marking
$v_3' = v_1^{k+1} \mjoin v_2^{k+1}$, where $v_1^{k+1} \ominus
v_0^{k+1} = u_1' \ominus u_0'$ and $v_2^{k+1} = v_2'$ with $(u_2',
v_2') \in \mathcal{R}$. Hence $(u_3', v_3') \in \mathcal{R}'$.

In other words $v_3 \Ltr[S]{W_3}{\ell} v_3'$ and $(u_3', v_3') \in
\mathcal{R}'$, as desired.
\end{proof}

\subsubsection{Weak (and Strong) Firing Bisimilarity}

\begin{thm}
  \label{th:congruence-firing}
  Strong and weak firing bisimilarity are congruences.
\end{thm}

\begin{proof}
  The proof remains essentially the same as for step bisimulation
  (Theorem~\ref{th:congruence-strong-step} and
  Theorem~\ref{th:congruence-weak-step}). Only some minor adaptations
  are required.

  Let us focus on weak bisimulation, which is the more general case.
  We use the same notation as in Theorem~\ref{th:congruence-weak-step}
  and define $\mathcal{R}'$ in the same way. In order to prove that
  $\mathcal{R}'$ is an $\mathsf{S}$-weak bisimulation we proceed as
  follows.

% We define relation $\mathcal{R'}$ as for step bisimulation. 
Let $(u_3, v_3) \in \mathcal{R}'$ and let $u_3 \wltr[F]{Z_3}{\ell} u_3'$. Then there must be a step
\begin{center}
  $u_3 \trans{\epsilon_3} u_3'$
\end{center}
such that $\epsilon_3 \in \bar{T}_{Z_3}$ and
$\lambda_{Z_3}(\epsilon_3) = \ell$.
We can project the step over $Z_2$, thus getting
\begin{equation}
  \label{eq:obis}
  u_2 \trans{A_2} u_2'
\end{equation}
The delicate case is the one in which $\epsilon_3 = t_3 \in T_{Z_3} -
\alpha_2(T_{Z_2})$. In fact, in this case, $A_2$ is in general a proper
multiset (of interactions with the environment) and thus we cannot
argue, as in the case of step bisimulation, that the transition $u_2
\ltr[F]{Z_2}{\monSub{\lambda}{{Z_2}}(A_2)} u_2'$ must be simulated by $W_2$, since
only single firings are simulated.

In order to proceed, we have first to linearise the step in~(\ref{eq:obis}) as
\begin{equation}
  u_2 \ltr[F]{Z_2}{-_{s_1}} \ldots \ltr[F]{Z_2}{-_{s_k}} \ltr[F]{Z_2}{+_{s_{k+1}}}  \ldots \ltr[F]{Z_2}{+_{s_{k+h}}} u_2'
\end{equation}
Interestingly, the joint effect of the projection and of the
linearization corresponds to the function $\psi$ used
in~\cite[page~96]{v:modular-petri} to project a firing in the combined
net to a firing sequence in the host net.
Now we can say that this is simulated in $W_2$ by 
\begin{center}
  $v_2 \Ltr[F]{W_2}{-_{s_1}} \ldots \Ltr[F]{W_2}{-_{s_k}} \Ltr[F]{W_2}{+_{s_{k+1}}}  \ldots \Ltr[F]{W_2}{+_{s_{k+h}}} v_2'$
\end{center}
namely
\begin{quote}
  $v_2 \Ltr[F]{W_2}{0} \ltr[F]{W_2}{-_{s_1}} \Ltr[F]{W_2}{0} \ldots$\\
  \mbox{} \ \ \ $\ldots \Ltr[F]{W_2}{0} \ltr[F]{W_2}{-_{s_k}} \Ltr[F]{W_2}{0} \Ltr[F]{W_2}{0} \ltr[F]{W_2}{+_{s_{k+1}}} \Ltr[F]{W_2}{0}$  \ldots\\
  \mbox{} \ \ \ $\ldots \Ltr[F]{W_2}{0} \ltr[F]{W_2}{+_{s_{k+h}}} \Ltr[F]{W_2}{0} v_2'$
\end{quote}
which in turn (since $-_{s_i}$ and $+_{s_j}$ firings can be clearly
postponed and anticipated, respectively) can be reorganised as
\begin{center}
  $v_2 \Ltr[F]{W_2}{0} \ltr[F]{W_2}{-_{s_1}} \ldots \ltr[F]{W_2}{-_{s_k}} \ltr[F]{W_2}{+_{s_{k+1}}}  \ldots \ltr[F]{W_2}{+_{s_{k+h}}} \Ltr[F]{W_2}{0} v_2'$
\end{center}
and thus finally to
\begin{center}
  $v_2 \Ltr[F]{W_2}{0} \Ltr[F]{W_2}{\ell_2} \Ltr[F]{W_2}{0} v_2'$
\end{center}
where $\ell_2 = (\bigoplus_{i=1}^k +_{s_1}) \oplus (\bigoplus_{i=1}^h
-_{s_i})$. Then we can proceed exactly as in the proof for step bisimilarity.
\end{proof}

\subsection{Comparison to CCS}
\ \\

\noindent
We now give some hints as to why weak (firing) bisimilarity is a congruence in
the case of open nets, but not in CCS~\cite{Mil:CCS}.  Remember that a
classical counterexample for CCS is as follows: $p_1 = \tau.a.0
\approx a.0 = p_2$, but $q_1 = \tau.a.0 + b.0 \not\approx a.0 + b.0 =
q_2$. The reason for the latter inequality is that $q_1$ can do a
$\tau$ and become $a.0$, while $q_2$ cannot mimic this step.

\begin{figure}[t]
  \begin{center}
    \includegraphics[scale = .22]{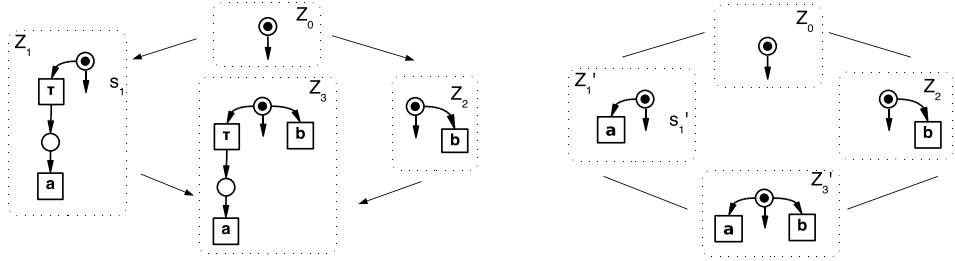}
  \end{center}
  \caption{Two pushouts of open nets for the comparison to CCS.}

  \label{fi:open-nets-ccs}
\end{figure}

Fig.~\ref{fi:open-nets-ccs} shows a similar situation of
nondeterministic choice for open nets, where $\tau$ is the only
unobservable label.  However, note that here the two nets $Z_1$
(corresponding to $\tau.a.0$) and $Z_1'$ (corresponding to $a.0$) are
\emph{not} weakly firing bisimilar. Whenever the $\tau$-transition is
fired in $Z_1$, resulting in the marking $m_1$, this can not be
mimicked in $Z_1'$ by staying idle, since then in $Z_1'$ a transition
with label $-_{s'_1}$ is possible, while a transition labelled
$-_{s_1}$ is not possible for the net $Z_1$ with marking $m_1$.  Also
note that the places $s_1$ respectively $s'_1$ must be output open in
order to allow composition with the net $Z_2$.

Roughly, this means that for open nets we are always able to observe
the first invisible action in an open component, which is reminiscent
of the definition of observation congruence 
% (denoted by $\approx^c$)
in CCS: two processes $p,q$ are called observation congruent if they
are weakly bisimilar, with the additional constraint that whenever the
first step of $p$ is a $\tau$-action, then it has to be answered by at
least one $\tau$-action of $q$ (and vice versa). In both settings it
is only the first $\tau$-action that can be observed but not the
subsequent ones.

\section{Some Proof Techniques for Bisimilarity}
\label{sec:proof-techniques}

We next present some properties of (strong and weak) bisimilarity,
which can help in bisimilarity proofs. We first show that the set of
open places can be uniformly reduced without altering the equivalence
of open nets. Then we provide an up-to technique for firing
bisimilarity.

%\subsection{Adding/removing tokens and open places}

We start by showing that given two bisimilar nets, if we ``close'' 
corresponding open places in both nets we still get two bisimilar nets.
Given an open net $Z$ and an open place $s \in O_Z^x$, let us denote
by $Z-(s,x)$ the open net obtained from $Z$ by closing place $s$,
i.e., $Z' = (N, O_{Z'})$, where $O_{Z'}^x = O_Z^x -\{s\}$.
The initial marking remains the same.

\begin{prop}[``closing'' open places]
  \label{pr:closing}
  Let $Z_1 \approx_\eta^\mathsf{x} Z_2$, with $\mathsf{x} \in \{
  \mathsf{F}, \mathsf{S} \}$. Let $s \in O_{Z_1}^x$ ($x \in \{ -, +\}$) be
  an open place in $Z_1$. Then the nets $Z_1-(s,x)$ and
  $Z_2-(\eta(s),x)$ are $\eta$-$\mathsf{x}$-bisimilar.
\end{prop}

\begin{proof}
  Let $Z_1' = Z_1-(s,x)$ and $Z_2' = Z_2-(\eta(s),x)$. Let
  $\mathcal{R} \subseteq \monSub{S}{1} \times \monSub{S}{2}$ be an
  $\eta$-$\mathsf{x}$-bisimulation such that $(\init{u}_1, \init{u}_2)
  \in \mathcal{R}$.
  Then $\mathcal{R}$ is a bisimulation between $Z_1'$ and
  $Z_2'$.  In fact, if $(u_1,u_2) \in \mathcal{R}$ and $u_1
  \wltr{Z_1'}{\ell} u_1'$ then clearly $u_1 \wltr{Z_1}{\ell} u_1'$. Since
  $\mathcal{R}$ is a bisimulation for $Z_1$ and $Z_2$ this implies
  that $u_2 \Ltr{Z_2}{\eta(\ell)} u_2'$ with $(u_1', u_2') \in
  \mathcal{R}$. Since $\ell$ is a label in $Z_1'$ where place $s$ has
  been closed, we are sure that $x_s \not\in \ell$, and thus $u_2
  \Ltr{Z_2}{\eta(\ell)} u_2'$ implies $u_2 \Ltr{Z_2'}{\eta(\ell)} u_2'$.
  % The case in which $u_1 \wltrtau{Z_1'} u_1'$ is treated analogously.
  Hence we get the desired result.
\end{proof}

%\subsection{An up-to technique}

We next provide a kind of \emph{up-to technique} for firing
bisimilarity. Given an open net $Z$, let us define the
\emph{out-degree} of a place $s \in S$  as
the maximum number of tokens that the firing of an extended event
can remove from $s$, formally: 

\begin{center}
  $\degree{s} = \max \left( \{ (\pre[t])(s) : t \in T_Z \} \cup
    \{ 1 : s \in O_Z^- \} \right)$ 
\end{center}

The idea, formalised by the notion of up-to bisimulation, is to allow
tokens to be removed from input open places, when they exceed the
out-degree of the place. More precisely, given a net $Z$ and a marking
$u \in \mon{S}$, let us say that a marking $v \in \mon{O_Z^+}$ is
\emph{subtractable} from $u$ if $\forall s \in O_Z^+.\ v(s) \leq
\max\{ u(s) - \degree{s}, 0 \}$. 
Note that when the number of tokens in a place $s$ does not exceed its
out-degree, i.e., $u(s) \leq \degree{s}$, then $v(s) =0$, i.e., no
token is subtractable from $s$. If instead, $u(s) > \degree{s}$, then
the tokens in $s$ which exceeds the out-degree of $s$ can be safely
subtracted from $s$.
It is clear that when $v$ is subtractable from $u$, all transitions
enabled in $u$ are also enabled in $u \ominus v$.
Note that the empty marking is subtractable from any other marking.

\begin{defi}[up-to firing bisimulation]
  \label{de:upto}
  Let $Z_1$ and $Z_2$ be open nets, and let $\eta : O_{Z_1}
  \leftrightarrow O_{Z_2}$ be a correspondence between $Z_1$ and $Z_2$.  A
  relation $\mathcal{R} \subseteq \monSub{S}{1} \times \monSub{S}{2}$
  between markings is called an \emph{up-to
    $\eta$-$\mathsf{F}$-bisimulation} if whenever $(u_1, u_2) \in
  \mathcal{R}$ then

  \begin{enumerate}[$\bullet$]
%   \item if $u_1 \wltrtau{Z_1} u_1'$, then there exists $u_2'$ such that
%     $u_2 \Ltr{Z_2}{0} u_2'$ and $v_1 \in \mon{(O_{Z_1}^+)}$ subtractable from $u_1'$
%     with $(u_1' \ominus v_1, u_2' \ominus \mon{\eta}(v_1)) \in
%     \mathcal{R}$.

  \item if $u_1 \wltr[F]{Z_1}{\ell} u_1'$, then there exist
    markings $u_2'$ such
    that $u_2 \Ltr[F]{Z_2}{\eta(\ell)} u_2'$, and $v_1 \in \mon{O_{Z_1}^+}$
    subtractable from $u_1'$,
    with  $(u_1' \ominus v_1, u_2' \ominus \mon{\eta}(v_1)) \in
    \mathcal{R}$;
    
  \item the symmetric condition holds.

  \end{enumerate}
\end{defi}

That is, the intuition behind up-to bisimulations is that some tokens
might be superfluous since they are not necessary to fire a
transition. Hence in the bisimulation game they can be removed in
the two successor markings.

A first technical lemma shows an invariance property of up-to
$\mathsf{F}$-bisimulations, with respect to adding tokens in open places.

\begin{lem}
  \label{le:up-to}
  Let $Z_1$ and $Z_2$ be open nets, let $\eta : O_{Z_1} \leftrightarrow
  O_{Z_2}$ be a correspondence between $Z_1$ and $Z_2$, and let
  $\mathcal{R}$ be an up-to $\eta$-$\mathsf{F}$-bisimulation between $Z_1$ and
  $Z_2$. Then
  \begin{enumerate}[(1)]

  \item given any $s \in O_{Z_1}^+$, the relation $\mathcal{R}^s =
    \mathcal{R} \cup \{ (u_1 \oplus s, u_2 \oplus \eta(s)) : (u_1, u_2) \in
    \mathcal{R} \}$ is an up-to $\eta$-$\mathsf{F}$-bisimulation.

  \item $\mathcal{R}' = \mathcal{R} \cup \{ (u_1 \oplus v_1, u_2
    \oplus \mon{\eta}(v_1)) : (u_1, u_2) \in \mathcal{R}\ \wedge\ v_1
    \in \mon{O_{Z_1}^+} \}$ is an up-to $\eta$-$\mathsf{F}$-bisimulation.
  \end{enumerate}
\end{lem}

\begin{proof}

  \emph{1}. 
  In order to simplify the notation, let us assume, without loss of
  generality, that $\eta$ is the identity (i.e., $O_{Z_1}^+ = O_{Z_2}^+$
  and $O_{Z_1}^- = O_{Z_2}^-$).

  Let $(u_1 \oplus s, u_2 \oplus s) \in \mathcal{R}^s$. Let us show
  that if $u_1 \oplus s \wltr[F]{Z_1}{\ell} u_1'$ then there exists $u_2
  \oplus s \Ltr[F]{Z_2}{\ell} u_2'$ and $v \in O_{Z_1}^+$ subtractable from
  $u_1'$ with $(u_1' \ominus v, u_2' \ominus v) \in \mathcal{R}^s$.
  The other cases are completely analogous.

  Observe that, since $s \in O_{Z_1}^+$, we have
  \begin{quote}
    $u_1 \wltr[F]{Z_1}{+_s} u_1 \oplus s$.
  \end{quote}
  By definition of $\mathcal{R}^s$, we have $(u_1,u_2) \in \mathcal{R}$
  and thus
  \begin{equation}
    \label{eq:upto1}
    u_2 \Ltr[F]{Z_2}{+_s} u_2'' \quad \text{and} \quad 
    (u_1 \oplus s \ominus v', u_2'' \ominus v') \in
    \mathcal{R}^s
  \end{equation}
  for a suitable $v' \in O_{Z_1}^+$ subtractable from $u_1' \oplus s$.
  Also notice that, since a $+_s$ can always be
  performed, we can assume that the firing sequence (\ref{eq:upto1})
  is of the kind
  \begin{equation}
    \label{eq:upto2}
    u_2 \ltr[F]{Z_2}{+_s} u_2 \oplus s \Ltr[F]{Z_2}{0} u_2''
  \end{equation}
  
  Now, if $u_1 \oplus s \wltr[F]{Z_1}{\ell} u_1'$, then, since $v'$ is
  subtractable from $u_1 \oplus s$, also $u_1 \oplus s \ominus v'
  \wltr[F]{Z_1}{\ell} u_1' \ominus v'$.
  Thus, by (\ref{eq:upto1})
  \begin{equation}
    \label{eq:upto3}
    u_2'' \ominus v' \Ltr[F]{Z_2}{\ell} u_2''' \quad \text{and} \quad (u_1' \ominus v' \ominus v'', u_2''' \ominus v'') \in
  \mathcal{R}^s
  \end{equation}
  for a suitable $v'' \in \mon{O_{Z_1}^+}$, subtractable from $u_1 \ominus v'$.

  Putting the above together with (\ref{eq:upto2}), we have that
  \begin{quote}
    $u_2 \oplus s \Ltr[F]{Z_2}{0} u_2'' \Ltr[F]{Z_2}{\ell} u_2''' \oplus v'$
  \end{quote}
  i.e., $u_2 \oplus s \Ltr[F]{Z_2}{\ell} u_2''' \oplus v'$ and, if we
  denote $u_2' = u_2''' \oplus v'$, $(u_1' \ominus v' \ominus v'',
  u_2' \ominus v' \ominus v'') \in \mathcal{R}^s$. It is immediate to
  see that $v' \oplus v''$ is subtractable from $u_1'$, and thus we
  conclude.

  \medskip
  
  \emph{2}.
  By an inductive reasoning, exploiting point~1, we can show that
  the relation $\mathcal{R}_n = \mathcal{R} \cup \{ (u_1 \oplus v_1,
  u_2 \oplus \mon{\eta}(v_1)) : (u_1, u_2) \in \mathcal{R}\ \wedge\
  v_1 \in \mon{O_{Z_1}^+}\wedge\ |v_1| \leq n \}$ is a
  $\eta$-$\mathsf{F}$-weak bisimulation up-to for any $n$. Then we
  exploit the fact that the union of weak bisimulations up-to is again
  a weak-bisimulation up-to.
\end{proof}

We can finally prove the soundness of the up-to technique.

\begin{prop}
  \label{pr:upto}
  Let $Z_1$ and $Z_2$ be open nets, and let $\eta : O_{Z_1}
  \leftrightarrow O_{Z_2}$ be a correspondence between $Z_1$ and $Z_2$.
  Let $\mathcal{R}$ be an up-to $\eta$-$\mathsf{F}$-bisimulation. Then for any
  $(u_1, u_2) \in \mathcal{R}$ we have that $(Z_1,u_1) \approx^{\mathsf{F}}_\eta
  (Z_2,u_2)$.
\end{prop}

\begin{proof}
  In order to simplify the notation, let us assume, without loss of
  generality, that $\eta$ is the identity (i.e., $O_{Z_1}^+ = O_{Z_2}^+$
  and $O_{Z_1}^- = O_{Z_2}^-$).

  Let us show that
  \begin{center}
    $\mathcal{R}' = \{ ( u_1 \oplus v, u_2 \oplus v) : (u_1, u_2) \in
    \mathcal{R}\ \land\ v \in \mon{(O_{Z_1}^+)} \}$
  \end{center}
  is an $\eta$-$\mathsf{F}$-bisimulation. Let $(u_1 \oplus v, u_2 \oplus v) \in
  \mathcal{R}'$, with $(u_1, u_2) \in \mathcal{R}$ and $v \in O_{Z_1}^+$,
  and assume that
  \begin{center}
    $u_1 \oplus v \wltr[F]{Z_1}{\ell} u_1'$.
  \end{center}

  By Lemma~\ref{le:up-to} we know that $\mathcal{R}'$ is an up-to
  bisimulation, and thus there exists a transition
  \begin{center}
    $u_2 \oplus v \Ltr[F]{Z_2}{\ell} u_2'$
  \end{center}
  and $v' \in \mon{O_{Z_1}^+}$, subtractable from $u_1'$, such that $(u_1'
  \ominus v', u_2' \ominus v') \in \mathcal{R}'$. However, by
  construction of $\mathcal{R'}$, this implies that
   \begin{center}
     $(u_1', u_2') \in \mathcal{R}'$
  \end{center}
  as desired.
\end{proof}

As it often happens with up-to techniques, the above result
might allow to show that two nets are firing bisimilar by exhibiting finite
relations (while bisimulations are typically infinite). E.g., consider
the open nets on the right, where label $a$ is observable. Then any\linebreak
\begin{minipage}[c]{0.60\linewidth}
firing bisimulation would include at least
the pairs \linebreak $\{ (k \cdot s, k
\cdot s) : k \in \mathbb{N} \}$, where $s$ is the only place. Instead,
according to the definition above $\{ (0, 0), (s, s) \}$ is an up-to
bisimulation.
\end{minipage}
\begin{minipage}[c]{0.4\linewidth}
  \vspace{1mm}
\begin{center}
  \scalebox{.28}{\includegraphics{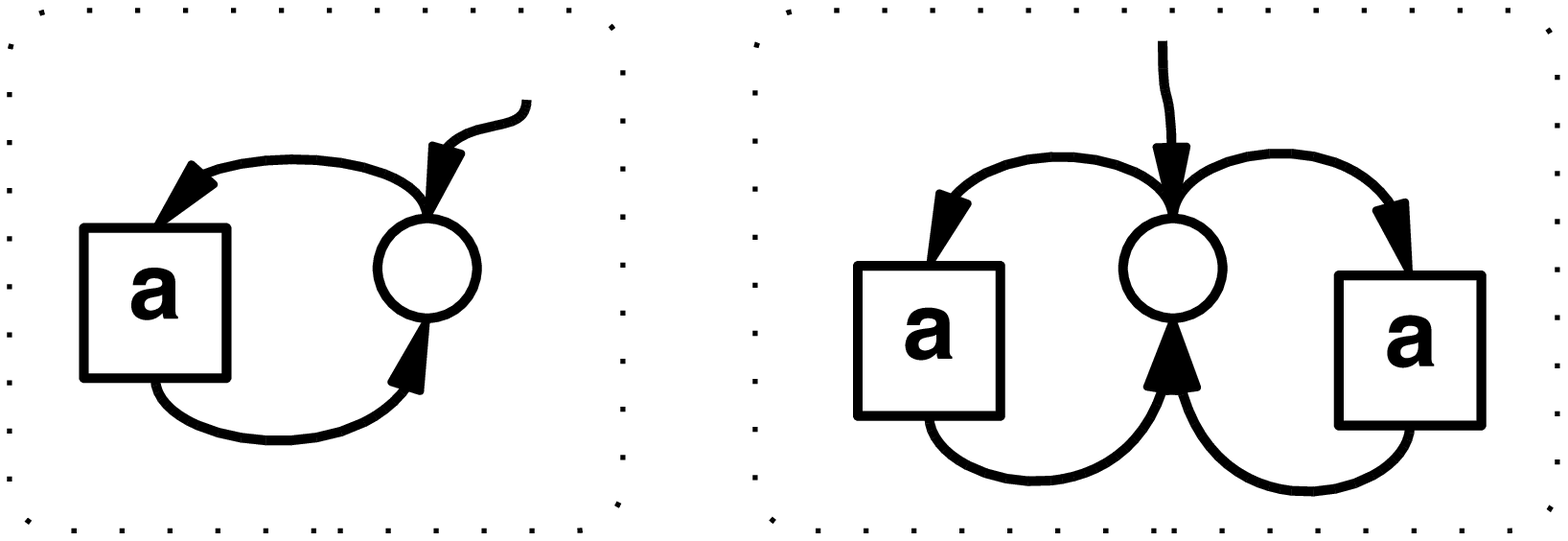}}
\end{center}
\end{minipage}

\smallskip

Note that, instead, the up-to technique does not extend to step
bisimilarity: since an unbounded number of tokens can be needed to
fire a parallel step there is no obvious generalisation of the notion of subtractable marking.

\section{Reconfigurations of Open Nets}

The results in the previous sections are used here to design a
framework where a system specified as a (possibly open) Petri net can
be reconfigured dynamically by transformation rules, triggered by the
state/shape of the system. The congruence results allows one to
characterise classes of reconfigurations which preserve the
observational behaviour of the system. 

\subsection{Behaviour Preserving Reconfigurations of  Open Nets}
\ \\

\noindent
The fact that the composition operation over open nets is defined in terms of a
pushout construction suggests naturally a way of reconfiguring open nets by
using the double-pushout approach to rewriting~\cite{Ehr:TIAA}.

A \emph{rewriting rule} over open nets consists of a pair of morphisms
in $\onet$:
\begin{center}
  $p = L_p \stackrel{l_p}{\leftarrow} K_p \stackrel{r_p}{\rightarrow} R_p$
\end{center}
where $L_p$, $K_p$, $R_p$ are open nets, called \emph{left-hand side},
\emph{interface} and \emph{right-hand side} of the rule $p$, and $l_p$, $r_p$
are open net embeddings.
Intuitively, the rule specifies that, given a net
$Z$, if the left-hand side $L_p$ matches a subnet of $Z$ then this can be
reconfigured into $Z'$ by replacing the occurrence of $L_p$ with the
right-hand side $R_p$, preserving the subnet
%the interface net 
$K_p$.

The notion of transformation is formally defined below.

\begin{figure}[t]
  \[
  \begin{array}{cc}
    \vcenter{
      \xymatrix@R=5mm{
        L_p \ar[d]_m & K_p \ar@{.>}[d]_n \ar[l]_{l_p} \ar[r]^{r_p} &
        R_p \ar@{.>}[d]^h \\
        Z & D \ar@{.>}[l]^d \ar@{.>}[r]_b & Z'
      }
    }
    & \hspace{-20mm}
    \vcenter{
      \includegraphics[scale = .24]{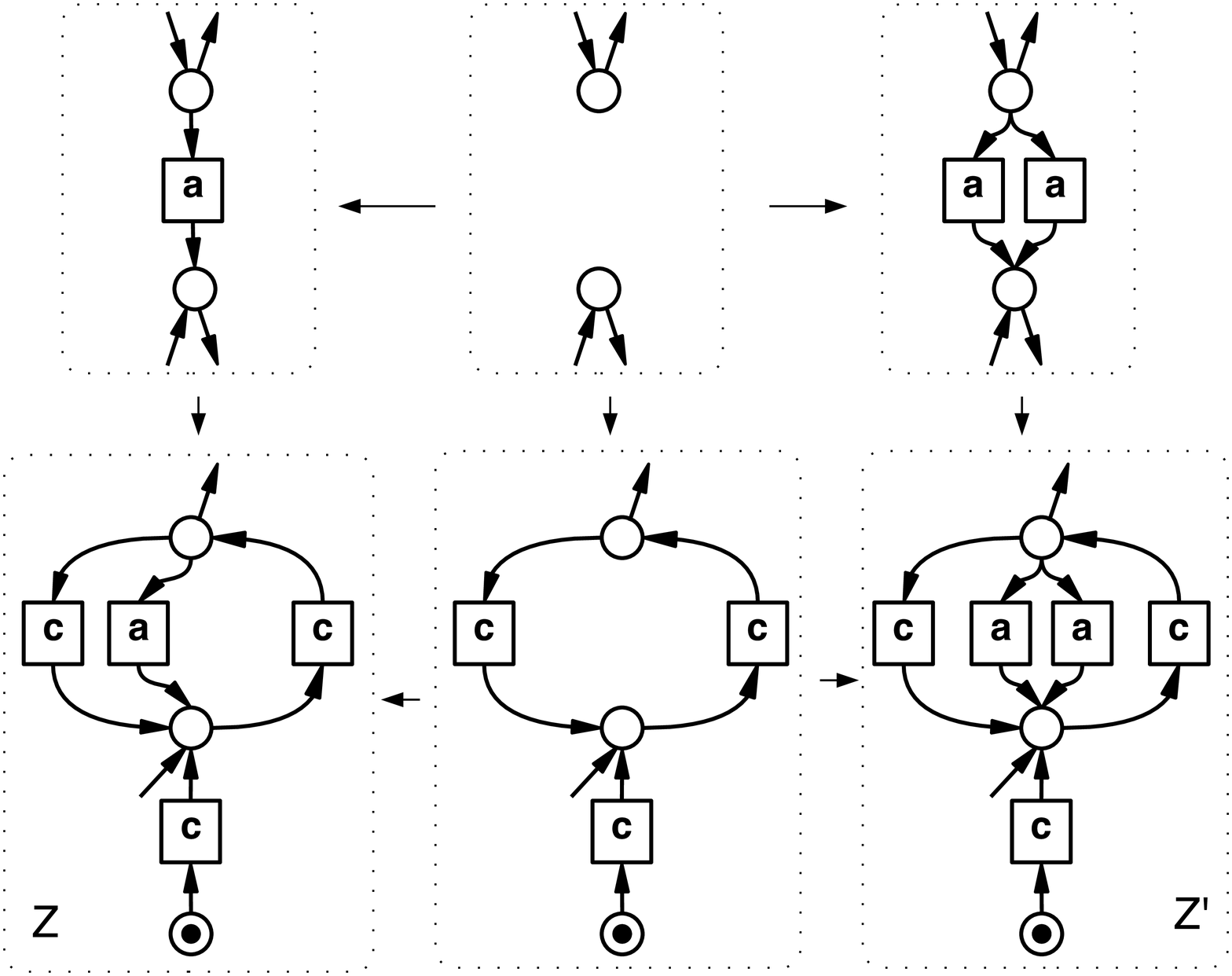}
    }
    \\
    (a) & (b)\ \ \ \ \ \ \ \ \ \ \ \ \ \ \ \ \
  \end{array}
  \]
\caption{Transforming open nets through DPO rewriting.}
\label{fi:open-dpo}
\end{figure}

\begin{defi}[open net transformation]
\label{de:transformation}
  Let $p$ be a rewriting rule over open nets, let $Z$ be an open net and let
  $m : L_p \to Z$ be a match, i.e., an open net embedding. We say that $Z$
  rewrites to $Z'$ using $p$ at match $m$, denoted $Z \rew{p,m} Z'$ or simply
  $Z \rew{p} Z'$, if the diagram of Fig.~\ref{fi:open-dpo}(a) can be
  constructed in $\onet$,
  where both squares are pushouts, and morphism $n$ is 
  composable with both $l_p$ and $r_p$.
\end{defi}
We stress that we are interested in transformations where the two pushout
squares are built from composable arrows (technically, this ensures that the
transformation can be performed in $\net$ and then ``lifted'' to $\onet$).

% \begin{figure}[t]
%   \[
%   \xymatrix@R=5mm{
%     L_p \ar[d]_m & K_p \ar@{.>}[d]_n \ar[l]_{l_p} \ar[r]^{r_p} &
%     R_p \ar@{.>}[d]^h \\
%     Z & D \ar@{.>}[l]^d \ar@{.>}[r]_b & Z'
%   }
%   \]

%   \caption{Double-pushout rewriting over open nets}
%   \label{fi:dpo-onet}
% \end{figure}

%With the given definition of transformation of open nets, t

We can now characterise the rules which do not alter the
observational behaviour of an open Petri net as the rules with
bisimilar left and right-hand side.

\begin{defi}[behaviour preserving rules]
  \label{de:beh-pres-rule}
  A $\mathsf{x}$-behaviour preserving rule ($\mathsf{x} \in \{
  \mathsf{F}, \mathsf{S} \}$) is an open net rewriting rule $p$ such
  that $L_p \approx^{\mathsf{x}}_\eta R_p$, where $\eta = (r_p \circ
  l_p^{-1})_{|O_{L_p}}$.
\end{defi}

Then the next result is an easy consequence of
Theorem~\ref{th:congruence}.

\begin{thm}[behaviour-preserving reconfigurations]
  \label{th:behaviour-preserving-reconfigurations}
  Let $p$ be a $\mathsf{x}$-behaviour preserving rule ($\mathsf{x} \in
  \{ \mathsf{F}, \mathsf{S} \}$).
    % Consider an open net rewriting rule $p$ such that $L_p
    % \approx_\eta R_p$, where $\eta = (r_p \circ l_p^{-1})_{|O_{L_p}}$.
    % 
    Given an open net $Z$, if $Z \rew{p,m} Z'$ via a 
    % proper 
    match $m :
    L_p \to Z$, then $Z \approx^{\mathsf{x}} Z'$. 
  \end{thm}

\begin{proof}% [Sketch]
  Just observe that, in the DPO diagram 
  of Figure~\ref{fi:open-dpo}(a),
  %% 
  %% \[
  %% \xymatrix@R=5mm{
  %%   L_p \ar[d]_m & K_p \ar@{.>}[d]_n \ar[l]_{l_p} \ar[r]^{r_p} &
  %%   R_p \ar@{.>}[d]^h \\
  %%   Z & D \ar@{.>}[l]^d \ar@{.>}[r]_b & Z'
  %% }
  %% \]
  since the arrows $l_p$, $n$ and $r_p$, $n$ are composable, we can apply
  Theorem~\ref{th:congruence} to conclude that $Z \approx^{\mathsf{x}}
  Z'$.
\end{proof}

For instance, consider the double-pushout diagram in
Fig.~\ref{fi:open-dpo}(b).  It can be easily seen that the left- and
right-hand sides of the applied rule are strongly (step) bisimilar.
Hence we can conclude that $Z$ and $Z'$ are strongly (step) bisimilar
as well.

\subsection{Applying Rules to Open Nets}
\ \\

\noindent
As it is common in the categorical approaches to (graph) rewriting,
the notion of open net transformation proposed in
Definition~\ref{de:transformation} is rather ``declarative'' in style,
because it requires the existence of two pushouts in category $\onet$,
without stating how they can be constructed, and under which
conditions.  A more explicit description of the conditions under which
a rule can be applied to an open net and of the way the resulting net
can be constructed, is clearly necessary for practical purposes.
Looking at Fig.~\ref{fi:open-dpo}(a), 
given a rule $p$ and a match $m : L_p \to Z$, in order to
build the open net transformation:

\begin{enumerate}[$\bullet$]
  
\item The \emph{pushout complement} of $l_p$ and $m$ must exist. The
  resulting arrows $n$ and $d$ must be such that $l_p$ and $n$ are
  composable.  A necessary condition for the existence of the pushout
  complement is a sort of \emph{dangling condition}: a place can be
  deleted only if all the transitions connected to this place are
  removed as well, otherwise the flow arcs of this transition would
  remain dangling. This ensures that the pushout complement exists and
  is unique in the underlying category $\net$, but, as discussed
  below, it is not sufficient, in general, to conclude the existence
  of the pushout complement in $\onet$.

  Additionally, there can be several pushout complements and in this case a
  canonical choice should be considered.

\item
  The resulting arrow  $n$ must be composable with $r_p$: then we know
  how to build $Z'$ by Proposition~\ref{pr:push-onet}.

\end{enumerate}

Unfortunately, although a general theory of DPO
rewriting has been developed recently in the framework of adhesive
categories~\cite{LS:AQC}, we cannot exploit it here
since the category of open nets falls outside the scope of the theory.

Next we analyse the conditions which ensure the applicability of open
net rules. We will first consider the case of general, possibly
non-behaviour preserving rules. Then we will instantiate the developed
theory to the setting of behaviour preserving rules, which turns out to
be simpler and more intuitive. The reader which is not interested in
the general case can safely skip it.

\subsubsection{Applying General Rules.}

\begin{figure}[t]
  \centering
  \subfigure[]{
    \scalebox{.28}{\includegraphics{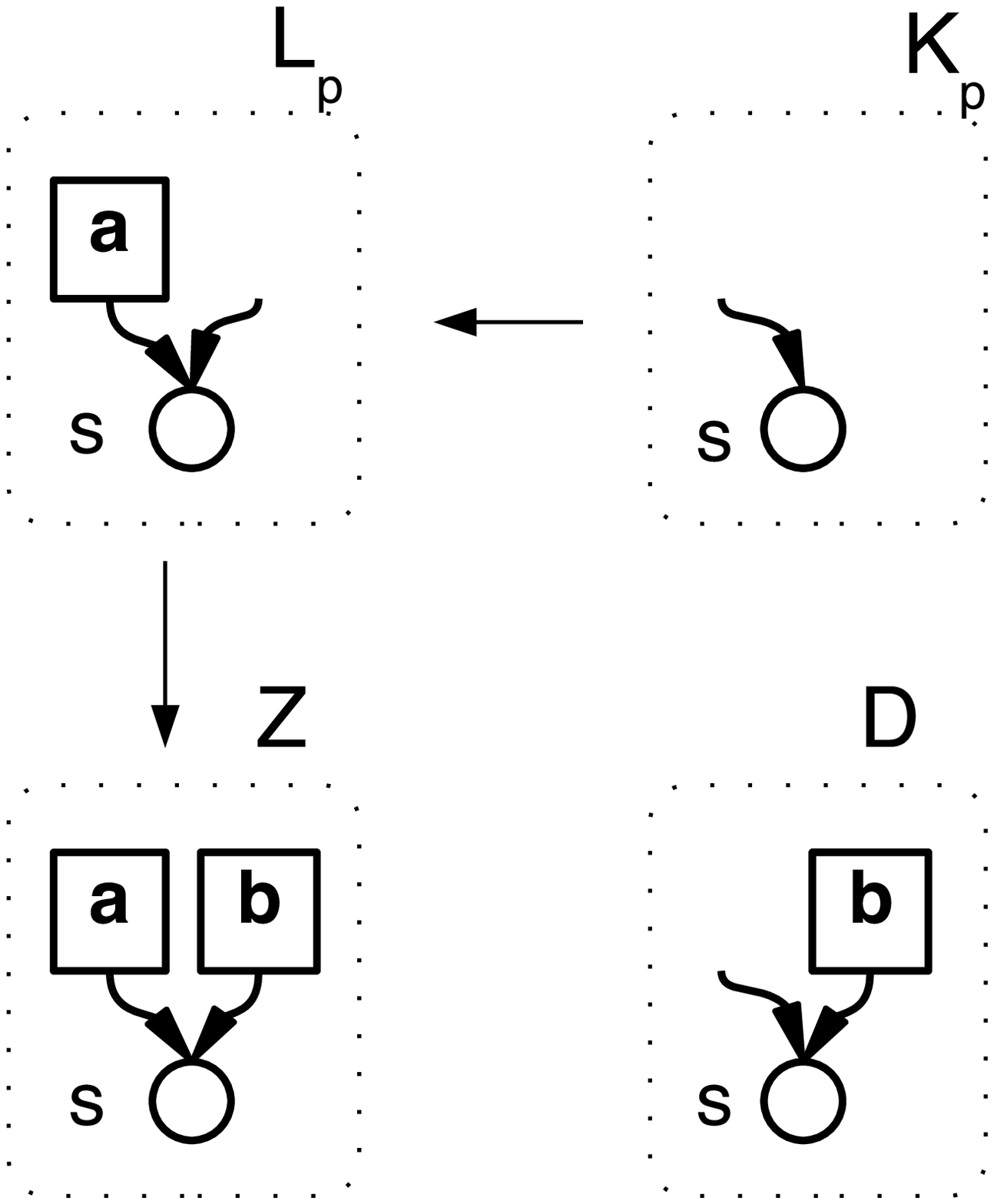}}
    \label{fi:no-poc}
  }
  \hspace{5mm}
  \subfigure[]{
    \scalebox{.28}{\includegraphics{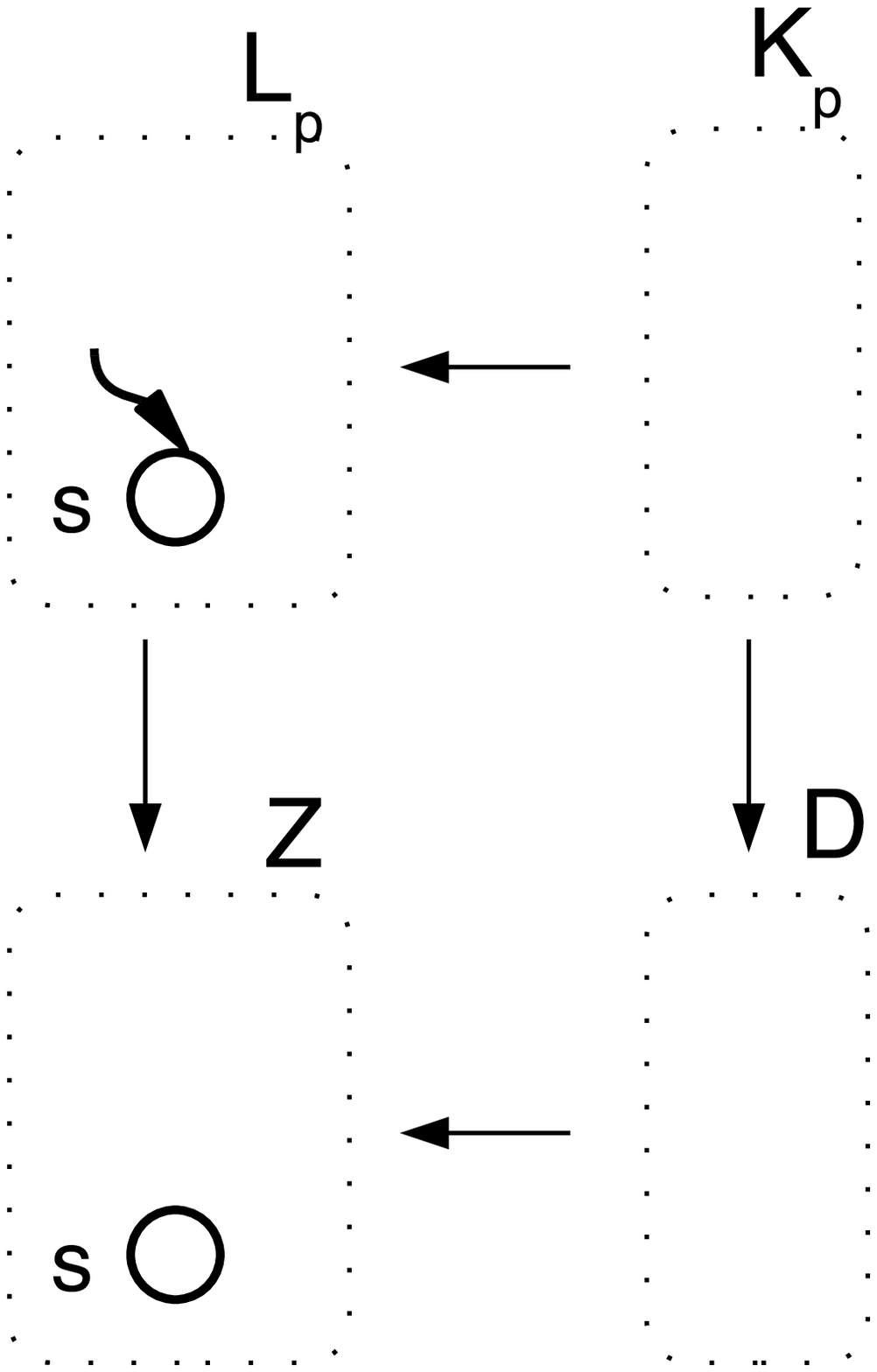}}
    \label{fi:no-poc-bis}
  }
  \hspace{5mm}
  \subfigure[]{
    \scalebox{.28}{\includegraphics{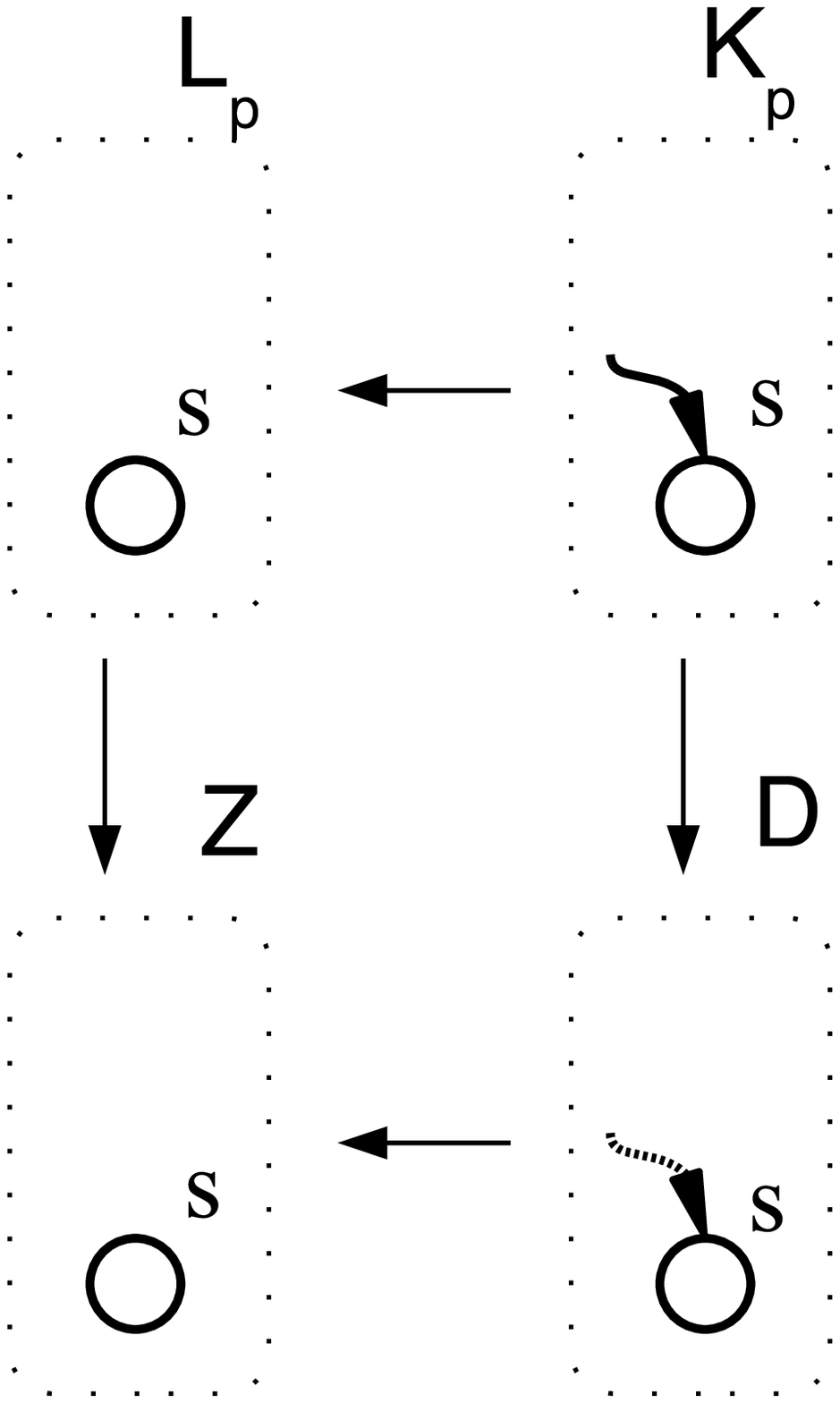}}
    \label{fi:two-poc}
  }

\caption{(a),(b) A pushout complement in $\net$ which cannot be lifted to
  $\onet$ and (c) A situation in which the pushout complement is not
  unique in $\onet$.}

\end{figure}

In this section we develop general results concerning the
applicability of a rewriting rule to an open net. Given an open net
$Z$, a rule $p$ and a match $m : L_p \to Z$, we first focus on the
existence of the pushout complement in $\onet$.
As mentioned above, a first necessary condition is a sort of
\emph{dangling condition}, which, however, in general, is not
sufficient.
Consider, for instance, the diagram in Fig.~\ref{fi:no-poc}. It is
easy to realise that the only place in $D$ must be input open since an
additional transition is attached to such place in $Z$. However, the
resulting diagram is not a pushout in $\onet$: since the places in $L_p$
and in $D$ are input open also their image in $Z$ should be input
open.
Similarly, the diagram Fig.~\ref{fi:no-poc-bis} is not a pushout in
$\onet$, although the underlying diagram is a pushout in $\net$, since
place $s$ of $Z$ should be input open.

Moreover, in the case of general rules, the pushout complement in
$\onet$ might not be unique. In fact, whenever, as in
Fig.~\ref{fi:two-poc}, there is an open place in $K_p$ whose image is
not open in $L_p$ (and thus neither in $Z$), then the corresponding
place in $D$ can be either open or not. For instance, the diagram in
Fig.~\ref{fi:two-poc} admits two possible pushout complements
consisting of an open net $D$ with a single place $s$ which can be or
not input open.

Under additional requirements it is possible to prove the existence of
a minimal pushout complement $D$, i.e., a pushout complement which
embeds into any other and which is taken as a canonical
choice. Roughly, the minimal pushout complement is the maximally open
one: whenever a place could be either open or not, it is taken to be
open (in Fig.~\ref{fi:two-poc}, this corresponds to take the pushout
complements $D$ with place $s$ input open).

% In order to simplify the notation, in the sequel we will assume, without loss
% of generality, that for any rule $p$, the arrows $l_p$ and $r_p$ are
% inclusions.

\begin{lem}[existence of the pushout complement]
  \label{le:po-complement}
  Let $p$ be a rewriting rule over open nets, let $Z$ be an open net and let
  $m : L_p \to Z$ be a match. Assume that
  \begin{enumerate}[\em(1)]

  \item for all places $s \in L_p - l_p(K_p)$ we have $\pre[m(s)], \post[m(s)]
    \subseteq m(L_p - l_p(K_p))$;

  \item
    $m(l_p(\inp{l_p}) \cap O_{L_p}^+) \subseteq O_Z^+$  
    and $m(l_p(\out{l_p}) \cap O_{L_p}^-) \subseteq O_Z^-$;

  \item $m(O_{L_p}^x - l_p(O_{K_p}^x)) \subseteq O_Z^x$ for $x \in \{+,-\}$.

  \end{enumerate}
  Then the pushout complement exists in $\net$, defined as $D = Z -
  m(L_p - l_p(K_p))$, componentwise over the place and transition
  sets, and it can be lifted to a minimal pushout complement in
  $\onet$ by taking as input open places:
  \begin{center}
     $O_D^+ = d^{-1}(O_Z^+) \cup n(O_{K_p}^+ - O_{L_p}^+)$
  \end{center}
  Output open places are defined analogously.
  The initial marking $\init{u}_D$ is defined by $\init{u}_D(s) = \init{u}_Z(d(s))$ for any place
  $s \in S_D$.
\end{lem}

\proof
  The proof is long, but straightforward. We have already motivated the
  dangling condition above.
  In order to understand condition~2, observe that, roughly,
  a place $s$ of $L_p$ is in   $l_p(\inp{l_p})$ if   applying the 
  rule $p$ the place is preserved but at least one
  transition in $\pre[s]$ is removed.
%
%%   $l_p(\inp{l_p})$ is the set of places $s$ in $L_p$, such that
%%   applying the rule $p$ the place $s$ is preserved but at least one
%%   transition in $\pre[s]$ is removed. 
  Since the rule deletes an input
  transition from $m(s)$ -- the image of $s$ in $Z$ -- the
  corresponding place in $D$ belongs to $\inp{d}$ and thus it must be
  input open.  Therefore if $s$ is open also in $L_p$, necessarily, by
  the construction of pushout in $\onet$, $m(s)$ must be open in $Z$.
  Similarly, for condition~3, if a place is open in $L_p$ and it is not in the
  image of $K_p$ then necessarily it will be open in $Z$.
  
  Formally we have to show that (a) the mappings $n$ and $d$ are well-defined
  open net morphisms, (b) $l_p$ and $m$ are composable and (c) $Z$ is the
  pushout.
  Minimality of the pushout complement then follows by construction.

  \begin{enumerate}[({a}.1)]

  \item $n$ is a well-defined open net morphism.\\
    Let us prove that $n^{-1}(O_D^+) \cup \inp{n} \subseteq O_K^+$
    (the condition on  output open places is analogous).  If $s \in
    n^{-1}(O_D^+)$ we have two possibilities according to the way
    $O_D^+$ is defined.
    \begin{enumerate}[$-$]
    \item If $n(s) \in d^{-1}(O_Z^+)$ then $d(n(s)) \in O_Z^+$. Since $d \circ
    n = m \circ l_p$ and $m \circ l_p$ is a well-defined open net morphism, we
    deduce that $s \in O_K^+$.

    \item If $n(s) \in n(O_K^+ - O_L^+)$, since $n$ is injective, we have that
    $s \in O_K^+ - O_L^+ \subseteq O_K^+$.
    \end{enumerate}
    If instead $s \in \inp{n}$ then $m(l_p(s)) \in \inp{m \circ
    l_p}$. Since $m \circ l_p$ is an open net morphism, we conclude $s
    \in O_{K_p}^+$, as desired.

    Concerning the initial marking, note that for any $s \in S_K$ we
    have $\init{u}_K(s) = \init{u}_Z(m(l_p(s)) = \init{u}_D(d(n(s)) =
    \init{u}_D(s)$, where the last equality holds by construction.

    \medskip
    
  \item $d$ is a well-defined open net morphism.\\
    Also in this case we only prove that $d^{-1}(O_Z^+) \cup \inp{d}
    \subseteq O_D^+$ (the condition on  output open places is
    analogous). If $s \in d^{-1}(O_Z^+)$ then $s \in O_D^+$ by
    definition. If, instead, $s \in \inp{d}$ then it is easy to see
    that there exists $s' \in S_K$ such that $s' \in \inp{l_p}
    \subseteq O_K^+$. Now, there are two subcases:
    \begin{enumerate}[$-$]
      
    \item If $l_p(s') \in O_L^+$ we have that $s' \in l_p(\inp{l_p}) \cap O_L^+$ and
      thus $m(s') \in m(l_p(\inp{l_p}) \cap O_L^+) \subseteq O_Z^+$ by
      condition~2. Since $d(s) = m(s')$ we deduce that $s \in d^{-1}(O_Z^+)
      \subseteq O_D^+$ by construction of $D$.

    \item If $l_p(s') \not\in O_L^+$ then $s' \in O_K^+ - O_L^+$, and
      thus $n(s') \in n(O_K^+ - O_L^+) \subseteq O_D^+$, by construction of $D$.
      
    \end{enumerate}
    The condition over the initial marking is trivially satisfied by
    construction.

    \medskip

  \item[(b)] $n$ and $l_p$ are composable.\\
    We show the two conditions for composability separately:
    \begin{enumerate}[$-$]

    \item $n(\inp{l_p}) \subseteq O_D^+$\\
      In fact, if $s \in \inp{l_p}$, then it is easy to see that $m(l_p(s)) \in
      \inp{d} \subseteq O_D^+$.  Now, $m(l_p(s)) = d(n(s))$ and, since $d$ is an
      open net morphism, it must reflect open places, and thus $n(s) \in
      O_D^+$.
  
    \item $l_p(\inp{n}) \subseteq O_L^+$\\
      If $s \in l_p(\inp{n})$ then, it is easy to see that $s \in \inp{m}
      \subseteq O_L^+$, as desired.
    \end{enumerate}
    
    \medskip

  \item[(c)] $Z$ is the pushout.\\
    We know that $Z$ is the pushout of $n$ and $l_p$ in $\net$. We have to prove
    that it is also a pushout in $\onet$. 
  
    Concerning the set of open places we have to show that
    \begin{center}
      $O_{Z}^x \supseteq \{ s \in S_Z : m^{-1}(s) \subseteq
      O_{L}^x\ \wedge\ d^{-1}(s) \subseteq O_{D}^x \}$.
    \end{center}
    Then the converse inclusion, and thus equality, follows from the
    fact that $m$ and $d$ are open net morphisms.

    Let $s \in S_Z$ such that there are $s' \in O_L^+$ and $s'' \in O_D^+$ such
    that $m(s') = s = d(s'')$. Thus, there is $s''' \in S_K$ such that
    $l_p(s''') = s'$ and $n(s''') = s''$.
    
    Since $s'' \in O_D^+$, then either $s'' \in d^{-1}(O_Z^+)$ or $s'' \in
    n(O_K^+ - O_L^+)$. Since $s' \in O_L^+$ and $l_p(s''') = s'$, the second
    possibility cannot arise. In the first case $s = d(s'') \in O_Z^+$, as
    desired.

    When $s$ is only in the image of $D$, the proof is analogous. When it is
    only in the image of $L_P$, we can use condition~3 in the hypothesis.\qed
  \end{enumerate}

\noindent Summarizing, condition~1 of Lemma~\ref{le:po-complement} is
a dangling condition.  By the remaining conditions, if a place $s$ in
$L_p$ is open, and the rule prescribes either the deletion of
incoming/outgoing transitions from such place (condition~2) or the
deletion of the place itself (condition~3), then the image of $s$ in
$Z$ must be open.  Examples of what fails when conditions~2 and~3 are
violated can be found in Fig.~\ref{fi:no-poc} and~\ref{fi:no-poc-bis}.

It is worth observing that in the case of rules $p$ such that morphism
$l_p$ preserves open places, i.e., $l_p(O_{K_p}^x) \subseteq
O_{L_p}^x$ for $x \in \{ +, -\}$, the above result ensures the
existence of a unique pushout complement.

Given a match $m : L_p \to Z$ as in the proposition above, the
transformation can be completed if $n : K_p \to D$ and $r_p : K_p \to
R_p$ are composable. For this we need to suitably restrict matches.

% \begin{defi}[proper rules]
%   An open net rule $p$ is called \emph{proper} if
%   $r_p(l_p^{-1}(O_L^+)) \subseteq O_R^+$.
% \end{defi}
% %
% Intuitively, a rule is proper if the open places of the left-hand side net, which
% are not deleted, remain open in the right-hand side net.

\begin{figure}
  \subfigure[]{
    \scalebox{.26}{\includegraphics{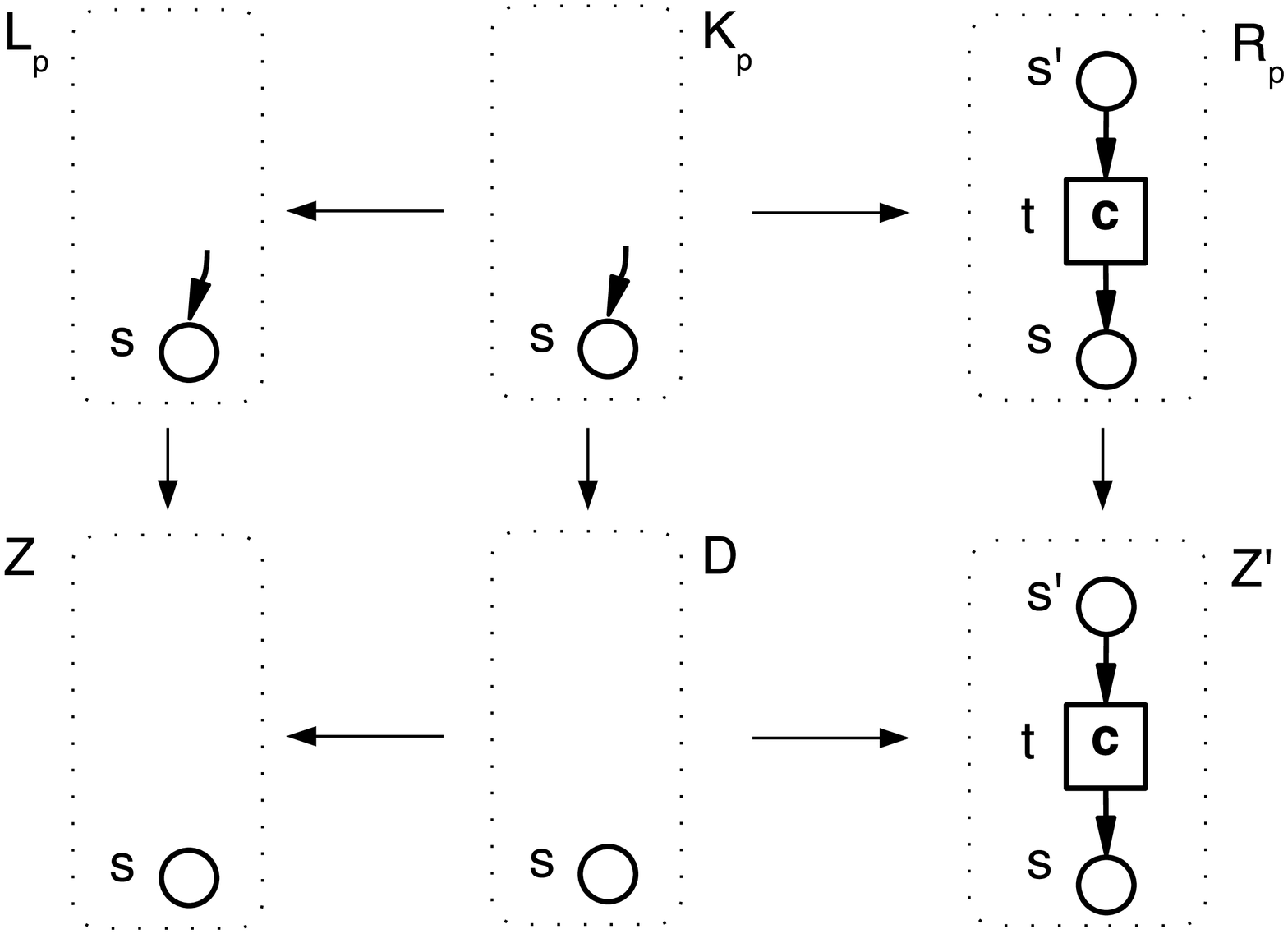}}
    \label{fi:cond4}
  }
  \hspace{1mm}
  \subfigure[]{
    \scalebox{.26}{\includegraphics{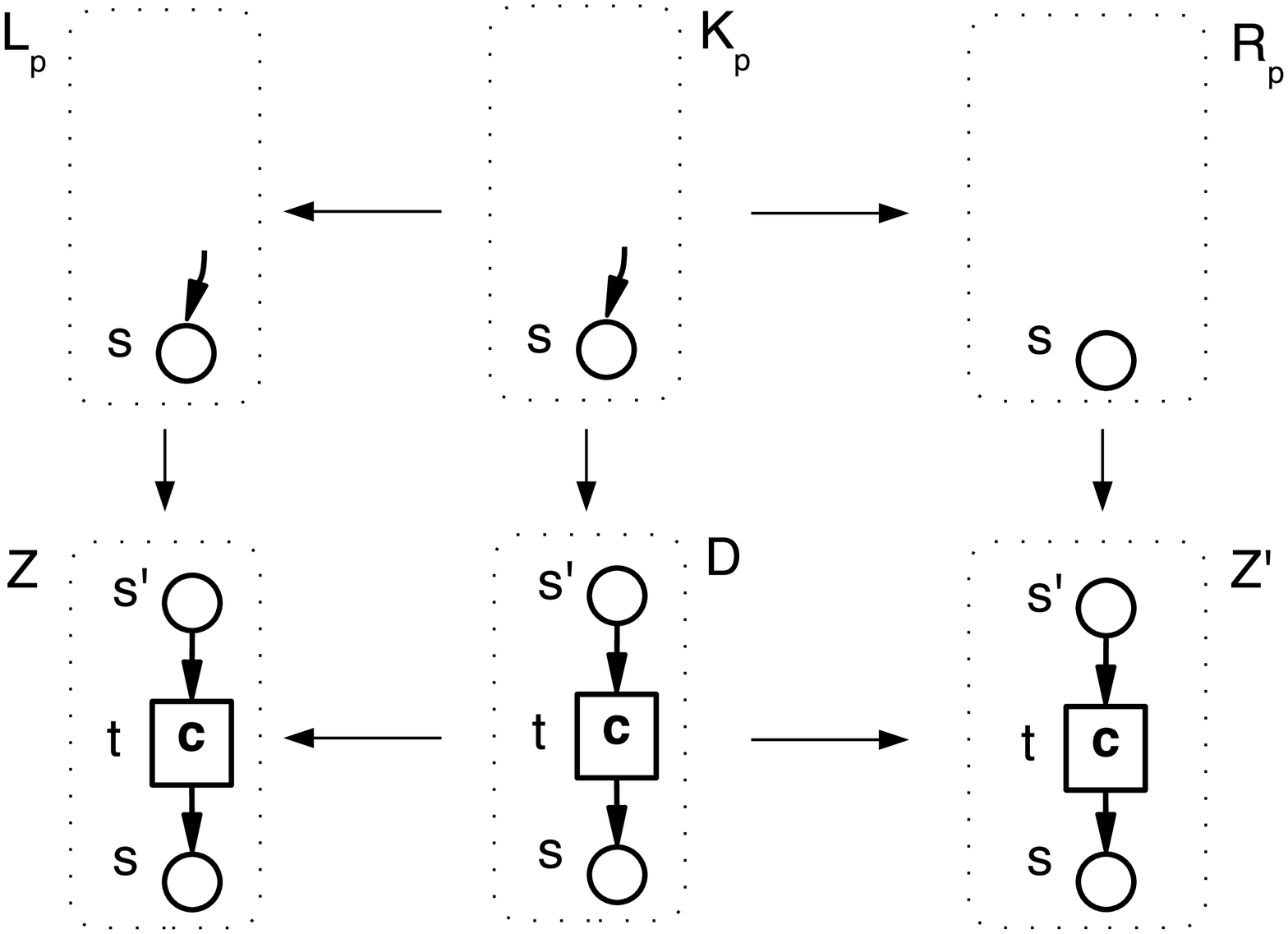}}
    \label{fi:cond5}
  }
  
  \caption{Examples of non-proper matches violating (a) condition~4 
    and (b) condition~5.}

\end{figure}

\begin{defi}[proper match]
  \label{de:proper-match}
  Let $p$ be a rewriting rule over open nets and let $Z$ be an open
  net. A match $m : L_p \to Z$ is called \emph{proper} if it satisfies
  conditions~1, 2, and 3 in Lemma~\ref{le:po-complement} and 

  \begin{enumerate}[(1)]
    \setcounter{enumi}{3}

  \item for any $s \in K_p$,  if $s \in \inp{r_p} - \inp{l_p}$ then $m(l_p(s)) \in O_Z^+$;
    
  \item 
    $r_p(l_p^{-1}(\inp{m})) \subseteq O_{R_p}^+$;
    
  \end{enumerate}
  plus the dual conditions on output places.
\end{defi}
Intuitively, a match is proper if whenever $s \in l_p(\inp{r_p})$,
i.e., the rule $p$ creates a new (ingoing) transition connected to
place $s$, then $m(s)$ is (input) open (condition 4). Additionally,
input (output) places for the match which are preserved by the rule
must be input (output) open in $R_p$.
An example in which condition~4 is violated can be found in~Fig.~\ref{fi:cond4}.
For place $s$ in $K_p$ we have $s \in \inp{r_p}$, since transition $t$ is added in $R_p$, but $s \not\in \inp{l_p}$. Note that the mapping from $D$ to $Z'$ is not a valid open net morphism, since place $s$ in $D$ is not open.
In Fig.~\ref{fi:cond5} instead is condition~5 which is violated. Place
$s$ of $L_p$ is in $\inp{m}$, it is preserved by the rule, but the
corresponding place in $R_p$ is not open. Again we cannot complete the
DPO step since the mapping from $R_p$ to $Z'$ is not a valid open net
morphism (place $s$ should be input open in $R_p$).

We finally arrive at the desired result.

\begin{lem}[applying general rules]
  \label{le:general-transformation}
  Let $p$ be a rule over open nets, let $Z$ be an open net and
  let $m : L_p \to Z$ be a proper match.
  Then there exists a transformation $Z \rew{p,m} Z'$.
\end{lem}

\proof
  Let $p$ be a rule over open nets, let $Z$ be an open net and
  let $m : L_p \to Z$ be a proper match.
  Then, by using Lemma~\ref{le:po-complement} we can construct the minimal
  pushout complement of $l_p$ and $m$, as in Fig.~\ref{fi:open-dpo}(a).
  
  In order to conclude, it suffices to show that $n$ and $r_p$ are
  composable. To this aim observe that by properness of the match:

  \begin{enumerate}[$\bullet$]
    
  \item
    $n(\inp{r_p}) \subseteq O_D^+$ (and the same condition holds for $\out{.}$)\\
    In fact, let $s \in \inp{r_p}$ We distinguish two
    possibilities. If $s \in \inp{l_p}$ then necessarily $n(s) \in
    \inp{d}$ and thus $n(s) \in O_D^+$, since $n$ is an open net
    morphism. If instead, $s \not\in \inp{l_p}$, then $s \in \inp{r_p}
    - \inp{l_p}$, hence, by condition 4 of
    Definition~\ref{de:proper-match}, $m(l_p(s)) \in O_Z^+$. Since
    $m(l_p(s)) = d(n(s))$ and $d$ is an open net morphism, we conclude
    that also in this case $n(s) \in O_D^+$.

  \item
    $r_p(\inp{n}) = r_p(l_p^{-1}(\inp{m})) \subseteq O_{R_p}^+$ (and the same condition holds for $\out{.}$)\\
    Immediate by condition 5 of Definition~\ref{de:proper-match}.\qed
  \end{enumerate}

\subsubsection{Applying Behaviour Preserving Rules.}

Sufficient hypotheses which ensure the applicability of behaviour
preserving rules are made explicit in the following statement. This is
a corollary of the general theory of transformations for open nets
developed before.  

% If Conditions~(a)--(c) below hold, then we can be sure that the
% pushout complement exists, i.e., the dangling condition is satisfied,
% and that the resulting arrow is composable with the right-hand part of
% the rule. Note that these conditions are only sufficient and can be
% simplified here since we consider only behaviour-preserving rules.  It
% is difficult to state sufficient \emph{and} necessary conditions in a
% concise way.

\begin{cor}[applying behaviour preserving rules]
  \label{co:proper}
  Let $p$ be a $\mathsf{x}$-behaviour preserving rule, let $Z$ be an
  open net and let $m : L_p \to Z$ be a match such that:
  \begin{enumerate}[a.]

  \item for all $s \in L_p - l_p(K_p)$ we have $\pre[m(s)] \cup \post[m(s)]
     \subseteq m(L_p - K_p)$;
 
  \item for all $s \in K_p$, if $s \in \inp{l_p}$ and $l_p(s) \in
    O_{L_p}^+$ then $m(l_p(s)) \in O_Z^+$;

  \item for all $s \in K_p$, if $s \in \inp{r_p} - \inp{l_p}$ 
    then $m(l_p(s)) \in O_Z^+$;

  \end{enumerate}
  and the dual of the last two conditions, obtained by replacing 
  $\inp{}$ by $\out{}$ and $+$ by $-$, hold.
  Then, there exists a transformation $Z \rew{p,m} Z'$.
\end{cor}

\begin{proof}
  This is an easy consequence of Lemma~\ref{le:general-transformation}.
  We need to show that conditions (a)-(c) ensure that the match $m$
  is proper, i.e., it satisfies conditions 1--5 of
  Lemma~\ref{le:po-complement} and Definition~\ref{de:proper-match}.
  
  Condition~1 is the same as condition (a), condition~2 is just a
  compact notation for condition (b) and condition~4 is exactly
  condition (c).
  Concerning condition~3, observe that, since $p$ is a behaviour
  preserving rule then $(r_p \circ l_p^{-1})_{|O_{L_p}}$ is a
  correspondence between the left- and right-hand side. 
  This means
  that for any place $s$ in $O_{L_p}^x$ there must be a place $s'$ in
  $K_p$ such that $l_p(s') = s$, and, by definition of open net
  morphism $s'$ must be open, i.e., $s' \in O_{K_p}^x$. Therefore
  $O_{L_p}^x \subseteq l_p(O_{K_p}^x)$ and thus condition~3 is
  trivially satisfied.  
  Similarly, for condition~5, observe that, by definition of open net
  morphisms, $\inp{m} \subseteq O_{L_p}^+$, and, thus
  \begin{quote}
    $r_p(l_p^{-1}(\inp{m})) \subseteq r_p(l_p^{-1}(O_{L_p}^+)) = O_{R_P}^+$.
  \end{quote}
  The last equality is justified by the fact that $p$ is behaviour
  preserving, and thus, as observed above, $(r_p \circ
  l_p^{-1})_{|O_{L_p}}$ is a correspondence between $L_p$ and $R_p$.
\end{proof}

The intuition underlying the conditions above is the following.
Condition~(a) is a typical \emph{dangling condition}, which we have already commented.
% a place can be deleted only if all the transitions connected to this
% place are removed as well, otherwise the flow arcs of this transition
% would remain dangling. 
%
Condition~(b) says that if $s \in \inp{l_p}$, i.e., if some
(ingoing) transitions are deleted from $s$ then the image of $s$ 
in $Z$ must be (input) open if so is its image in $L_p$.
Finally, by condition~(c), if $s \in \inp{r_p}-\inp{l_p}$, i.e., the rule
$p$ creates a new (ingoing) transition connected to place $s$,
without replacing any old one, then the image of $s$ in $Z$
must be (input) open.
%
% Technically, conditions 2 and 3 (and their dual) ensure the existence
% of a \emph{minimal} pushout complement $D$, i.e., a pushout complement
% which embeds into any other, which is the one that we choose to define
% the transformation; the conditions also guarantee the composability of
% $n$ with both $l_p$ and $r_p$.
% %
% The net underlying the minimal pushout complement is $D = Z - m(L_p -
% l_p(K_p))$ (with set difference componentwise on places and
% transitions), and the open places of $D$ are given by $O_D^x =
% d^{-1}(O_Z^x)$ for $x \in \{ +, -\}$.  The initial marking
% $\init{u}_D$ is defined as $\init{u}_D(s) = \init{u}_Z(d(s))$ for any
% place $s \in S_D$.

As an example, consider again the DPO diagram in Fig.~\ref{fi:open-dpo}(b).
It is not difficult to see that the rule and the match
satisfy the conditions of Corollary~\ref{co:proper}. Hence we can
complete the double-pushout construction transforming  
$Z$ into $Z'$, as depicted in the same figure.

\subsection{Modeling Dynamic Reconfigurations of Services}
\ \\

\noindent
Open nets allow us to specify a system as built out of smaller
components. Then, its behaviour is captured by the firing or step
behaviour of the open net. However, for highly dynamic systems, as
mentioned in the introduction, it can be useful to have the
possibility of specifying that, under suitable conditions, some
structural changes or reconfigurations of the system can take place.
For instance
% , a fault in a component could trigger a
% reconfiguration which replaces the component, or 
the invocation of a service
could trigger a rule which provides an implementation of the required service.

\begin{figure}[t]
  \begin{center}
    \scalebox{.38}{\includegraphics{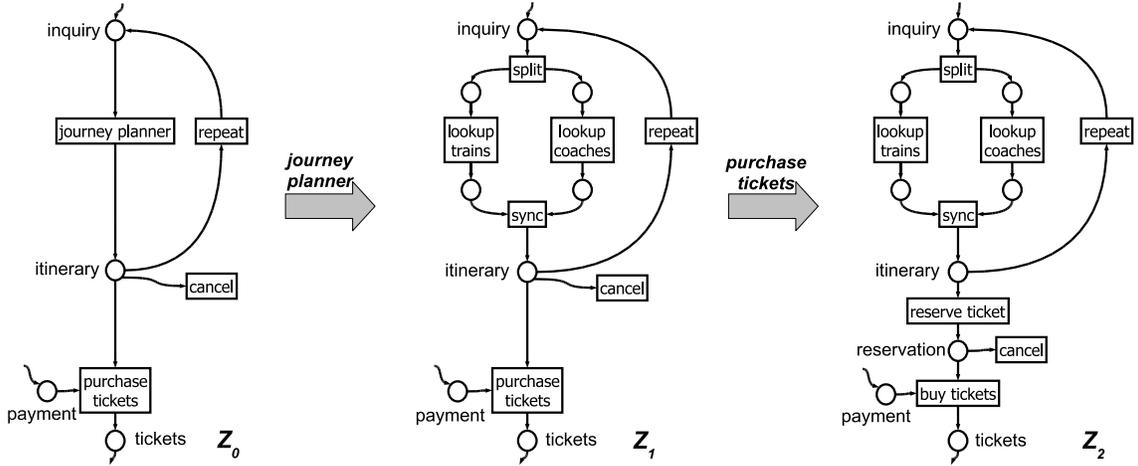}}
  \end{center}

  \caption{Transformation of open nets representing a travel agent's portal.}
  \label{fi:portal}
\end{figure}

The theory of open net reconfigurations can do the job.
% In particular, according to
% Theorem~\ref{th:behaviour-preserving-reconfigurations}, if the left-
% and right-hand sides of a rule are bisimilar, a reconfiguration using
% such rule  ``preserves'' the observable behaviour of the system.
As an example, consider net $Z_0$ in  Fig.~\ref{fi:portal}
which models the view of a traveller on the journey planning and
ticket purchase services offered through a travel agency portal.

\begin{figure}[t]
  \begin{center}
    \scalebox{.4}{\includegraphics{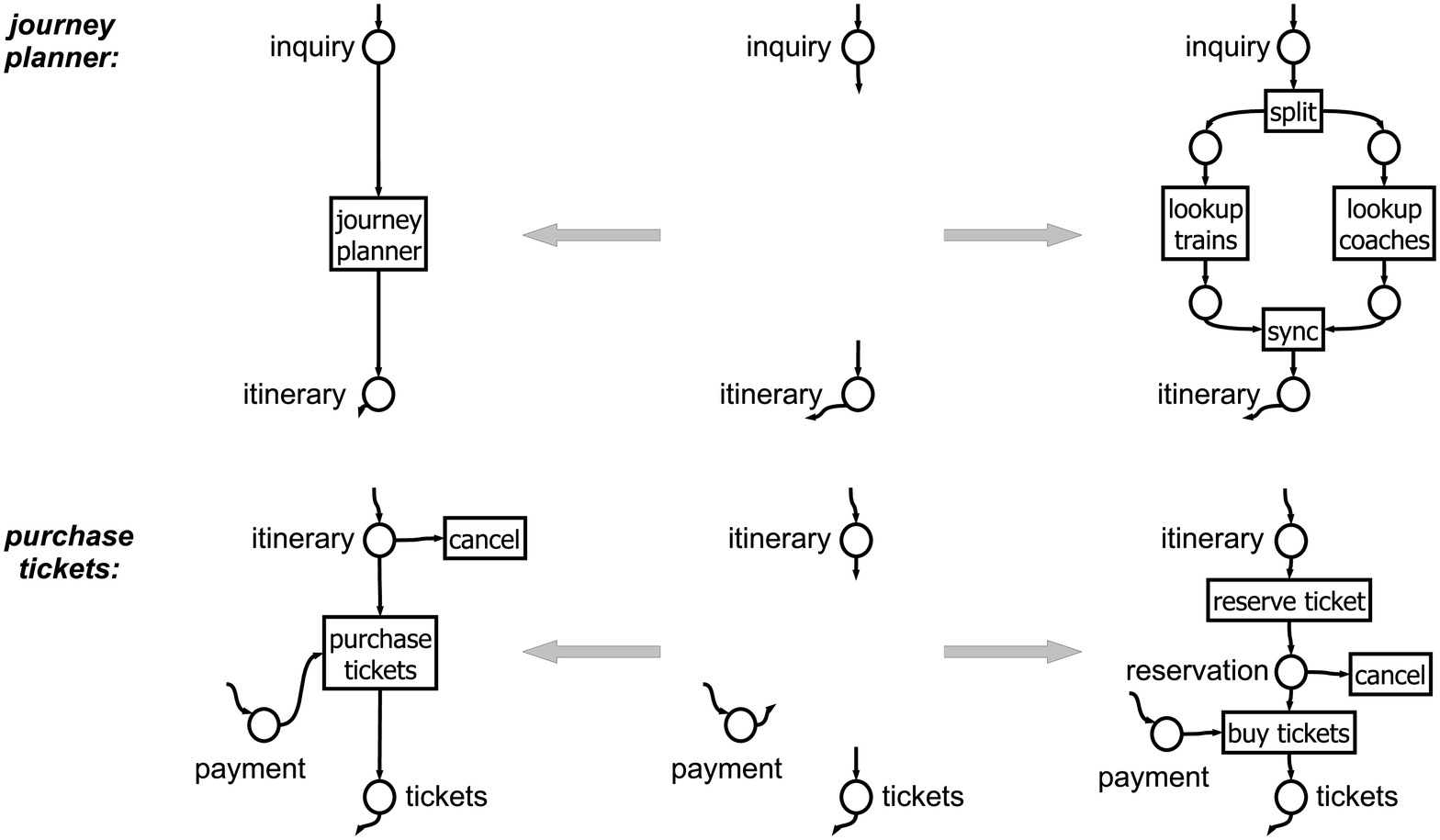}}
  \end{center}

  \caption{Rules}
  \label{fi:rules}
\end{figure}

We distinguish \emph{abstract transitions} representing services that
should be provided elsewhere and \emph{concrete transitions}
representing local services and control flow actions. The invocation
of an external service can be seen at different levels of
abstraction. From the point of view of the client process it is just
the firing of an abstract transition. At a lower level of abstraction,
%the execution of the service
it is captured by a rule such as the one at
the top of Fig.~\ref{fi:rules}. An application of this rule, replacing
the abstract transition by a new open net, models the discovery and
binding of the concrete services required.
The left- and right-hand sides of the rule are weakly firing (actually,
also step) bisimilar if we observe only the interactions at the open
(interface) places, i.e., if we take $\Lambda_\tau = \Lambda$. This
can be seen as a proof of the fact that the bound service meets the
requirements: both in the abstract transition and in its concrete
counterpart any inquiry will produce a corresponding itinerary.

The rule at the bottom of Fig.~\ref{fi:rules} represents a case where
a simple pattern is replaced by a richer one. On the left we say that,
given an itinerary, we can either purchase the required tickets or
cancel the processes. On the right the transaction is refined, adding
a prior reservation phase, while keeping the option to cancel. As
above, the rule has weakly firing (and step) bisimilar left- and right-hand
sides, ensuring that the visible effect of the abstract and concrete
transitions at the interfaces is the same.

A possible sequence of transformations is shown in
Fig.~\ref{fi:portal}.
By Theorem~\ref{th:behaviour-preserving-reconfigurations}, we are sure
that the transformations do not change the observable behaviour
of the system, i.e., the start and end nets are weakly bisimilar, a
fact that can be interpreted as a proof of conformance of the provided
service with respect to the abstract specification.

%\begin{figure}[t]
%  \begin{center}
%    \scalebox{.35}{\includegraphics{trafo.eps}}
%  \end{center}
%
%  \caption{Trafo.}
%  \label{fi:trafo}
%\end{figure}

% As they are given, rules could be applied either statically,
% assembling the process at design time, or dynamically at runtime.
% Notice that we could also enrich the rules by markings. For example,
% if rule \emph{journey planner} was to contain a token at the inquiry place
% in its left- and right-hand sides, the binding of the concrete
% service, modeled by the application of the rule, 
% could only happen on demand at runtime.

We have shown only a small example application, however, we believe
that this technique can be applied to larger case studies, such as the
banking scenario studied in~\cite{bbcg:web-services-equ}. In order to do this
automatically, it would be necessary to implement mechanized
bisimulation checking procedures. For finite state spaces, this is
quite straightforward, for infinite state spaces we could resort to
the techniques presented in~\cite{h:proving-up-to}. In any case the
up-to technique presented in Section~\ref{sec:proof-techniques} will
be very useful for practical case studies.

Another relevant question is the following: which kind of 
bisimilarity should be used? While strong firing bisimilarity
is conceptually the simplest behavioural equivalence, practical examples
usually require weak bisimulations in order to abstract from internal
or silent
moves. Finally, step bisimulation is able to distinguish processes
that differ with respect to the degree of concurrency. This can be
relevant if the observer is able to distinguish different degrees of
parallelism or if we take into account efficiency questions.

\section{Conclusions and Related Work}
\label{se:conclusion}

Open nets, introduced in~\cite{BCEH:CMRS,BCEH:CSOP}, are
a reactive extension of standard Petri nets which allows to model
systems interacting with an unspecified environment. 

As mentioned in the introduction there is a vast related literature. A
close conceptual relationship exists with the early studies on modular
construction and refinement techniques (see,
e.g.,~\cite{v:petri-stepwise-refine,SM83,Mul85,Vog87}) and on
composition operators and compositional semantics for Petri
nets (see, e.g.,~\cite{And83,Ber87,Bau88,Vos87}).
The last class comprises also the algebraic approaches to Petri nets
which view the class of Petri nets as a category and, characterising
the semantics of interest as a universal constructions, automatically
deduce the compositionality for suitably defined
operators~\cite{Win:ES,Win:PNAM,MM:PNM}.

% instance, the
% paper~\cite{Win:ES} proposes a category of Petri nets and defines an
% unfolding semantics, characterised as a right adjoint. As a
% consequence the semantics is compositional with respect to operations
% on nets defined in terms of categorical limits. Notably, among such
% operations we can find a form of parallel composition with
% synchronisation~\cite{Win:PNAM}, which, roughly speaking, join two
% nets by forcing the synchronisation of transitions with the same
% label. The same paper studies also an operation of composition based
% on colimits, which, in particular, as it happens in our approach,
% allows one to merge two nets along a chosen subnet.
% %
% The algebraic view is pushed forward in another seminal
% paper~\cite{MM:PNM}, where the deterministic process
% semantics (in the sense of Best-Devillers) is characterised as the
% free algebra (up to suitable axioms) over such a signature.
% %
% Being obtained as a free construction, which in categorical terms
% provides a left adjoint, in this case the semantics is compositional
% with respect to operations defined in terms of colimits.

More recent approaches, which focus more explicitly on the definition
of notion of module and interface and where the reactive aspects are
taken into account in the semantics can be classified roughly into two
classes.
Some approaches aim at defining a ``calculus of nets'', where a set
of process algebra-like operators allow one to build complex nets
starting from a set of predefined basic components. In this family,
the papers~\cite{NPS:CBCP,PW:UATC} propose an algebra of (labelled)
Petri nets with interfaces, consisting of public (input) places and
(output) transitions, with operators which allow e.g., to add new
transitions and places, to connect existing public transitions and
places by new arcs, to hide items in the net.
We also recall the Petri Box
calculus~\cite{BDH:BCCA,KEB:OSPB,KB:ODSBC}, where a special class of
safe nets, called \emph{plain boxes}, provides the basic components,
which are then combined by means of (refinement-based) composition
operators.
Another family of approaches can be classified as
``component-oriented'': the emphasis, rather than on the algebraic
aspects, is put on the mechanisms which allow one to build larger
systems by combining nets with clearly identified interfaces.
For instance the book~\cite{v:modular-petri} proposes a technique for
inserting a net, called daughter net, into a so-called host net. The
composition is realised by joining the two nets along a predefined set
of places, playing the role of open places. The distinction between
input and output open places, absent in~\cite{v:modular-petri},
instead is later considered in~\cite{v:efficiency-asynchronous}. A
compositionality result is proved for language equivalence and a
notion of bisimilarity, very close to ours, is defined.
Interestingly, the same book also focuses on an alternative approach to
net composition, based on an operation of synchronised parallel
product in the style of~\cite{Win:PNAM}. Such operation, roughly
speaking, joins two nets by forcing the synchronisation of transitions
with the same label.
Other members of the ``component-oriented'' family are, for example,
the \emph{Petri net components}~\cite{Kin:CPOS} and the nets with
\emph{pins}~\cite{Bas:phd}.
We also recall \emph{workflow nets}~\cite{Aal:APNW} which have been
proposed as a formal model for the description of workflows,
i.e., business processes specified in terms of tasks and shared
resources. Workflow nets are special Petri nets satisfying suitable
conditions, like the existence of one initial and one final place:
tokens in such places characterise the start and the end,
respectively, of the represented process. The model has been extended
for the specification of \emph{interorganisational
  workflows}~\cite{Aal:IWAB}, represented as a set of workflow nets
connected through additional places for asynchronous communication and
synchronisation requirements on transitions.
Additional references, as well as a detailed comparison between the
approaches to Petri net composition and reactivity just cited and the
open net model can be found in~\cite{BCEH:CSOP}.

In this paper, firstly we have generalised the theory of open nets, 
including the characterisation of net composition using pushouts, to
the case of marked nets. 
Next we have introduced several natural notions of bisimilarity over
open nets, showing that weak bisimilarities, arising in the presence
of unobservable actions, and, as a particular case, also strong
bisimilarities are congruences with respect to the colimit-based
composition operation over open nets.
The considered notions of bisimilarity differ for the choice of the
observations. These can be single firings, thus leading to what we
called firing bisimilarity, a standard notion of interleaving
equivalence, capable of capturing the branching structure of
computations. Alternatively, we can observe parallel steps, thus
obtaining step bisimilarity, which allows to capture, to some extent,
the degree of parallelism that is possible in a component. This can be
useful, e.g., when a component is replaced by another one since we
might be interested in taking a replacement that exhibits at least the
same concurrent behaviour and is hence equally efficient.

In recent years, reactive extensions of Petri nets have been obtained
by exploiting a general theory of reactive systems developed for
automatically deriving bisimulation congruences. Specifically, an
encoding of Petri nets as bigraphical reactive systems has been
proposed in~\cite{Mil:BFPN}, while~\cite{SS:CPN} proposes an encoding
of nets as reactive systems in the cospan category over an adhesive category.
Our results about strong firing bisimilarity can be seen as a
generalisation of those in~\cite{Mil:BFPN,SS:CPN}, which essentially
are developed for a special kind of open nets, where there is no
distinction between  input  and output open places. Furthermore the
composition operation used in the cited papers does not allow
synchronisation of transitions (technically, the interface net does
not contain transitions).

Concerning weak step bisimilarity, some connections seem to exist with
the work on action refinement, which goes back to
\cite{v:petri-stepwise-refine}. For example, in~\cite{Vog:BAR} (weak) step
bisimilarity is shown to be a congruence with respect to a refinement
operation which allows to replace a single event with a deterministic
finite event structure.
Although the setting is different and a direct comparison is not
possible, we observe that, compared to refinement-based approaches,
where single transitions are refined by a subnet, the theory presented
here works for general reconfigurations, in which both the left- and
right-hand sides can be general, arbitrarily large nets.

Weak (step) bisimilarity for Petri nets is studied also
in~\cite{NPS:CBCP}. They observe that such an equivalence is not a
congruence in general, but for Petri nets satisfying a suitable condition on
the labelling of the public transitions (\emph{well-labelled} nets), a
context closure allows one to get a congruence which is then
characterised by means of a universal context.
The setting is
different from ours since the issue of net composition is tackled at a
finer level of granularity: the basic components of a net are assumed
to be transitions with empty pre- and post-set and single places,
which are then combined by means of constructors that allow one to
connect places and transitions.
Still it would be interesting to understand if a formal relation can be
established, e.g., trying to internalise the pushout-based
composition operation in the algebra of connectors of~\cite{NPS:CBCP}.

Similarities exist also with the problem studied in~\cite{BBCG:BCWS}, where
a reactive Petri net model which admits a compositional behavioural
equivalence is exploited, in the framework of web-services, to provide
a theoretical basis to service composition and discovery. This
technique is then used in a case study for checking the correctness of
service specifications and the replaceability of services in a banking
scenario~\cite{bbcg:web-services-equ}.
Disregarding the technical differences, such as the fact that the
mentioned paper deals with C/E nets and the use of read arcs, the kind
of nets of interest for this approach are essentially a subclass of
open Petri nets, satisfying some structural requirements (all labels
are invisible and the interface consists of a single input and a
single output place, plus some read places).
Generally speaking, compositional Petri net models appears to be
promising as a formalism for the specification of control and
composition in service oriented architectures as suggested, e.g.,
in~\cite{BB:PNBMWS,Mar:AWS,AH:YAWL,MRS:OGA}.
Investigating possible applications of (reconfigurable) open Petri
nets, along the lines of the presented example, in the setting of
web-service specification and analysis represent a stimulating
direction of future research.

In the second part of the paper we have proposed a rewriting-based
framework for Petri nets with reconfigurations. We have shown how our
congruence results
% for the observational semantics 
can be used to identify classes of reconfigurations which do not alter
the observational behaviour of the system.
This is applied to a small case study of a workflow-like model of a
travel agency, where we showed how abstract services can be replaced
by more concrete implementations and how we can ensure that the
behaviour of the full net is preserved under such operations.

Action refinement of Petri nets (see,
e.g.,~\cite{v:petri-stepwise-refine,SM83,Mul85,Vog87}), that we
already mentioned above, can be seen as a special form of
reconfiguration.
The idea of using rewriting techniques for providing a reconfiguration
mechanism for Petri nets has been already explored in the literature
(see, e.g., reconfigurable nets of~\cite{BLO:MCSRN,LO:ISDC} and
high-level replacement systems applied to Petri nets
in~\cite{PER:HLRA}).
In this approaches, however, the emphasis is more on rewriting as a
computational mechanism, rather than on the study of the way the
behaviour of the system is affected by the reconfigurations.
In future work, besides deepening the relationships between these
approaches and ours, we will continue studying the notion of
reconfigurable open nets and describe in more detail how
reconfigurations can be triggered by the net itself, for example by
reaching certain markings or by firing certain transitions, following
an intuition similar to that of dynamic nets~\cite{BS:HLPNTTJC}.

Finally, it would be worth studying whether a formal duality 
can be established between our morphisms and standard simulation
morphisms for Petri nets. Viewing our morphisms as inverses of
(partial) simulation morphisms would allow to get a precise
correspondence between our pushout-based composition and
pullback-based synchronisation of Petri nets.
Surely by simply taking Winskel's morphisms~\cite{Win:ES} this does
not work (technically because when they are undefined on a transition
they must be undefined on the corresponding pre- and post-set). Also
more general morphisms for Petri nets, like those proposed
in~\cite{Vog:ENPS,BB:GMPN}, would not provide an immediate solution.
Still, it looks feasible to identify generalisations of such morphisms
to the context of open Petri nets allowing to develop a dual
theory based on simulations.

\medskip

\noindent\textbf{Acknowledgement:} We would like to thank the 
referees for their insightful and detailed comments.

\bibliographystyle{plain}
%\bibliography{Personal,Biblio,MyPetri}
\bibliography{LMCS}

\begin{thebibliography}{10}

\bibitem{And83}
C.~Andr\'e.
\newblock The behaviour of a {P}etri net on a subset of transitions.
\newblock {\em RAIRO}, 17:5--21, 1983.

\bibitem{BLO:MCSRN}
E.~Badouel, M.~Llorens, and J.~Oliver.
\newblock Modeling concurrent systems: Reconfigurable nets.
\newblock In H.~R. Arabnia and Y.~Mun, editors, {\em Proceedings of PDPTA'03},
  volume~4, pages 1568--1574. CSREA Press, 2003.

\bibitem{BCEH:CMRS}
P.~Baldan, A.~Corradini, H.~Ehrig, and R.~Heckel.
\newblock Compositional modeling of reactive systems using open nets.
\newblock In K.G. Larsen and M.~Nielsen, editors, {\em Proceedings of
  CONCUR'01}, volume 2154 of {\em LNCS}, pages 502--518. Springer Verlag, 2001.

\bibitem{BCEH:CSOP}
P.~Baldan, A.~Corradini, H.~Ehrig, and R.~Heckel.
\newblock Compositional semantics for open {P}etri nets based on deterministic
  processes.
\newblock {\em MSCS}, 15(1):1--35, 2005.

\bibitem{Bas:phd}
T.~Basten.
\newblock {\em In terms of nets: {S}ystem design with {Petri} nets and process
  algebra}.
\newblock PhD thesis, Eindhoven University of Technology, 1998.

\bibitem{Bau88}
B.~Baumgarten.
\newblock On internal and external characterisation of {PT}-net building block
  behaviour.
\newblock In G.~Rozenberg, editor, {\em Advances in Petri nets}, volume 340 of
  {\em LNCS}, pages 44--61. Springer, 1988.

\bibitem{BB:GMPN}
M.~A. Bednarczyk and A.~M. Borzyszkowski.
\newblock General morphisms of {Petri} nets (extended abstract).
\newblock In J.~Wiedermann, P.~van Emde~Boas, and M.~Nielsen, editors, {\em
  Proceedings of ICALP'99}, volume 1644, pages 190--199. Springer Verlag, 1999.

\bibitem{BB:PNBMWS}
B.~Benatallah and R.~Hamadi.
\newblock A {P}etri net-based model for {W}eb service composition.
\newblock In K.-D. Schewe and X.~Zhou, editors, {\em Australasian Database
  Conference, Conferences in Research and Practice in Information Technology},
  volume~7, pages 191--200. Australian Computer Society, 2003.

\bibitem{Ber87}
G.~Berthelot.
\newblock Transformations and decompositions of nets.
\newblock In W.~Brauer, editor, {\em Petri Nets: Central models and their
  properties}, volume 254 of {\em LNCS}, pages 359--376. Springer, 1987.

\bibitem{BDH:BCCA}
E.~Best, R.~Devillers, and J.~G. Hall.
\newblock The {Petri} box calculus: a new causal algebra with multi-label
  communication.
\newblock In G.~Rozenberg, editor, {\em Advances in Petri Nets}, volume 609 of
  {\em LNCS}, pages 21--69. Springer Verlag, 1992.

\bibitem{BBCG:BCWS}
F.~Bonchi, A.~Brogi, S.~Corfini, and F.~Gadducci.
\newblock A behavioural congruence for web services.
\newblock In {\em Proceedings of FSEN '07}, volume 4727 of {\em LNCS}, pages
  240--256. Springer, 2007.

\bibitem{bbcg:web-services-equ}
F.~Bonchi, A.~Brogi, S.~Corfini, and F.~Gadducci.
\newblock Compositional specification of web services via behavioural
  equivalence of nets: A case study.
\newblock In {\em Proc. of Petri Nets '08}, pages 52--71, 2008.

\bibitem{BS:HLPNTTJC}
M.G. Buscemi and V.~Sassone.
\newblock High-level {Petri} nets as type theories in the join calculus.
\newblock In {\em Proceedings of FoSSaCS'01}, volume 2030 of {\em LNCS}, pages
  104--120. Springer, 2001.

\bibitem{Ehr:TIAA}
H.~Ehrig.
\newblock Tutorial introduction to the algebraic approach of graph-grammars.
\newblock In H.~Ehrig, M.~Nagl, G.~Rozenberg, and A.~Rosenfeld, editors, {\em
  Proceedings of the 3rd International Workshop on Graph-Grammars and Their
  Application to Computer Science}, volume 291 of {\em LNCS}, pages 3--14.
  Springer Verlag, 1987.

\bibitem{h:proving-up-to}
D.~Hirschkoff.
\newblock Automatically proving up to bisimulation.
\newblock In {\em Proc. of MFCS '98 Workshop on Concurrency}, number~18 in
  ENTCS, 1998.

\bibitem{Kin:CPOS}
E.~Kindler.
\newblock A compositional partial order semantics for {P}etri net components.
\newblock In P.~Azema and G.~Balbo, editors, {\em Application and Theory of
  Petri Nets}, volume 1248 of {\em LNCS}, pages 235--252. Springer Verlag,
  1997.

\bibitem{KB:ODSBC}
M.~Koutny and E.~Best.
\newblock Operational and denotational semantics for the box algebra.
\newblock {\em Theoretical Computer Science}, 211(1--2):1--83, 1999.

\bibitem{KEB:OSPB}
M.~Koutny, J.~Esparza, and E.~Best.
\newblock Operational semantics for the {Petri} box calculus.
\newblock In B.~Jonsson and J.~Parrow, editors, {\em Proceedings of
  CONCUR~'94}, volume 836 of {\em LNCS}, pages 210--225. Springer Verlag, 1994.

\bibitem{LS:AQC}
S.~Lack and P.~Soboci\'{n}ski.
\newblock Adhesive and quasiadhesive categories.
\newblock {\em RAIRO -- Theoretical Informatics and Applications},
  39(3):511--555, 2005.

\bibitem{LM:DBCRS}
J.~Leifer and R.~Milner.
\newblock Deriving bisimulation congruences for reactive systems.
\newblock In C.~Palamidessi, editor, {\em Proceedings of CONCUR'00}, volume
  1877 of {\em LNCS}, pages 243--258. Springer Verlag, 2000.

\bibitem{LO:ISDC}
M.~Llorens and J.~Oliver.
\newblock Introducing structural dynamic changes in {Petri} nets:
  Marked-controlled reconfigurable nets.
\newblock In F.~Wang, editor, {\em Proceedings of ATVA'04}, volume 3299, pages
  310--323. Springer Verlag, 2004.

\bibitem{Mar:AWS}
A.~Martens.
\newblock Analyzing {W}eb service based business processes.
\newblock In M.~Cerioli, editor, {\em Proceedings of FASE'05}, volume 3442 of
  {\em LNCS}, pages 19--33. Springer, 2005.

\bibitem{MRS:OGA}
P.~Massuthe, W.~Reisig, and K.~Schmidt.
\newblock {An Operating Guideline Approach to the SOA}.
\newblock {\em Annals of Mathematics, Computing \& Teleinformatics},
  1(3):35--43, 2005.

\bibitem{MM:PNM}
J.~Meseguer and U.~Montanari.
\newblock Petri nets are monoids.
\newblock {\em Information and Computation}, 88:105--155, 1990.

\bibitem{Mil:CCS}
R.~Milner.
\newblock {\em A {C}alculus of {C}ommunicating {S}ystems}, volume~92 of {\em
  LNCS}.
\newblock Springer Verlag, 1980.

\bibitem{Mil:BRS}
R.~Milner.
\newblock Bigraphical reactive systems.
\newblock In K.~G. Larsen and M.~Nielsen, editors, {\em Proceedings of
  CONCUR'01}, volume 2154 of {\em LNCS}, pages 16--35. Springer Verlag, 2001.

\bibitem{Mil:BFPN}
R.~Milner.
\newblock Bigraphs for {Petri} nets.
\newblock In J.~Desel, W.~Reisig, and G.~Rozenberg, editors, {\em Lectures on
  Concurrency and Petri Nets}, volume 3098 of {\em LNCS}, pages 686--701.
  Springer, 2003.

\bibitem{Mul85}
K.~M\"uller.
\newblock Constructable {P}etri nets.
\newblock {\em Elektr. Inf. Kybern.}, 21:171--199, 1985.

\bibitem{NPS:CBCP}
M.~Nielsen, L.~Priese, and V.~Sassone.
\newblock Characterizing {B}ehavioural {C}ongruences for {P}etri {N}ets.
\newblock In {\em Proceedings of CONCUR'95}, volume 962 of {\em LNCS}, pages
  175--189. Springer Verlag, 1995.

\bibitem{NT:DNDC}
M.~Nielsen and P.~S. Thiagarajan.
\newblock Degrees of non-determinism and concurrency: A {P}etri net view.
\newblock In M.~Joseph and R.~Shyamasunda, editors, {\em Proceedings of
  FSTTCS'84}, pages 89--117. Springer Verlag, 1984.

\bibitem{PER:HLRA}
J.~Padberg, H.~Ehrig, and L.~Ribeiro.
\newblock High level replacement systems applied to algebraic high level net
  transformation systems.
\newblock {\em Mathematical Structures in Computer Science}, 5(2):217--256,
  1995.

\bibitem{p:basic-category-theory}
B.~Pierce.
\newblock {\em Basic Category Theory for Computer Scientists (Foundations of
  Computing)}.
\newblock MIT Press, 1991.

\bibitem{PW:UATC}
L.~Priese and H.~Wimmel.
\newblock A uniform approach to true-concurrency and interleaving semantics for
  {Petri} nets.
\newblock {\em Theoretical Computer Science}, 206(1--2):219--256, 1998.

\bibitem{Rei:PNI}
W.~Reisig.
\newblock {\em Petri {N}ets: {A}n {I}ntroduction}.
\newblock EATCS Monographs on Theoretical Computer Science. Springer Verlag,
  1985.

\bibitem{SS:CPN}
V.~Sassone and P.~Sobocinski.
\newblock A congruence for {P}etri nets.
\newblock In {\em Proceedings of PNGT'04}, volume 127(2) of {\em Electronic
  Notes in Theoretical Computer Science}, pages 107--120. Elsevier Science,
  2005.

\bibitem{SM83}
I.~Suzuki and T.~Murata.
\newblock A method for stepwise refinement and abstraction of {Petri} nets.
\newblock {\em Journal of computer and system sciences}, 27:51--76, 1983.

\bibitem{v:petri-stepwise-refine}
R.~Valette.
\newblock Analysis of {Petri} nets by stepwise refinements.
\newblock {\em Journal of Computer and System Sciences}, 18(1):35--46, 1979.

\bibitem{Aal:APNW}
W.~van~der Aalst.
\newblock The application of {P}etri nets to workflow management.
\newblock {\em The Journal of Circuits, Systems and Computers}, 8(1):21--66,
  1998.

\bibitem{Aal:IWAB}
W.~van~der Aalst.
\newblock Interorganizational workflows: An approach based on message sequence
  charts and {P}etri nets.
\newblock {\em System Analysis and Modeling}, 34(3):335--367, 1999.

\bibitem{AH:YAWL}
W.M.P. van~der Aalst and A.H.M. ter Hofstede.
\newblock Yawl: yet another workflow language.
\newblock {\em Information Systems}, 30(4):245--275, 2005.

\bibitem{Vog87}
W.~Vogler.
\newblock Behaviour preserving refinement of {P}etri nets.
\newblock In G.~Tinhofer and G.~Schmidt, editors, {\em Proceedings of WG'86:
  Graph theoretic concepts in computer sciencea}, volume 246 of {\em LNCS},
  pages 82--93. Springer, 1987.

\bibitem{Vog:ENPS}
W.~Vogler.
\newblock Executions: A new partial-order semantics of {P}etri nets.
\newblock {\em Theoretical Computer Science}, 91(2):205--238, 1991.

\bibitem{v:modular-petri}
W.~Vogler.
\newblock {\em Modular Construction and Partial Order Semantics of Petri Nets},
  volume 625 of {\em LNCS}.
\newblock Springer, 1992.

\bibitem{Vog:BAR}
W.~Vogler.
\newblock Bisimulation and action refinement.
\newblock {\em Theoretical Computer Science}, 114(1):173--200, 1993.

\bibitem{v:efficiency-asynchronous}
W.~Vogler.
\newblock Efficiency of asynchronous systems that communicate asynchronously.
\newblock In {\em Proc. of ICATPN '00}, volume 1825 of {\em LNCS}, pages
  424--444, 2000.

\bibitem{Vos87}
K.~Voss.
\newblock Interface as a basic concept for system specification.
\newblock In K.~Voss, editor, {\em Concurrency and Nets}, pages 585--604.
  Springer, 1987.

\bibitem{Win:ES}
G.~Winskel.
\newblock Event {S}tructures.
\newblock In {\em Petri Nets: Applications and Relationships to Other Models of
  Concurrency}, volume 255 of {\em LNCS}, pages 325--392. Springer Verlag,
  1987.

\bibitem{Win:PNAM}
G.~Winskel.
\newblock Petri nets, algebras, morphisms, and compositionality.
\newblock {\em Information and Computation}, 72(3):197--238, 1987.

\end{thebibliography}

\end{document}